\pdfoutput=1
\documentclass[structabstract]{aa}  
\usepackage{amsmath,amssymb}
\usepackage{graphicx}
\usepackage{txfonts}
	\usepackage{natbib}           
        \bibpunct{(}{)}{;}{a}{}{,} 
	\usepackage{graphicx}         
	\usepackage{lscape}           
        \usepackage{subfig}           
  	\usepackage{color}            
	\usepackage{longtable}
	\usepackage{multirow}
	\usepackage[version=3]{mhchem}
        \usepackage{verbatim}
        \usepackage{aalongtable} 
        \usepackage{hyperref}
        \usepackage[normalem]{ulem}
        
\begin{document}
   \title{Molecular line survey of the high-mass star-forming region NGC~6334I with Herschel/HIFI and the SMA\thanks{{\it Herschel} is an ESA space observatory with science instruments provided by European-led Principal Investigator consortia and with important participation from NASA.}}

   \author{A. Zernickel\inst{1}, P. Schilke\inst{1}, A. Schmiedeke\inst{1}, D. C. Lis\inst{3}, C. L. Brogan\inst{4}, C. Ceccarelli\inst{5}, \\ 
          C. Comito\inst{2}, M. Emprechtinger\inst{3}, T. R. Hunter\inst{4}, and T. M\"oller\inst{1}}

   \institute{I. Physikalisches Institut der Universit\"at zu K\"oln, Z\"ulpicher Stra\ss e 77, 50937 K\"oln, Germany\\
              \email{zernickel@ph1.uni-koeln.de}
         \and
             Max-Planck-Institut f\"ur Radioastronomie, Auf dem H\"ugel 69, 53121 Bonn, Germany
         \and
             California Institute of Technology, MC 301-17, Pasadena, CA 91125 USA
         \and
             NRAO, 520 Edgemont Road, Charlottesville, VA 22903, USA
         \and
            IPAG, Universit\'e Joseph Fourier, CNRS, BP 53, 38041 Grenoble Cedex 9, France
             }

   \date{Received 12 June 2012 / Accepted 25 August 2012}

 
  \abstract
  {} 
   {We aim at deriving the molecular abundances and temperatures of the hot molecular cores in the high-mass star-forming region NGC~6334I and consequently deriving their physical and astrochemical conditions.}
   {In the framework of the Herschel guaranteed time key program CHESS (Chemical Herschel Surveys of Star Forming Regions), NGC~6334I is investigated by using the Heterodyne Instrument for the Far-Infrared (HIFI) aboard the Herschel Space Observatory. A spectral line survey is carried out in the frequency range 480--1907~GHz, and further auxiliary interferometric data from the Submillimeter Array (SMA) in the 230~GHz band provide spatial information for disentangling the different physical components contributing to the HIFI spectrum. The spectral lines in the processed Herschel data are identified with the aid of former surveys and spectral line catalogs. The observed spectrum is then compared to a simulated synthetic spectrum, assuming local thermal equilibrium, and best fit parameters are derived using a model optimization package.}
    {A total of 46 molecules are identified, with 31 isotopologues, resulting in about 4300 emission and absorption lines. High-energy levels ($E_u>1000$~K) of the dominant emitter methanol and vibrationally excited HCN ($\nu_2=1$) are detected. The number of unidentified lines remains low with 75, or $<$ 2\% of the lines detected. The modeling suggests that several spectral features need two or more components to be fitted properly. Other components could be assigned to cold foreground clouds or to outflows, most visible in the \ce{SiO} and \ce{H2O} emission. A chemical variation between the two embedded hot cores is found, with more N-bearing molecules identified in SMA1 and O-bearing molecules in SMA2.}
   {Spectral line surveys give powerful insights into the study of the interstellar medium. Different molecules trace different physical conditions like the inner hot core, the envelope, the outflows or the cold foreground clouds. The derived molecular abundances provide further constraints for astrochemical models.}

   \keywords{Submillimeter: ISM -- Line: identification -- Astrochemistry -- ISM: molecules -- ISM: abundances -- Stars: formation}

   \titlerunning{Molecular line survey of NGC~6334I}
   \authorrunning{A. Zernickel, P. Schilke, A. Schmiedeke et al.}
   \maketitle
%

\section{Introduction}
\label{intro}
The astrochemical composition of the interstellar gas depends on the physical environment and the evolutionary state. In the cold dense regions molecules are unsaturated, while in contrast saturated molecules are much more abundant in hot cores \citep{herbst}. Unsaturated molecules are carbon rich with long carbon chains, like radicals as \ce{C_n H} (n=2--8)  or cyanopolyynes \ce{HC_n N} (n=3,5,7,9,11), whereas saturated organic molecules are hydrogen rich with single bonds of carbon like \ce{CH3OCH3}. Molecular ions are predominantly unsaturated in the cold phase, because most hydrogenation reactions are endothermic or hindered by potential barriers \citep{herbst2}. Ices on dust grains begin to form in the cold phase and are observed by broad absorption bands at mid-infrared wavelengths. Sequential hydrogenitation of oxygen or nitrogen forms water or ammonia, respectively. Embedded in the water-ices are relevant amounts of CO, and through further hydrogenation \ce{CO -> HCO -> H2CO -> H2COH -> CH3OH}. Time-dependent astrochemical models suggest that abundances of certain molecules, such as \ce{H2S} and \ce{SO2}, might be used as an indicator of the evolutionary phase \citep{Nomura,Herpin}, and greater chemical complexity can be expected for longer evolution timescales \citep{Garrod}. Approximately 165 molecules have been detected in the ISM or circumstellar shells. \footnote{A current list can be found at \url{http://www.astro.uni-koeln.de/cdms/molecules}}

\Citet{Dishoeckhifi} emphasized the importance of spectral line surveys for the research of star-forming regions. Line surveys give the possibility to obtain a census of all atoms and molecules and give insights into their thermal excitation conditions and dynamics by studying line intensities and profiles, which allows seprating different physical components. The discovery of new species or new excited lines from known molecules is possible. Furthermore, the wide range of energy levels and eventually opacities of one species allows a better determination of the temperature and abundance. The uniform recording with the same spectrometer in one observation reduces the calibration uncertainties. 

The giant molecular cloud \object{NGC~6334}, also known as the ``Cat's Paw Nebula'', lies in the Carina-Sagittarius Arm of the Milky Way at a distance of $(1.7 \pm 0.3)$~kpc \citep{distance}.
Studies at infrared wavelengths revealed sites of massive  star formation \citep{persi}. \object{NGC~6334I} was first explored by \citet{emerson} as the brightest infrared source in the northern part of the cloud, containing an OH maser and an ultracompact \ion{H}{II} region. According to \citet{emerson}, the pumping of the OH maser requires an energy source of a massive collapsing protostar ($M>30~M_{\odot}$). \ion{H}{II} regions are associated with high-mass stars (spectral classes O and B) where the UV radiation is strong enough to ionize hydrogen. \citet{Russeil} studied the densest cores in NGC~6334 in molecular line and dust continuum emission. For NGC~6334I, they derive a total mass of $M$=1206~$M_{\odot}$ from the continuum flux density at 1.2~mm, an average density of $n_{\rm H_2}$=$1.2 \times 10^6$~$\rm cm^{-3}$ and an upper limit for the virial mass of $M_{\rm vir}$=525~$M_{\odot}$. The total mass estimate is relatively high, because a low value of the dust temperature of only 20~K was assumed, as compared to the value of T=100~K derived by \citep{Sandell}, which leads to a factor of 6 difference of the mass estimate. While an average dust temperature of T=25~K is appropriate for most cores in the NGC~6334 complex \citep{ngcdust}, the dust temperature is known to be much higher in NGC~6334I.

Interferometric studies revealed that NGC~6334I consists of several cores, labeled SMA1 to SMA4 \citep{Hunter}. The separation distance between SMA1 and SMA2 is about 3.5$''$ or 6000~AU. SMA1 and SMA2 show rich spectra of molecular transitions and SMA3 coincides with the \ion{H}{II} region which exhibits free-free and dust emission and is excited by the near-infrared source IRS 1E \citep{buizer}. SMA4 only shows dust emission but no line emission, which is still unexplained.
Table~\ref{tNGC} summarizes characteristic values for NGC~6334I derived from the dust continuum emission at different wavelengths and, which classifiy the source as a low-luminosity high-mass hot core. The overall size of NGC~6334I is 10$''$ $\times$ 8$''$, where 1$''$ equals to 1700~AU.
\begin{table}
  \centering
  \caption{Mass and temperature estimates for NGC~6334I}
  \label{tNGC}
    \begin{tabular}{cccccccc}
    \hline \hline
     Source & $M (M_{\odot})$ & $T_{\rm d}$ (K) & $\beta$ & n ($\rm cm^{-3}$) & $L_{\rm bol}$ ($L_{\odot}$) \\
     \hline
     Total\tablefootmark{a} & 200 $\pm$ 100 &   100    &   1.7 $\pm$ 0.3  &  $1.2\times 10^7$ &   $2.6 \times 10^5$  \\
    \hline
   SMA1\tablefootmark{b} & 17 $\pm$ 50\% &   100    &     &   &      \\ 
   SMA2\tablefootmark{b} & 11 $\pm$ 50\% &   100    &     &   &       \\
   SMA3\tablefootmark{b} & 33 $\pm$ 50\% &   60    &     &   &      \\
   SMA4\tablefootmark{b} & 36 $\pm$ 50\% &   33    &     &   &      \\
\hline
 \end{tabular}
\tablefoot{Values taken from \tablefoottext{a} \citet{Sandell} and \tablefoottext{b} \citet{Hunter}. $M$ is the mass, $T_d$ the average dust temperature, $\beta$ the dust emissivity index, $n$ the hydrogen density and $L_{\rm bol}$ the bolometric luminosity.}
\end{table}
SMA1 and SMA2 are so-called hot molecular cores, where certain attributes apply according to \citet{cesaroni}: a high temperature ($T\ge$100~K), sizes smaller than 0.1~pc, large masses ($10-1000$~$M_{\odot}$) and luminosities exceeding $10^4$~ $L_{\odot}$. Furthermore they are often associated with water masers and ultracompact \ion{H}{II} regions. Hot cores are heated by high-mass protostars which are embedded in a dusty gas envelope. Their rich molecular spectra originate from the evaporation of ice mantles covering the dust grains, which release at $T\sim$100~K molecules to the gas phase. Molecules released from grain mantles then begin to drive a fast, high-temperature gas-phase chemistry, forming complex organic species like acetone, beginning from precursors like methanol. Their rotational, vibrational and radiative excitation leads to a characteristic emission spectrum. This explains why the spectrum is poorer in earlier stages like the colder protostellar core \object{NGC~6334I(N)} \citep{Brogan,Walsh}, 2$'$ or 1~pc north of NGC~6334I. 

Multiple spectral line surveys have been carried out for NGC~6334I: \citet{McCutcheon} 334--348~GHz, \citet{Schilke} 459--461~GHz and 817--819~GHz, \citet{Thorwirth} 88--115~GHz and 218--267~GHz, \citet{Walsh} 84--116~GHz and the latest from \citet{Kalinina} in the range 81--242~GHz.
Besides, several maser lines have been observed from \ce{NH3} \citep{beuther}, \ce{CH3OH} \citep{walshm} and \ce{H2O} \citep{Migenes} as well as bipolar outflows in CO \citep{outflow,Qiu} and SiO \citep{sio}. This source is part of the CHESS program (Chemical Herschel Surveys of Star Forming Regions), a key program of the Herschel Space Observatory \citep{Ceccarelli}. The aim is to study eight different sites and phases of star formation varying in mass, luminosity, evolutionary state, astrochemical composition etc. by conducting and comparing their spectra. In this article, the entire spectrum (500--1900~GHz) of NGC~6334I is analyzed which was observed by Herschel in 2010. Previously, some small sections of the spectrum were studied to find transitions of new molecules not observed before in the interstellar medium (ISM): Oxidaniumyl (\ce{H2O+}) by \citet{Ossenkopf} and chloronium (\ce{H2Cl+}) by \citet{Lis}. A detailed analysis of water has been made by \citet{water} and of methylidyne (CH) by \citet{Wiel}.

\begin{figure}
 \resizebox{\hsize}{!}{\includegraphics[angle=270,width=1\textwidth]{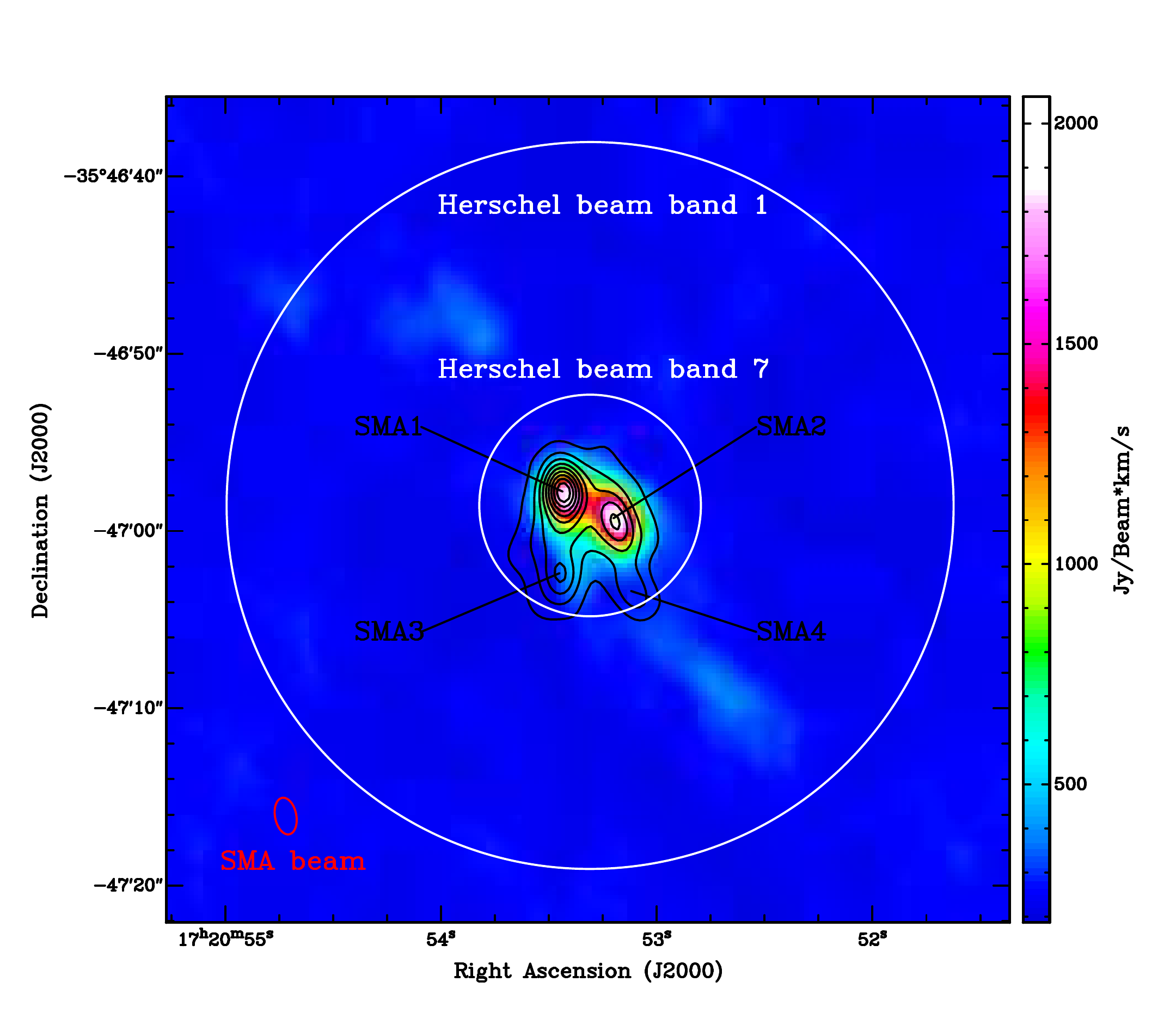}}
 \caption[smaall]{Integrated SMA line intensity map over all species. Overlaid in black contours is the continuum emission. The contour levels are from 90\% to 10\% of the maximum in steps of 10\%. The large and small white circles represent the FWHM Herschel beam size for band 1 and band 7, respectively.}
\label{fsma}
\end{figure}

One difficulty is that if a large beam encompasses a complex, unresolved source structure, the spectrum can become complex due to different overlapping components. The spatial information from the SMA is complementary. The Herschel beam is larger than NGC~6334I so that the source is spatially unresolved, while the SMA beam can resolve the individual cores. The SMA maps thus reveal the source morphology and give insights into the distribution of various molecules. This helps to decompose the contributions from different components present in the HIFI spectrum and to differentiate between the two hot cores SMA1 and SMA2, the extended emission and outflows, see Fig.~\ref{fsma}.
\begin{figure*}
 \resizebox{\hsize}{!}{\includegraphics[trim=5cm 0cm 0cm 0cm,clip,angle=270,width=1\textwidth]{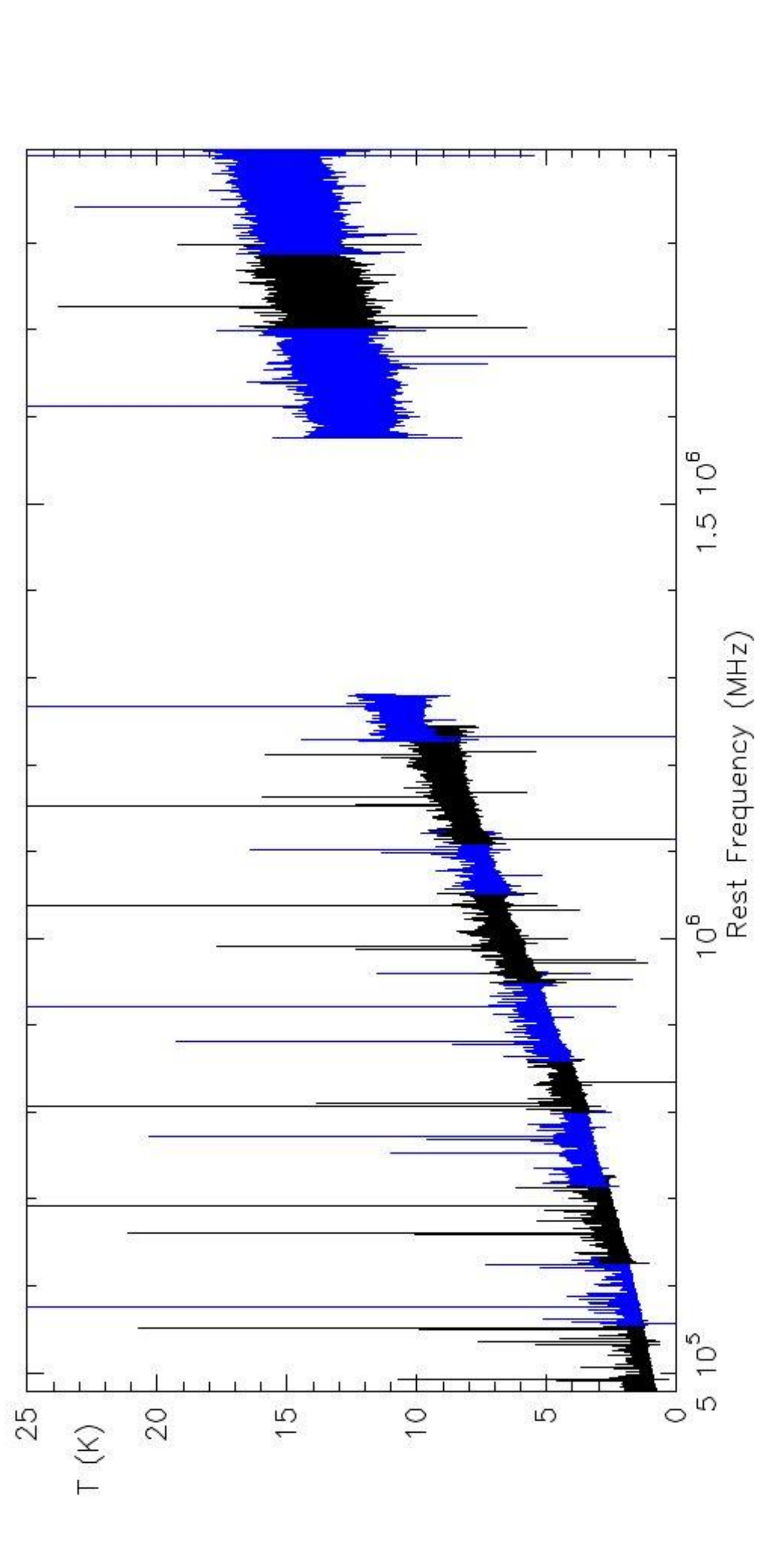}}
  \caption{Complete HIFI spectrum of NGC~6334I. B-bands are coloured in blue and bands 5--7 are smoothed by averaging over 5--10 channels.}
  \label{allbands}
\end{figure*}
\section{Observations}
\subsection{Observations with Herschel}

Observations were conducted with the high spectral resolution receiver system HIFI (Heterodyne Instrument for the Far-Infrared; \citealt{HIFI}) aboard the Herschel Space Observatory \citep{HSO}.
HIFI is divided into seven bands with different receivers which overlap at the band edges. Each band is separated again into two parts a and b using different polarizations and local oscillator chains. The full frequency range is from 480--1907~GHz, with a gap between 1280 and 1410~GHz. The first five bands employ SIS (superconductor-insulator-superconductor) mixers with an intermediate frequency (IF) bandwidth of 4~GHz. Bands 6 and 7 are based on Hot-Electron Bolometer (HEB) mixers and have an IF bandwidth of 2.4~GHz.
The HPBW of the 3.5~m Herschel telescope varies with wavelength from 41$''$ to 12$''$. 
\begin{table}
  \centering
  \caption{Overview over all HIFI bands for the observation of NGC~6334I}
   \label{tHIFI}
    \begin{tabular}{crcccr}
    \hline\hline
    Band  & Freq. range& Beam size  & Date & Obs. time & rms \\
           & (GHz)           &      ($''$)    &  & (h)  & (K) \\
    \hline
    1a    & 480-560 & \multirow{2}{*}{41}  & 28.02.2010 & 3.6   & 0.015 \\
    1b    & 554-636 &       & 02.03.2010 & 2.2   & 0.03 \\
    2a    & 626-726 &  \multirow{2}{*}{29}     & 09.10.2010 & 3.7   & 0.04 \\
    2b    & 714-801 &      & 19.03.2010 & 3.3   & 0.05 \\
    3a    & 799-860 &  \multirow{2}{*}{25}    & 09.10.2010 & 1.8   & 0.05 \\
    3b    & 858-961 &      & 19.03.2010 & 3.9   & 0.07 \\
    4a    & 949-1061&  \multirow{2}{*}{21}     & 09.10.2010 & 4.8   & 0.09 \\
    4b    & 1052-1122&        & 03.03.2010 & 2.9   & 0.10 \\
    5a    & 1108-1244&  \multirow{2}{*}{18.5}     & 13.10.2010 & 4.2   & 0.37 \\
    5b    & 1227-1280&        & 12.03.2011 & 1.7   & 0.37 \\
    6a    & 1410-1575&  \multirow{2}{*}{14.5}    & 14.10.2010 &  0.9     &  \\
    6b    & 1575-1700&        & 14.10.2010 & 3.6   & 1.20 \\
    7a    & 1703-1799&   \multirow{2}{*}{12.5}     & 14.10.2010 & 2.6   & 1.27 \\
    7b    & 1788-1907&        & 13.10.2010 & 3.2   & 1.27 \\
    \hline
    \end{tabular}%
\end{table}%
Table~\ref{tHIFI} gives a log of the HIFI observations of NGC~6334I and Fig.~\ref{allbands} shows the whole calibrated spectrum. The continuum level  increases continuously with frequency, except for band 5b, where a small offset is seen. The line emission in band 1a contributes only 3\% to the total flux because the HPBW at 520~GHz is four times bigger than the source size so that in addition the continuum from the surrounding area is observed. The emission lines with the highest peak intensity, $T>$15~K, are those of CO and \ce{^{13}CO}. The telescope was pointed toward the coordinates $\alpha$(J2000): 17$^{\rm h}$20$^{\rm m}$53.32$^{\rm s}$, $\delta$(J2000): --35\degr46\arcmin58.5\arcsec, between SMA1 and SMA2. 

The observations were carried out mainly in two sessions in March and October 2010, in dual beam switch mode with an internal chopper mirror switching between the source and two reference positions 3$^\prime$ away from the target to provide flat baselines and calibrate the continuum. The  total observing time was $\sim$42~h. The wide band spectrometer provided a spectral resolution of 1.1~MHz, resulting in a velocity resolution of 0.7--0.2~km\,s$^{-1}$. The double sideband spectra were processed with HIPE (Herschel Interactive Processing Environment; \citealt{Ott}) version 5, deconvolved to single-sideband spectra following the algorithms described in \citep{cs}, and exported to CLASS (see Sect. \ref{xclass}) for subsequent analysis.
All spectra are corrected for the beam efficiency and the intensity scale is the main-beam brightness temperature in K. The spectra presented here are equally weighted averages of the horizontal and vertical polarizations, in order to reduce the noise. Band 4a has some sinusoidal baseline features from standing waves. For band 6a only a small portion around a \ce{HCl+} transition was observed, but the data are corrupted and not discussed here. The band coverage is unbiased, meaning that all frequency ranges are included with the band edges where the noise increases. 
Different sideband gains can lead to discrepancies between line intensities in the two bands in the overlap region, see Fig.~\ref{fig:n2h+} for an example in the $J=6-5$ transition of \ce{N2H+} at 558.966~GHz. The discrepancies
amount to 10--20\% in the bands 1--3 and become negligible in the higher bands. In the further analysis, this uncertainty can be avoided and has marginal effects when several lines of a given molecule are present. 

\subsection{Auxiliary observations from the Submillimeter Array}
Supporting SMA data at 230~GHz are kindly provided by C.~Brogan and T.~Hunter. The continuum map is identical to that published in \citet{Hunter}, whereas the spectral image cubes are unpublished. Detailed information about the observation and calibration can be found therein. 
The data were calibrated again in 2007 in Miriad (described in \citealt{Brogan}). The separation of the line and continuum emission was performed in the uv plane. For the purpose of modelling, the interferometric line and continuum emission were recombined in CLASS for the portion of the spectrum containing the foreground CN absorption feature. The observed frequency range covers two 2~GHz wide bands, 216.6--218.6~GHz in the lower sideband and 226.6--228.6~GHz in the upper sideband, with a frequency resolution of $\sim$0.8~MHz. The map covers a $64''\times64''$ field and
the synthesized beam is $2.2''\times1.3''$ in the USB. The line-to-continuum ratio is 12\% for SMA1 and 21\% for SMA2 in the given bands. 

\section{Analysis: line identification and modeling}
\subsection{Simulating spectra with XCLASS}
\label{xclass}
A spectral line is identified by comparing the center frequency with a measured frequency of a known chemical species (atom, molecule, ion) in the laboratory. 
Two large databases relevant for molecular astrophysics, from the radio to the far-infrared region, have been built up in the last decades: The Cologne Database for Molecular Spectroscopy (CDMS) \citep{CDMS} and the Jet Propulsion Laboratory (JPL) catalogue \citep{JPL}. If the exact source velocity is unknown or several candidates are present for the identification of a spectral line, secondary criteria such as the upper energy level or the Einstein A coefficient must be taken into account. Although many lines in the spectra of hot cores are blended, the identification in spectral surveys is more confident because more lines of a species are present, which can then be checked for consistency, i.e. all lines of a molecule which are expected to be there should be found with the expected line strength, line shape and velocity.

The main program used in this work is XCLASS, which was described in \citet{Comito} and applied to a ground-based Orion KL line survey. It is an extension of CLASS (Continuum and Line Analysis Single-dish Software), which is part of the GILDAS\footnote{\url{http://www.iram.fr/IRAMFR/GILDAS}} package. XCLASS includes entries from the CDMS, JPL and a private catalogue, which contain all necessary information: rest frequency $\nu_l$, integrated intensity, lower state energy $\rm E_l$, upper state degeneracy $\rm g_{up}$, quantum numbers and the partition function $Q$. With this database, it is possible to simulate a spectrum of a molecule, by providing parameters such as the excitation temperature and abundance, and compare the simulated spectrum with the observed one. Similar software tools developed in the last years are {\it Weeds} \citep{WEEDS}, a package for CLASS, and CASSIS \citep{cassis}.
The line identification for NGC~6334I, which transfers to a molecule identification in XCLASS, is made easier because 20 detected molecules are reported in the line survey from \citet{Kalinina} at lower frequencies and in other surveys mentioned in Sect. \ref{intro}.

The function which models a spectrum in XCLASS is a solution of the radiative transfer equation for an isothermal object in one dimension, called detection equation \citep{stahler}. Furthermore, the finite source size and dust attenuation are considered. The model function is:
\begin{equation}
T(\nu)=\sum\limits_{m} \sum\limits_{c} \eta(\theta_{m,c})\left[J(T_{\rm ex}^{m,c})-J(T_{\rm bg})\right](1-e^{-\tau(\nu)^{m,c}}) \cdot e^{-\tau_{\rm d}} \ ,
 \label{eqnde}
\end{equation}
with the optical depth
\begin{align}
\tau(\nu)^{m,c} &= \sum\limits_{l} \tau(\nu)^{m,c}_l \\ 
\tau(\nu)^{m,c}_l &= \frac{c^3}{8 \pi \nu^3} A_{ul} N_{\rm tot}^{m,c}\frac{g_l e^{-E_l/kT_{\rm ex}^{m,c}}}{Q(m,T_{\rm ex}^{m,c})} (1-e^{-hv/kT_{\rm ex}^{m,c}}) \phi(\nu)^{m,c} \ ,
\end{align}
the dust optical depth 
\begin{equation}
 \tau_{\rm d} = N_{\rm H} \cdot \kappa_{1.3\rm mm} \cdot \left(\frac{\nu}{230 \ \rm GHz}\right)^{\beta} \cdot 3.3\times10^{-26} \ ,
\label{eqdust}
\end{equation}
the Gaussian line profile 
\begin{equation}
\phi(\nu)^{m,c}=\frac{2\sqrt{ln2}}{\sqrt{\pi} \Delta v^{m,c}} \exp \left(-\frac{(\nu-(\nu_l+\nu_{\rm LSR}^{m,c}))^2}{2\sigma^2} \right)
\end{equation}
and the beam filling factor
\begin{equation}
	\eta(\theta_{m,c})=\frac{\theta_{m,c}^2}{\theta_{m,c}^2+\theta_{\rm t}^2} \ .
\end{equation}
The radiation temperature $J(T)$ is defined as $J(T)=\frac{hv}{k} \frac{1}{e^{h\nu/kT}-1}$. The sum goes over the indices $m$ for molecule, $c$ for component and $l$ for spectral line transition. $T_{bg}$ is the background continuum temperature and $\theta_{\rm t}$ the telescope beam size. The line profile function is normalized so that $\int_0^{\infty} d\nu \phi(\nu)=1$.
The factor $e^{-\tau_{\rm d}}$ takes the dust attenuation into account where the dust mass opacity $\kappa$ is taken from \citet{ohdust}. For NGC~6334I, we adopt a value of $\beta=2$ and $\kappa_{1.3\rm mm}=0.42$~cm$^2 \,$g$^{-1}$, which is between the values for non-coagulated dust grains with thin ice mantles ($\kappa_{1.3\rm mm}=0.51$~cm$^2 \,$g$^{-1}$) and without ones ($\kappa_{1.3\rm mm}=0.31$~cm$^2 \,$g$^{-1}$). The total hydrogen column density $N_{\rm H}=N(\rm H)$+$2N(\rm H_2)$  is $3 \times 10^{24}$ cm$^{-2}$, which is derived from the average hydrogen density given in Table~\ref{tNGC} and a source size of 10$''$ and is in agreement with values from previous articles \citep{Wiel,Lis} considering the uncertainties.

Overall, there are five input parameters for one component of a molecule:
   \[
      \begin{array}{lp{1\linewidth}}
         \theta_{m,c}  & source size in arcsec    \\
         T_{\rm ex}  & excitation temperature in K                    \\
         N_{\rm tot} & total column density in cm$^{-2}$            \\
         \Delta v & velocity width (FWHM) in km\,s$^{-1}$     \\
         v_{\rm LSR}     & source velocity in km\,s$^{-1}$.\\
      \end{array}
   \]
Components can be identified as spatially distinct sources within the beam, such as clumps, hot dense cores, colder envelopes or outflows, and can usually be distinguished by different radial velocities. They do not interact with each other radiatively, but are simply superposed in the model. However, for absorption lines the emission from other components is considered first by calculating the background emission spectrum, which is then used as a new ``continuum'' for absorption lines. By fitting all species and their components at once, line blending and optical depth effects are taken into account. 
Because of the high densities in the hot cores ($n_{\rm H_2}>10^7$cm$^{-3}$), it is assumed that most molecules are thermalized at the gas temperature, so that the assumption of LTE is reasonable. With the LTE condition, every population of a level $i$ is known and given by the Boltzmann distribution \(n_i=\frac{N}{Q} g_i e^{-E_i/(kT)}\). The modeling of a spectrum speeds up line identification enormously, since several hundred lines of a molecule like methanol are predicted at once in a HIFI band.

The modeling can be done together with isotopologues (and higher vibrational states) of a molecule assuming a conversion factor stored in a separate database. All parameters are expected to be the same except the abundance.  
For carbon, \citet{Milam} derive an average gradient of $^{12}$C/$^{13}$C=$6.21(1.00)\cdot D_{GC}+18.71(7.37)$, where $D_{GC}$ is the distance to the Galactic center in kpc. The carbon isotopic ratio for NGC~6334I would be 57(10). Eight isotopic species are found in this survey, besides the main isotopes. The ratios used for NGC~6334I are: [\ce{^{35}Cl}]/[\ce{^{37}Cl}]=3, [\ce{^{32}S}]/[\ce{^{34}S}]=23,~[\ce{^{32}S}]/[\ce{^{33}S}]=127,~[\ce{^{12}C}]/[\ce{^{13}C}]=60, [\ce{^{14}N}]/[\ce{^{15}N}]=300,~[\ce{^{16}O}]/[\ce{^{18}O}]=500 and [\ce{^{16}O}]/[\ce{^{17}O}]=1500. Deuterium is found also, but here no common ratio is taken because the chemical fractionation is very high. The ortho to para ratio for some molecules is assumed to be the statistical equilibrium value (high temperature limit).

\subsection{Model fitting with MAGIX}
\label{subsec:fitting}
Once an initial guess is set up, the parameters can be fitted using the program MAGIX (Modelling and Analysis Generic Interface for eXternal numerical codes; \citealt{MAGIX,moeller}). It delivers a package of different optimization algorithms and an interface to XCLASS. For every of the five fit parameters a range can be set over which to search for the best fit, or the parameter can be kept constant. Used here is the Levenberg-Marquardt (LM) algorithm with the method of least squares, which tries to minimize the squares of the difference between the fit function $y(x_i|a_0\ldots a_{M-1})$ with the M fit parameters a and the data. The sum over all n data points $(x_i,y_i)$ is defined as: 
\begin{equation}
	\chi^{2}=\sum \limits_{i=0}^{n-1} \frac{\left(y_i-y(x_i|a_0\ldots a_{M-1})\right)^2}{\sigma_i^2} \ .
\end{equation}
The uncertainty for $y_i$ is $\sigma_i$. 
The iteration stops when a minimum is found and the algorithm has converged, or when the change in $\chi^{2}$ from one iteration step to another is very small, e.g. $\Delta \chi^{2}<10^{-6}$. Since this function is a priori unknown, it becomes clear that the algorithm can converge into a local and not a global minimum. 
A bad fit for all parameter values can be a hint for non-LTE populations, like HCN lines with high critical densities, or maser activity.
The automated fitting has to be corrected manually sometimes, since different problems can occur, which are described in the following.

\textbf{Starting values:} The choice of the initial values for the parameters, is fundamentally important in the LM algorithm. Bad or random starting values lead to an unnecessary amount of computation time. The starting values are set manually by the user and should be physically reasonable. 
Because of their fit behavior, the five fit parameters in XCLASS are divided into two groups: $\theta, T_{\rm ex},N$ and $\Delta v, v_{\rm LSR}$. The
last two are generally well fixed and do not change anymore, so they are held constant. For high opacities, the line width is broadened 
which is known as optical depth broadening. To check for a local or global minimum, MAGIX was iterated again with another set of starting values
and then checked if it still converged to the same point. The disadvantage is that this takes computation time with several repetitions and the variation possibilities increase dramatically with more velocity components for a given molecule. This check has been done regularly with molecules with one component, but infrequently for more complicated cases. The best fit parameters for the species with multiple overlapping components may thus not be uniquely determined.

\textbf{Calculation time:} The time increases with the number of parameters and data points. To reduce the calculation time, especially for ``weeds'',
not all transitions from every band are included but a variety of optical thin and thick lines, lowest and highest frequencies and accordingly a big range of different energy states. Suitable for this are Q-branches of methanol or dimethyl ether. Weeds is the colloquial term for interstellar molecules, which show a plethora of spectral lines. They are all organic asymmetric rotors with a methyl group (-\ce{CH3}), except of \ce{SO2}. The classification is not strict, but these five are considered to be the most important ones: methanol, dimethyl ether, ethanol, methyl formate and sulfur dioxide. In methanol, \ce{CH3OH}, the methyl group can tunnel the potential barrier at every 120$^\circ$ relatively to the hydroxyl group (-OH), leading to an internal rotation or torsion about the C-O bond. The torsional levels are split into sub-levels (A and E symmetry), and the interaction of torsional and rotational motions yields multiple energy states and a complex spectrum. Furthermore, heavier and larger molecules have a larger moment of inertia, which is inverse to the rotational constants, so that they produce denser spectra.
The determination of the excitation temperature with XCLASS is much faster than the so called Boltzmann plot or rotational diagram method, where the upper state abundance $N_{u}$ of a line transition can be derived from the integrated area in the optical thin limit. This method has the disadvantages that it is unpractical for many lines and misleading for line blends and high optical depths.

\textbf{Uniqueness:} The model function Eq. \ref{eqnde} has two limits where degeneracy occurs: in the optically thick limit $\tau\gg1$, the last term vanishes and $J(T_{\rm ex})$ and $\eta(\theta)$ enter linearly into the equation; in the optically thin limit $\tau\ll1$, $(1-e^{-\tau})\approx\tau$ and $\eta(\theta)$ and $N$ enter linearly. This results in the fact that these two parameter pairs cannot be determined independently. In the fitting procedure
this can be noticed in the behavior that $\chi^2$ changes only slightly while one of the parameter increases and the other decreases.
It is not always obvious if a spectral line is optically thick, especially when several components add together. A solution to the problem of degeneracy are isotopologues, if present. With the ratios mentioned above, a consistent model can be fitted with constraints on $N$ and also for $\Delta v$. 
In some cases the rare isotopologues are fitted well, but the main isotopologue is optically thick and too weak to explain the high observed intensities; this is the case for \ce{CH3OH}, SO, \ce{H2CO} and \ce{HCO+}. This can be a hint that a second component is present when no signs for that could be deduced from the line shape. However, every additional emission component should be justified, since this makes the model more complex and increases the possibility of redundancy. In the case of NGC~6334I, two near overlapping components are justified by the SMA observations which show that both cores SMA1 and SMA2 show line emission at different velocities.  
\begin{figure*}
\resizebox{\hsize}{!}{\includegraphics[trim=5cm 0cm 0cm 0cm,clip, angle=270, width=1\textwidth]{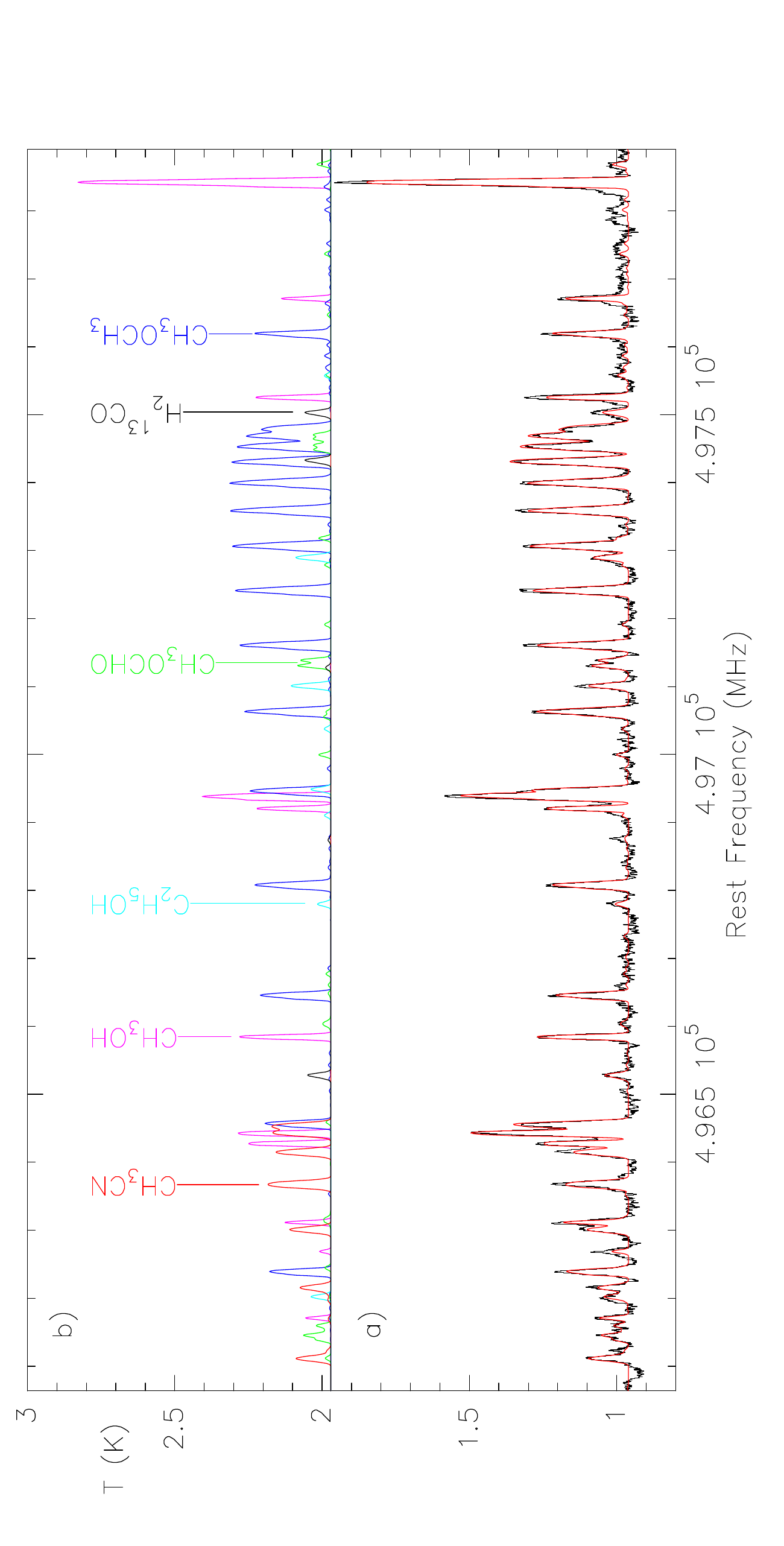}}
\caption{(a) Fit example around 497~GHz showing the data (black) and the fit (red). The decomposition into the contributions of individual species is shown above in (b).}
\label{fitex}
\end{figure*}  

\textbf{Weighting:} $\chi^2$ is the sum over all modeled spectral lines. One weighting could be done by the uncertainty of $y_i$, $\sigma_i$, which is currently not implemented in MAGIX. But this is not a severe limitation, because the noise or rms is nearly constant across a given HIFI band. It is relevant when the line intensity of the main isotopologue is much higher than for the less abundant isotopologue. Stronger lines are then more weighted compared to weak lines because the change in $\chi^2$ is larger. This can lead to a model that is biased toward fitting the stronger lines. Another problem is the fitting in higher-frequency bands with larger noise (band 5--7). Here the algorithm tends to increase the line width because it smooths the line wings better. Line widths from transitions in lower bands reveal then that this is an overestimation. To fix this, the line width must be held constant and the best fit value is taken from transitions in lower bands.

\textbf{Error estimate and confidence limits:} It is assumed that in general the standard deviation of $\theta, T$ and $N$ is at least 10\%, because a small higher or lower value does not change the fit significantly. Additionally, there are other errors not accounted for in the fit like calibration uncertainties, uncertain isotopic ratios, $N(\rm H_2)$, etc., so that in the worst case only the magnitude of $N$ is roughly known.
\paragraph{} 
Due to the fact that lines of one species are blended with lines of other molecules, the fitting procedure is done in three major steps: 
\\
\textbf{1. Fitting intense lines and weeds.} By modeling weeds one filters them out and takes into account the line blendings. For intense lines of diatomic molecules like SO, the effect of line blending is small. Blended lines are first excluded from the fitting range. 
\\
\textbf{2. Fitting weaker and absorption lines.} For weak lines, the noise becomes important so that the fitting procedure is less reliable. 
Too weak lines, below 2$\sigma$, of a molecule are therefore excluded from the fitting range, except for cases where they can serve to give an upper limit for $T$ or $N$. Absorption lines are not expected to be associated with the hot cores in NGC~6334I because they come mainly from cold foreground gas.
Apart from that are ground state transitions with low energy levels from \ce{H2O} or \ce{NH3}.  
\\ 
\textbf{3. Repeating all in a second iteration.} The difference now is that the simulated spectrum of all molecules are used as a new background. The parameters are saved
globally in a file and read in by MAGIX, but only the molecule of interest is fitted again. This time, blended lines of a species are taken into account
and more components added, if necessary. A simultaneous fit of all parameters  of all components of all molecules is in principle possible, but far too time consuming and inefficient. In the end, the result is an improved and consistent fit, see the extract in Fig.~\ref{fitex}. This could be done iteratively several times, but it was noticed that the changes are marginal and larger only for molecules with very few transitions present.    
\subsection{Analysis of interferometric data}
The presence of several components with small radial velocity differences which overlap makes the assignment to one certain physical source difficult. The advantage of interferometer data is having the spectral and spatial information together. On the other hand, the SMA frequency range of 4~GHz is small compared to HIFI and extended emission can be missed, because it is filtered out by the interferometer. Two spectra were extracted, one from the peak emission toward SMA1 and the other toward the peak emission of SMA2.
It must be noticed that for some molecules like methanol, the emission overlaps from both cores. A direct comparison between them reveals that firstly SMA1 is constantly red-shifted compared to SMA2, secondly the line width in SMA1 is broader and thirdly some lines are only present in SMA1 or SMA2. The intensities in Jy/beam are converted to Kelvin by using the formula 
\begin{equation}
T_{\rm mb}(\rm K)=\frac{1.22 \times 10^6 \cdot F (\rm Jy)}{[\nu(\rm GHz) \cdot \theta ('') ]^2} \ ,
\end{equation}
where $\theta$ is the beam size. The conversion factor is then about 12.5~K/Jy for the LSB and 8.5~K/Jy for the USB. The rms is about 2.5~K in the LSB and 2~K in the USB.
Maps of the integrated intensity over one spectral line of all detected molecules were made with the toolkit Karma \citep{karma} and the source sizes were derived from Gaussian fits. 
The whole procedure explained in Sect. \ref{subsec:fitting} was performed again with the SMA spectra. The model derived from the HIFI spectrum was extrapolated to the lower frequencies of the SMA bands and compared with the observed one. Deviations can be explained by the fact that the HIFI model includes the outer parts like the envelope and is an average over a range of temperatures, whereas the SMA spectra correspond to the inner hot core with higher temperatures.     

The new derived source sizes $\theta$ were adapted for the HIFI spectrum. For some molecules, the source sizes derived from the HIFI spectrum were already in good agreement with the one from the SMA data. Small deviations ($\leq$ 2\(^{\prime\prime}\)) are permitted when they lead to a lower $\chi^2$ value. 
The source size is likely to change when excitation conditions differ, so that the high-$J$ transitions in the HIFI spectra may come from a more compact region
than that observed in the SMA maps. For the cases where the lines are optically thin or thick, $\theta$ is held fixed in order to cancel out the ambiguity.  

\section{Results}
\subsection{General results}
A total of 41 molecules with their isotopologues are identified in the HIFI spectrum. The best fit results are given in Table~\ref{tXCLASS}. The whole survey would need several hundred pages of figures and tables with line lists. Therefore, to give an impression, only selected fits for each molecule are attached in Appendix~B. The modeled spectrum for each molecule and the full observed and modeled spectrum for NGC~6334I will be made available online, after reprocessing the HIFI spectra with HIPE 8. See Appendix~B for further details.
The average source velocity (considering all components in the range --5 to --10~km\,s$^{-1}$ associated with NGC~6334I) is --7.3~km\,s$^{-1}$ and average line width 4.2~km\,s$^{-1}$ with a standard deviation of about $\pm$ 1~km\,s$^{-1}$ for both spectral features. The velocity width is dominated by turbulence, since the thermal Doppler width contributes less than 1~km\,s$^{-1}$ for typical values of $T$=100~K and $m$=20~amu. 
Some molecules show strong broad wings which are produced by outflows: CO, CS, SiO, \ce{H2O} and HCN. \ce{H2CS} has an asymmetric line profile
with a stronger redshifted wing. Especially the component at --17  to --20~km\,s$^{-1}$, observed in C, \ce{NH3} and \ce{H2S} with $T=10-30$~K is presumably the blue outflow lobe. Two cold ($\sim$5~K), extended foreground clouds are detected consistently with components around --2~km\,s$^{-1}$ and +7~km\,s$^{-1}$ in the following species: C and \ce{C+}, CH and \ce{CH+}, \ce{H2O} and \ce{H2O+}, HF, \ce{H2Cl+} and \ce{OH+}. This is in agreement with an observed OH absorption component at +6~km\,s$^{-1}$ towards several NGC~6334 cores reported by \citet{Brooks}. 
\begin{table*}
  \caption{Fit results for the HIFI spectrum of NGC~6334I}
   \centering
\label{tXCLASS}
\begin{tabular}{lrrrrrr|lrrrrrr}
\hline\hline
    Species & $\Theta$   & $T_{ex}$ & $N$        & $\Delta v$ & $v_{LSR}$ & Notes  &Species & $\Theta$   & $T_{ex}$ & $N$        & $\Delta v$ & $v_{LSR}$ & Notes  \\
            &  (\arcsec) &  (K)     & (cm$^{-2}$) & (km\,s$^{-1}$)    & (km\,s$^{-1}$)           &        & &  (\arcsec) &  (K)     & (cm$^{-2}$) & (km\,s$^{-1}$)    & (km\,s$^{-1}$)           &  \\
\hline
     C     & 40    & 56    & $2.4 \times 10^{18}$ & 4.3   & -7.3  &  			 & \ce{HCS+}  & 4.0   & 68    & $7.6 \times 10^{14}$ & 4.3   & -7.1  &   a  \\
          & 29    & 68    & $2.1 \times 10^{17}$ & 4     & 0.2   &  			 &      NO    & 3.5   & 50    & $2.7 \times 10^{17}$ & 3.9   & -7.5  &      \\
          & ext   & 6     & $2.3 \times 10^{17}$ & 1     & -2    &   			 & SiO   & 3.4  & 30    & $3.0 \times 10^{15}$ & 5     & -7.4  &       \\
          & ext   & 2.7     & $4.7 \times 10^{15}$ & 1.3     & 6.5   			 &         & HF    & ext   & 2.7   & $1.2 \times 10^{14}$ & 8.6   & -5    &        \\
          & 48   & 26     & $4.0 \times 10^{16}$ & 3     & -20.6   	 &  b       &         & ext   & 2.7   & $1.2 \times 10^{13}$ & 1.9   & 0.3   &         \\
  \ce{C+} & ext   &  112  & $1.8 \times 10^{18}$ & 6     & -9    &    		 &            & ext   & 2.7   & $2.5 \times 10^{13}$ & 2.2   & 7     &         \\
          & ext   & 4     & $6.0 \times 10^{17}$ & 3     & -0.5  &   			 & HCl   & 22    & 39    & $1.9 \times 10^{14}$ & 4.3   & -6    & 6    \\
          & ext   & 5   & $6.0 \times 10^{17}$ & 4.6   & 6.6   &         		 & & ext   & 11    & $1.5 \times 10^{14}$ & 8.6   & -10   &        \\
   \ce{CH+}   & ext   & 13    & $1.5 \times 10^{14}$ & 5     & -9.8   & 1   		 & \ce{H2Cl+} & ext   & 7     & $4.9 \times 10^{13}$ & 11    & -1.2  & 7   \\
          & ext   & 8   & $1.6 \times 10^{14}$ & 6.6   & -2    &       			  & \ce{SH+}   & ext   & 2.7  & $2.5 \times 10^{12}$ & 4     & -5.7    &        \\
          & ext   & 2.7   & $2.9 \times 10^{13}$ & 4.4   & 7.8   &    			   &  \ce{H2S}   & 3     & 95   & $2.5 \times 10^{17}$ & 4     & -6.1  & 8    \\
    CH    & 13.4  & 95    & $1.9 \times 10^{14}$ & 2.9   & -8.2  &    			   &             & ext    & 20    & $1.0 \times 10^{15}$ & 5     & -7.8  &         \\
          & 64    & 4   & $6.7 \times 10^{13}$ & 1.9   & 6.7   &     			    &       & 18.5   & 11     & $8.6 \times 10^{14}$ & 17    & -17.5 &    b    \\
          & 38    & 5   & $3.0 \times 10^{13}$ & 6     & -2.3  &       			  & CS & 2.4   & 87    & $3.6 \times 10^{17}$ & 4.5     & -6.8  & 9   \\
    NH    & 39    & 2.7  & $6.0 \times 10^{13}$ & 2     & -6    & 			 &     &	2.2  & 73    & $1.4 \times 10^{17}$ & 3.8     & -7.8  &     \\
          & ext   & 2.7  & $9.0 \times 10^{13}$ & 10    & -4.6  &         & 			& ext  & 103    & $6.5 \times 10^{13}$ & 6     & -6.8  &  \\
   \ce{NH2}    & 3.4   & 108   & $3.6 \times 10^{15}$ & 4     & -8.5  &    		   & NS    & 3     & 89    & $3.4 \times 10^{15}$ & 5     & -7    &   a     \\
          & ext   & 12    & $2.9 \times 10^{13}$ & 2     & -6.1  &        		 & OCS   & 2.5     & 102   & $1.2 \times 10^{18}$ & 4     & -7.5  &       \\
          & ext   & 2.7  & $1.0 \times 10^{14}$ & 30    & -13   &       		 & SO   & 2.7   & 70    & $2.1 \times 10^{17}$ & 4.1     & -7.4  &   10  \\
    \ce{NH3}   & 8  & 28    & $2.3 \times 10^{16}$ & 4.2     & -7.7  & 2               &     & 5.8   & 63    & $7.6 \times 10^{15}$ & 4.0     & -8.0  &       \\
          & ext   & 9     & $6.0 \times 10^{13}$ & 2.9     & -6.9    &                    &\ce{SO2}   & 1.8   & 100   & $9.5 \times 10^{16}$ & 4.7   & -7.8  &        \\
          & ext   & 9     & $2.2 \times 10^{14}$ & 16     & -17   &    b             & \ce{H2CS}  & 3.4   & 106   & $2.7 \times 10^{16}$ & 5     & -7.9  & 11    \\
    \ce{NH2D}  & 8   & 28    & $2.3 \times 10^{14}$ & 3.8   & -8.6  &    	   & \ce{CH3OH} & 2.7   & 99   & $1.4 \times 10^{19}$ & 2.9     & -7.9  & 12    \\
    \ce{N2H+}  & 31    & 29    & $4.3 \times 10^{13}$ & 2.5   & -6.8  &    	  & & 5.3   & 61    & $1.3 \times 10^{17}$ & 2.8     & -8.3  &       \\
          & 20    & 35    & $3.5 \times 10^{13}$ & 2.5   & -9.5  &       	  & \ce{CH3OCH3} & 2.1   & 87    & $3.0 \times 10^{18}$ & 3.2   & -7.4  & 13    \\
    \ce{OH+}   & ext   & 2.7   & $2.0 \times 10^{13}$ & 4     & 3.3   &        	& & 5.0     & 69    & $7.0 \times 10^{16}$ & 2.7   & -8.1    &        \\
          & ext   & 2.7   & $2.8 \times 10^{13}$ & 6.5   & -2    &         	& \ce{CH3OCHO} & 3     & 115   & $7.1 \times 10^{17}$ & 3.5   & -7.65  &       \\
    \ce{H2O} & 4.2   & 59  & $2.1 \times 10^{18}$ & 4.3   & -8.3  & 3  & \ce{C2H5OH}  & 2     & 117   & $1.4 \times 10^{17}$ & 3.9     & -8    &      \\
    HDO   & 4.4   & 440   & $1.2 \times 10^{15}$ & 5.3   & -7.7  &    a    & \ce{H2CO}  & 2.7   & 78    & $1.5 \times 10^{17}$ & 4.5   & -7.4  & 14    \\
    \ce{H2O+}  & ext   & 24    & $3.7 \times 10^{13}$ & 8.3   & -0.7  &   &  & 5.5   & 52    & $1.2 \times 10^{16}$ & 2.8   & -8.5  &        \\
    \ce{C2H}   & 8.7   & 32    & $4.3 \times 10^{15}$ & 5     & -6.9    &       & \ce{CH3CN} & 1.9   & 154   & $2.8 \times 10^{16}$ & 4.7   & -7.3  &       \\
    CN    & 14.8  & 22    & $1.1 \times 10^{15}$ & 5     & -6.4  &    & \ce{C2H4O}  & 2.1   & 80    & $5.4 \times 10^{16}$ & 3     & -7.6  &       \\
    CO & 32    & 60     & $3.7 \times 10^{19}$ & 4.6   & -6.8  & 4     &  \ce{H2CCO} & 3     & 117   & $6.0 \times 10^{15}$ & 3     & -7.7  &    a   \\
    HCN & 3.5   & 32    & $2.6 \times 10^{18}$ & 6     & -5.6  & 5   &  \ce{HCO+}  & 5.1   & 73    & $1.3 \times 10^{15}$ & 4.7   & -7.5  & 15  \\
        & 5.0   & 48    & $3.0 \times 10^{16}$ & 5     & -7.5  &     &               & 31    & 35    & $1.4 \times 10^{14}$ & 4.9   & -7.2  &         \\
    DCN   & 3.5   & 54  & $3.7 \times 10^{14}$ & 6     & -5.6  &     &     \ce{CH2NH} & 2.4   & 75    & $6.0 \times 10^{15}$ & 5     & -7.0    &   a    \\
    HNC   & 3.5   & 50  & $1.1 \times 10^{15}$ & 6     & -5.6  &     &       HNCO     & 2.0   & 284   & $1.5 \times 10^{16}$ & 5     & -7.7  &      \\
          & 5.0   & 40  & $3.3 \times 10^{14}$ & 4.4   & -8.2  &     &  \ce{NH2CHO} & 2     & 117   & $4.7 \times 10^{15}$ & 4.2     & -7.7 &       \\
          &       &     &                      &       &       &     &   \ce{HC3N}  & 2.8     & 134   & $4.2 \times 10^{15}$ & 9     & -5.6  &       \\       
\hline
\end{tabular}
  \tablefoot{Several rows for one species denote multiple components. Extended source sizes are abbreviated with ext.
  (1) including \ce{^{13}CH+}. (2) including \ce{^{15}NH3}. (3) only isotopologues \ce{H2^{18}O}, \ce{H2^{17}O}. 
   (4) only isotopologues \ce{^{13}CO}, \ce{C^{18}O}, \ce{^{13}C^{18}O}, \ce{C^{17}O}. (5) only isotopologues \ce{H^{13}CN}, \ce{HC^{15}N}, \ce{H^{13}C^{15}N}. (6) including \ce{H^{37}Cl}.
    (7) including \ce{H2^{37}Cl+}. (8) including \ce{H2^{34}S}. (9) only isotopologues \ce{C^{34}S}, \ce{C^{33}S}, \ce{^{13}CS}, \ce{^{13}C^{34}S}. (10) including \ce{^{34}SO}, \ce{^{33}SO}.
    (11) including \ce{H2C^{34}S}. (12) including \ce{^{13}CH3OH}. (13) including \ce{^{13}CH3OCH3}. (14) including \ce{H2^{13}CO}. (15) including \ce{H^{13}CO+}. (a) $\tau$ small. (b) outflow component. }
  \label{fitresults}%
\end{table*}%
\begin{table*}
  \centering
  \caption{List of all detected chemical species in NGC~6334I}
    \begin{tabular}{llllllllll}
    \hline\hline
    1 atom & 2 atoms & 3 atoms & 4 atoms & 5 atoms & 6 atoms & 7 atoms & 8 atoms & 9 atoms & 10 atoms\\
    \hline
    C     & SO, NO & \ce{SO2}, \ce{HCO+} & \ce{H2CO} & \ce{CH2NH} & \ce{CH3OH} & \ce{C2H4O}   & \ce{CH3OCHO} & \ce{CH3OCH3} & \ce{CH3CH3CO} \\
\ce{C+}   & CS, SiO & OCS, \ce{H2O}      & \ce{H2CS} & \ce{HC3N}  & \ce{CH3CN} &  \ce{C2H3CN} &             & \ce{C2H5OH} &   \ce{(CH2OH)_2}  \\
          & NS, \ce{OH+} & HCN, \ce{NH2}      &  HNCO     & HCOOH      & \ce{NH2CHO}& \ce{CH3CCH}  &             &\ce{C2H5CN} &                    \\
          & CO, \ce{CH+} & HNC, \ce{HCS+}      & \ce{NH3}  & \ce{H2CCO} &            &              &             &           &                     \\
          & CN, CH      & \ce{H2S}, \ce{H2O+} &           &            &            &              &             &           &                     \\
          & \ce{HCl}, HF & \ce{C2H}, \ce{H2Cl+} &       &       &       &  &&&\\
          & NH, \ce{SH+}    & \ce{N2H+} &       &  &&&&& \\
    \hline
    \end{tabular}%
  \label{tlist}%
\end{table*}%

The best fit results for the SMA spectra are presented in Table~\ref{tsmaf}, the list of detected lines in Table~\ref{smail} and the spectra for the LSB and USB in Fig.~\ref{smafits} in Appendix~A. Overall, 20 molecules are identified, of which 15 are also present in the HIFI observations. The additional five molecules are all organic: acetone (\ce{CH3CH3CO}), ethylene glycol (\ce{(CH2OH)_2)}), formic acid (HCOOH), acrylonitrile (\ce{C2H3CN}) and propionitrile (\ce{C2H5CN}). The latter two only have low intensity features ($<3\sigma$) in the HIFI band 1a spectrum and are therefore considered to be not detected. Also, 3 isotopologues of OCS are found. In general, the fits give higher temperatures than those derived from the HIFI data, up to 200--300~K, because the emission originates from the center of the hot cores, observed in a smaller beam.  Although not true for all molecules, SMA2 has on average higher excitation temperatures, lower column densities, abundances and smaller line widths than SMA1.
On average, the source velocity for SMA1 is $-5.6 \pm 0.1$~km\,s$^{-1}$ and for SMA2 $-7.6 \pm 0.1$~km\,s$^{-1}$, so that the relative radial velocity difference is 2~km\,s$^{-1}$.
When applied e.g. to ethanol (\ce{C2H5OH}), which is only seen toward SMA2, the source velocity derived from the SMA spectra is consistent with that from HIFI. In the case of \ce{C2H4O} or \ce{NH2CHO}, a small discrepancy  of 0.8~km\,s$^{-1}$ is seen, which can be explained by the fact that in the SMA bands only a few transitions are detected, some of which are blended, so that the precise velocity determination is hindered. 

The chemical differences between both cores are determined from the SMA maps and the velocity components in the HIFI spectra. Those molecules that show dominant emission or only emission
in SMA2 are HCOOH, \ce{(CH2OH)_2)}, \ce{NH2CHO}, \ce{C2H4O} and \ce{C2H5OH}. For SMA1, the dominant species are \ce{HC3N}, \ce{C2H3CN}, \ce{C2H5CN}, \ce{H2S}, CN, HCN and HNC.
It is noticeable that more oxygen-bearing species, including water \citep{Emprechtinger}, arise in SMA2 and more nitrogen-bearing species in SMA1. This variation in the peak abundance is also observed in other star-forming regions, such as Orion KL between the hot core and the compact ridge \citep{Comito}.

\subsection{Statistics}
A spectral line is considered to be detected when the peak intensity exceeds the 3$\sigma$ level of the random noise.
Furthermore, the line width should be comparable to that of all other detected lines. In this way one excludes single channel spikes and sets a lower limit for
the integrated line intensity. A lower limit of 2$\sigma$ would be acceptable when this limit is not taken for a single line of a molecule but rather for many lines 
since the probability that this is a coincidence decreases when having many lines observed and a good model fit. Nevertheless, some weak features are included
in certain cases in the fitting range because they can give useful upper limits for the model. 

\begin{table}
  \centering
  \caption{Integrated spectral line intensities of all detected chemical species of the HIFI spectrum for NGC~6334I}
    \begin{tabular}{p{1.25cm}rr|p{1.25cm}rr}
    \hline \hline	
    Species &   $\int T {\rm d} v$     & Lines & Species &  $\int T {\rm d} v$       & Lines \\
            &  (K$\cdot$km\,s$^{-1}$)    &          &         & (K$\cdot$km\,s$^{-1}$)      &          \\
    \hline	
    \ce{CH3OH} & 6016  	         & 2063  & \ce{CH3CN} & 50    & 48 \\
    CO & 4715  & 31    			 & HCl   & 40    & 3 \\
    \ce{CH3OCH3} & 1486  & 637           &  \ce{N2H+}  & 23    & 4 \\
    \ce{H2O} & 1257  & 31    		 & OCS   & 22    & 16 \\
    \ce{CH3OCHO} & 582   & 558   	 & \ce{H2O+} & 16    & 1 \\
    \ce{H2CO}  & 428   & 105   		 & NS    & 16    & 12 \\
    CS & 280   & 45    			 & CN    & 15    & 6 \\
    \ce{C+}    & 262   & 1     		 & HNC   & 15    & 4 \\
    \ce{C2H5OH}  & 227& 134              & \ce{CH2NH} & 14    & 17 \\
    \ce{NH3}   & 226   & 13   		 & CH    & 14    & 1 \\
    \ce{SO2}  & 207   & 109   		 & \ce{OH+}  & 12    & 2 \\
    HCN & 205   & 27   			 & NO    & 11    & 8 \\
    \ce{NH2}    & 181   & 6     	 & \ce{H2Cl+} & 11    & 2 \\
    \ce{H2S}   & 168   & 32    		 & \ce{C2H}   & 11    & 6 \\
    \ce{HCO+}  & 152   & 10    		 & HNCO     & 10    & 25 \\
    HF    & 146   & 1     		 & \ce{H2CCO} & 4     & 9 \\
    \ce{H2CS}  & 141   & 78   		 & \ce{HCS+}  & 4     & 4 \\
    SO    & 119    & 51                  & \ce{NH2CHO} & 2     & 8 \\
    C     & 115   & 2     		 & \ce{HC3N}  & 2     & 4 \\
    \ce{CH+}   & 108    & 3    		 & SiO   & 2     & 1 \\
    NH    & 103   & 3    		 & \ce{SH+} & 1 & 1 \\
    \ce{C2H4O}  & 67 & 94	 &                     & & \\		 
    \hline	
    Sum    & 17485    &  4217   &    &      &  \\
    \hline
    \end{tabular}
\tablefoot{Included are all transitions from isotopologues (except \ce{CH3OD} and \ce{CH3^{18}OH}) and torsionally or vibrationally excited states.}
  \label{tstat}
\end{table}
Overall, 49 chemical species have been found in NGC~6334I so far, listed in Table~\ref{tlist}, ranging from fine structure transitions of C, through over mostly di- and triatomic molecules, to organic molecules with 10 atoms. About half of them were found in previous line surveys. Methyl acetylene (\ce{CH3CCH}) has been identified by \citep{Kalinina}, but is too weak in HIFI ($<1\sigma$). We find 31 isotopologues in our data.
The number of lines in all HIFI bands derived from the XCLASS fits is 4217, see Table~\ref{tstat}. Hyperfine structure splitting is not considered and counted as one line. For the number of unidentified lines see next section. An automated line counting on the observational spectra gives the same number in orders of magnitude, but is too inaccurate at the band edges and for blended lines. The integrated line intensities are derived from the simulated spectrum, and for bad fits like CO, CS, HCN etc. from Gaussian fits. Because the fits are not perfect, the presented values should be viewed with caution. 

The line density is 5.5/GHz in the first 5 HIFI bands, ranging from 17/GHz in 1a to $<$1/GHz in 5a. Bands 6 and 7 have a high rms and less than 10 lines each. It is evident that the five weeds mentioned in Sect. \ref{subsec:fitting} account for over 75 \% of all lines, and that methanol is the dominant emitter besides  CO. Methyl formate has a very dense spectrum, but is less intense at HIFI frequencies than in both SMA bands where it has the highest number of lines.
\subsection{Unidentified features}
Table~\ref{ulines} lists the frequencies and strengths of the 75 unidentified lines in the HIFI spectra and the 27 unidentified lines in the SMA spectra. The strength of the lines are derived from Gaussian fits. 211 unidentified features were detected initially, going up from band 1a to 4b, and none in the HEB Bands. All of them are emission lines except of one weak absorption line at 507\,686~MHz below 5$\sigma$. Most lines are relatively weak. The highest peak-to-noise ratio of 14 has the line at 492\,325~MHz.
Erroneous features are not listed here, but can be checked by comparing the bands in H and V polarization or at the edges where the bands overlap.
In this way, 5 false lines (called ghosts) at 720~GHz are noticed which are present in 2a, but not 2b. They arose presumably from a feedback in the LO and are no longer present after calibration with a newer version of HIPE. Notable are patterns like branches where several lines (5--10) appear at once in a narrow frequency range (1~GHz), namely at 510, 532, 537, 602, 641, 695, 720, 1009 and 1069~GHz. Other clear patterns like a recurring difference between lines could not be discovered.
\begin{table*}[!]
\centering
  \caption{List of unidentified lines in the SMA and HIFI spectra for NGC~6334I}
\begin{tabular}{lllll|lllll}
  \hline \hline	
  Frequency  & Area  & I  & I/rms  & Width & Frequency  & Area  & I  & I/rms  & Width \\
  (MHz) & (K km\,s$^{-1}$) & (K) &  & (km\,s$^{-1}$) & (MHz) & (K km\,s$^{-1}$) & (K) &  & (km\,s$^{-1}$)  \\
  \hline 
  SMA &  & &  &  & &  &  &  & \\
  \hline
  216671 & 41.2 & 8.1 & 3.2 & 4.8 & 227128 & 68.0 & 11.3 & 5.7 & 5.7 \\ 
  216715 & 71.7 \tablefootmark{b} & 13.5 & 5.4 & 5.0 & 227239 & 26.0 \tablefootmark{b} & 8.1 & 4.1 & 3.0 \\ 
  216913 & 48.0 & 12.4 & 5.0 & 3.6 & 227244 & 32.3 \tablefootmark{b} & 7.0 & 3.5 & 4.3 \\ 
  216919 & 82.8 & 24.1 & 9.6 & 3.2 & 227534 & 61.0 & 14.2 & 7.1 & 4.0 \\ 
  217000 & 54.9 & 11.4 & 4.6 & 4.5 & 227728 & 43.7 & 12.8 & 6.4 & 3.2 \\ 
  217056 & 90.0 & 20.8 & 8.3 & 4.1 & 227803 & 41.6 & 11.4 & 5.7 & 3.4 \\ 
  217262 & 91.5 \tablefootmark{b} & 22.3 & 8.9 & 3.9 & 227823 & 47.8 \tablefootmark{b} & 7.4 & 3.7 & 6.1 \\ 
  217427 & 38.2 & 11.6 & 4.6 & 3.1 & 227841 & 99.0 \tablefootmark{b} & 21.1 & 10.6 & 4.4 \\ 
  217496 & 70.9 & 18.8 & 7.5 & 3.5 & 228001 & 53.0 & 7.6 & 3.8 & 6.6 \\ 
  217593 & 88.3 & 24.5 & 9.8 & 3.3 & 228141 & 112 \tablefootmark{b} & 17.0 & 8.5 & 6.2 \\ 
  217678 & 43.5 & 9.6 & 3.8 & 4.3 & 228230 & 24.5 & 7.5 & 3.8 & 3.1 \\ 
  217689 & 119 & 27.4 & 11.0 & 4.1 &  &  &  &  &  \\ 
  217710 & 120 & 27.7 & 11.1 & 4.1 &  &  &  &  &  \\ 
  217738 & 76.1 & 18.0 & 7.2 & 4.0 &  &  &  &  &  \\ 
  218158 & 116 \tablefootmark{b} & 18.8 & 7.5 & 5.8 &  &  &  &  &  \\ 
  218600 & 58.8 & 11.8 & 4.7 & 4.7 &  &  &  &  &  \\ 
\hline
HIFI &  &  &  &  &  &  &  &  &  \\
\hline
492313 & 0.88 & 0.30 & 14.3 & 2.8   & 1004810 & 2.13 & 0.50 & 4.2 & 4.0\\
501501 & 0.40 \tablefootmark{b}  & 0.06 & 4.0 & 6.3 & 1007185 & 2.50 & 0.44 & 3.1 & 5.3 \\
504512 & 0.14 \tablefootmark{b}  & 0.05 & 3.3 & 3.0 & 1008602 & 2.10 & 0.33 & 2.5 & 6.1\\
505339 & 0.62 & 0.10 & 6.3 & 6.2    &   1008962 & 3.16 & 0.47 & 3.4 & 6.3\\
507167 & 0.72 \tablefootmark{b}  & 0.07 & 4.7 & 9.8 &  1009245 & 1.61 & 0.34 & 2.4 & 4.4\\
507673 & 0.33 & 0.07 & 4.3 & 4.8    &   1009460 & 1.85 & 0.54 & 4.2 & 3.2\\
515619 & 0.25 & 0.06 & 4.3 & 3.6    &   1009618 & 2.31 & 0.62 & 5.6 & 3.5\\
519093 & 0.70 \tablefootmark{b}  & 0.043 & 5 & 9.1 &  1009728 & 2.78 & 0.51 & 4.3 & 5.1 \\
528091 & 0.33 & 0.06 & 4.1 & 5.0   &  1009804 & 2.90 & 0.53 & 4.4 & 5.1 \\
530769 & 0.28 \tablefootmark{b} & 0.09 & 6.0 & 2.8 &  1009850 & 2.74 & 0.58 & 4.8 & 4.5\\
531303 & 0.36 \tablefootmark{b} & 0.062 & 4.1 & 5.4 & 1009881 & 3.20 & 0.50 & 4.2 & 6.0\\
531386 & 0.65 & 0.09 & 6.0 & 6.8   &   1017993 & 4.00 & 0.71 & 7.6 & 5.2\\
534943 & 0.63 \tablefootmark{b} & 0.049 & 3.3 & 12 &   1024147 & 2.78 & 0.60 & 6.0 & 4.3\\
536632 & 0.14 & 0.06 & 3.7 & 2.4   &   1040932 & 1.86 & 0.52 & 3.7 & 3.3\\
539134 & 0.36 & 0.08 & 5.3 & 4.2   &   1045510 & 2.30 & 0.54 & 6.0 & 4.0\\
556311 & 0.44 & 0.10 & 6.7 & 4.1 &   1047493 & 2.00 & 0.47 & 3.6 & 4.1 \\
602910 & 0.53 & 0.10 & 3.1 & 4.8 &   1047982 & 3.34 & 0.62 & 4.1 & 5.0\\ 
649022 & 0.51 & 0.12 & 4.0 & 4.0 &  1052768 & 4.50 \tablefootmark{b} & 0.75 & 3.0 & 5.7\\
649093 & 0.47 & 0.12 & 4.0 & 3.7 &  1065139 & 3.90 & 0.60 & 4.6 & 6.2\\ 
650170 & 0.74 & 0.13 & 3.8 & 5.4 &  1066454 & 2.00 & 0.49 & 4.9 & 3.8\\  
692016 & 0.39 & 0.10 & 2.5 & 3.7 &  1069053 & 1.45 & 0.30 & 3.0 & 4.5\\
693184 & 0.60 & 0.13 & 3.3 & 4.4 &  1069207 & 1.00 & 0.32 & 3.2 & 2.9\\ 
695857 & 0.41 & 0.13 & 3.3 & 2.9 &   1069495 & 1.35 & 0.34 & 3.4 & 3.7 \\
697351 & 0.95 & 0.12 & 3.0 & 7.7 &  1069625 & 1.35 & 0.32 & 3.2 & 3.9\\
698919 & 1.60 \tablefootmark{b} & 0.27 & 6.8 & 5.6 &  1069741 & 1.48 & 0.34 & 3.4 & 4.1\\ 
706552 & 0.55 & 0.16 & 4.0 & 3.2 &  1069851 & 2.70 \tablefootmark{b} & 0.43 & 3.9 & 5.9\\ 
717678 & 0.74 & 0.24 & 5.3 & 2.9 &  1076527 & 3.80 & 0.85 & 7.7 & 4.2 \\ 
741290 & 1.24 & 0.15 & 3.0 & 7.9 &   1078549 & 2.63 & 0.51 & 4.6 & 4.8 \\
743318 & 0.81 & 0.15 & 3.0 & 5.2 &   1088059 & 2.00 & 0.55 & 4.6 & 3.4\\
781479 & 1.40 & 0.22 & 4.4 & 6.0 &  1088302 & 1.65 & 0.44 & 3.7 & 3.4\\ 
785042 & 0.85 & 0.18 & 3.6 & 4.4 &   1095012 & 2.86 & 0.69 & 6.3 & 3.9\\
790927 & 3.30 & 0.52 & 5.8 & 6.0 &   1095149 & 1.70 & 0.45 & 4.1 & 3.5\\
791560 & 1.00 & 0.27 & 3.4 & 3.4 &  1098546 & 2.15 & 0.45 & 4.5 & 4.4\\
795572 & 1.60 & 0.26 & 3.7 & 5.7 &   1099864 & 2.90 & 0.62 & 6.2 & 4.4\\
840812 & 0.80 & 0.21 & 4.2 & 3.6 &   1099903 & 1.50 & 0.32 & 3.2 & 4.4 \\ 
885659 & 1.05 & 0.31 & 4.4 & 3.2 &   1108349 & 1.96 & 0.51 & 3.6 & 3.6 \\ 
972428 & 1.70 & 0.40 & 3.6 & 4.0 &   1112291 & 4.00 & 0.48 & 3.7 & 7.7 \\ 
1000929 & 3.90 & 0.67 & 7.4 & 5.4 &        &      &      &     &     \\
\hline
\end{tabular}
\tablefoot{Given are the center frequency in the velocity rest frame of SMA2, the line area, the peak intensity, the corresponding signal-to-noise ratio, and the line width (FWHM). \tablefoottext{b}: blended line.}
\label{ulines}
\end{table*}
If there are several candidates for an unidentified line, the simulation with XCLASS helps to constrain them. Any anticoincidence, which means that a simulated
line in other quantum numbers in other frequency bands does not match with the observed spectrum, is an indication for a false identification. That is why complete line surveys are better suited for confident line identifications than small windows or single line searches. Another constraint from interferometric data is that all lines of a species originate from the same source.
Since weeds have the highest number of lines, it is most probable that the unidentified lines originate from them, or one of their isotopologues not yet included in the database. Two papers were checked with reported line frequencies of deuterated methanol \ce{CH3OD} \citep{dmethanol} and the isotopologue 
\ce{CH3^{18}OH} \citep{18methanol}. 8 lines could be assigned to deuterated methanol and 128, especially most of the former reported branches, to
\ce{CH3^{18}OH}. Subtracting them, the remaining number of unidentified lines is 75 lines (including the 1009~GHz and 1069~GHz branches)
or about 2\% of the total of 4428 lines.
In the SMA bands, 144 transitions are counted of which 27 are unidentified. That makes 19\% and is much more than in HIFI frequency range.
It is probable that they originate from organic molecules in higher excited states. All unidentified lines in the SMA spectra are detected in both cores, but it is remarkable that, with the exception of two, they are all stronger in SMA2 than SMA1.

\section{Discussion}
\subsection{General remarks}
\citet{Schilke} analyzed spectra of NGC~6334I with APEX at 460 and 810~GHz with a bandwidth of 2.5~GHz. 12 molecules were identified and roughly fitted,
of which 6 show an acceptable agreement (same order of magnitude in $N$) with our results: \ce{CH3CN}, \ce{CH3OCH3}, \ce{CH3OCHO}, HNCO, \ce{N2H+} and \ce{SO2}. The deviation for other molecules can be explained by an over- or underestimation of the source size and by setting the temperature in steps of 50~K. 
In general, there is a good agreement between the HIFI and SMA column densities. The temperatures given in Table~\ref{tsmaf} are
not well determined, because for many molecules only one transition is present. 

Because no error bars are given, it should be pointed out that the given parameters for some molecules are less reliable than for others. 
The spectra of strong lines (CO, \ce{N2H+}, HCN, \ce{C+}...) could be affected by emission in the off-beam position.
In some cases, the assumption of LTE does not hold, e.g. for HCN, or the ambiguity leads to an uncertainty. Reliable values are these which are checked consistently with the SMA and have isotopologues, so that better constraints are available for $\theta$ and $N$. 
\begin{figure}
\resizebox{\hsize}{!}{\includegraphics[width=1\textwidth]{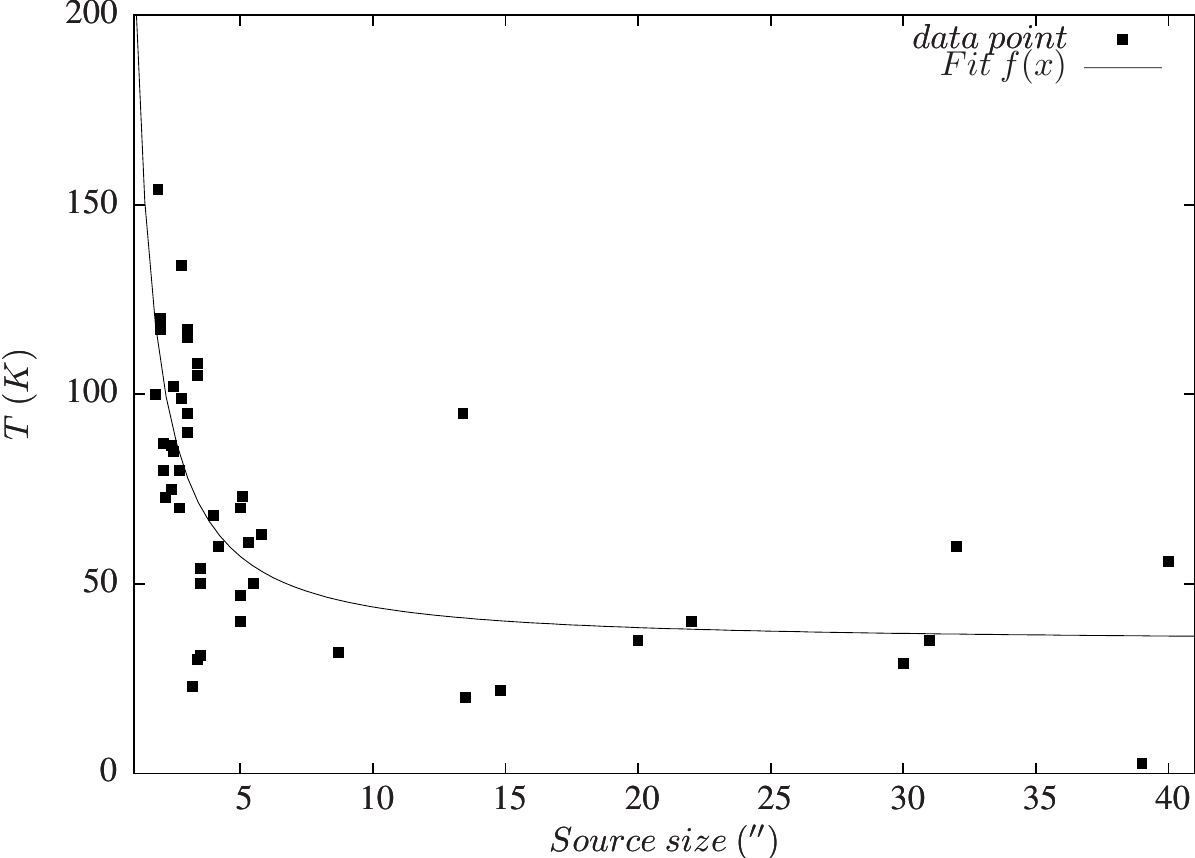}}
\caption{Derived excitation temperatures in dependence of the source size for molecules associated with NGC~6334I.}
\label{temp}
\end{figure}

In Fig.~\ref{temp}, the excitation temperature is plotted as a function of the source size for all molecules associated with NGC~6334I. The extended source sizes ($\theta \gg$ beam size) are not included. The data points can be roughly fitted by a function $f(x)=a \cdot x^{b}+c$ with an exponent of $b=-1.3 \pm 0.5$ and a constant $c=(35 \pm 11)$~K for $\theta>$6\(^{\prime\prime}\) in the outer parts or envelope where the temperature gradient becomes flat. 
This is consistent with models of hot cores which predict a steep increase inwards \citep{Rolffs}. The profile could be affected by the fit function, where in the optical thick limit, $T$ and $\theta$ are inversely proportional to each other with $T\propto 1/ \theta$, but this only concerns few molecules.

\begin{table}
  \centering
  \caption{Comparison of the fractional abundances in NGC~6334I with IRAS~16293-2422 and the Hot Core (HC) and Compact Ridge (CR) of Orion~KL}
    \begin{tabular}{lllll}
    \hline \hline	
 Molecule  & NGC & IRAS& Orion KL& Orion KL \\
   & 6334I &16293-2422$^c$& HC$^e$ & CR$^e$ \\
    \hline	
   CO      &     1.2(-5)           &       4.0(-5)$^d$      &    9.0(-5)     &      1.1(-4)     \\
   \ce{CH3OH}      &     4.7(-6)   &       3.0(-7)          &    1.4(-7)     &       4.0(-7)    \\
   \ce{CH3OCH3}      &   1.0(-6)   &       2.4(-7)          &     8.0(-9)    &       1.9(-8)    \\
   HCN      &    8.7(-7)           &         --             &       3.0(-7)  &       --    \\
   \ce{H2O}      &   6.7(-7)       &         --             &        5.3(-6) &      1.0(-4)     \\
   OCS      &     4.0(-7)          &       2.5(-7)          &     1.1(-8)    &       3.0(-8)    \\
   \ce{CH3OCHO}      &      2.4(-7)&       2.0(-7)          &      1.4(-8)   &          3.0(-8) \\
   CS      &     1.2(-7)           &       3.0(-9)$^d$              &   6.0(-9)     &      1.0(-8)      \\
   SO      &     8.3(-8)           &       2.5(-7)          &     1.9(-7)    &      3.0(-7)     \\
   \ce{H2S}      &  8.3(-8)        &       9.0(-8)          &      2.7(-6)$^f$   &        --   \\
   \ce{H2CO}      &     5.0(-8)    &       6.0(-8)          &     7.0(-9)    &     4.0(-8)      \\
   \ce{SO2}      &      3.3(-8)    &       1.0(-7)          &   1.2(-7)     &   1.6(-7)         \\
   HCOOH      &     1.8(-8)        &       6.2(-8)          &      8.0(-10)   &   1.4(-9)        \\   
  \ce{C2H5CN}  &    1.4(-8)        &       1.2(-8)          &      3.0(-9)   &    5.0(-9)       \\ 
  \ce{H2CS}      &     1.0(-8)    &       5.5(-9)          &        8.0(-10)  &   1.2(-9)       \\
  \ce{CH3CN}      &    9.3(-9)     &      1.0(-8)          &     4.0(-9)     &    5.0(-9)      \\
   HNCO      &       5.0(-9)       &      9.0(-9)          &      5.8(-9)    &       --   \\
   \ce{C2H}      &       1.4(-9)    &      2.2(-10)$^d$      &    5.5(-10)    &    5.3(-9)        \\
    \ce{HC3N}      &    1.4(-9)    &      1.0(-9)          &     1.8(-9)    &      6.0(-9)     \\
     SiO     &      1.0(-9)         &     4.5(-9)           &    6.0(-9)    &      8.0(-9)   \\
   \ce{HCO+}     &      4.3(-10)   &      1.4(-9)$^d$        &    1.1(-9)   &      1.0(-9)       \\
   CN      &      3.7(-10)         &      8.0(-11)$^d$        &     8.0(-10) &      8.0(-10)        \\
   \hline
    \end{tabular}
\tablefoot{Notation a(−b) indicates $\rm a\times 10^{\rm -b}$. (c) References: \citep{Cazaux}, \citep{Schoier}. (d) referring to the cooler, less dense envelope.
           (e) References: \citep{Sutton}, \citep{Caselli} and references therein. (f) from \citep{Persson}.}
  \label{abunv}
\end{table}

The fractional abundances $X$=$N_{\rm Species}/N_{\rm H}$ are distributed widely, ranging over eight orders of magnitude from the nitrogen hydrides ($X$=$10^{-12}$) to CO ($X$=$10^{-5}$, derived from the isotopologues). The hydrogen column density for NGC~6334I is $N_{\rm H}=3 \times 10^{24}$ cm$^{-2}$, see Sect. \ref{xclass}. The most abundant molecules with the highest column densities after CO are \ce{CH3OH} and \ce{CH3OCH3} (both $X$=$10^{-6}$), HCN and \ce{H2O} (both $X$=$10^{-7}$). 
The hydrogen column density can be estimated from the derived CO column density of $N=3.7\times10^{19}$ cm$^{-2}$ by using a standard fractional abundance of [CO]/[H]=$5\times 10^{-5}$. The result is $N_{\rm H}=7.4\times10^{23}$ for a source size of $\theta=32''$ and is lower compared to the value given above where
$\theta=10''$ is assumed. The following differences and similarities are evident when comparing NGC~6334I with an example of a low-mass protostar \object{IRAS 16293-2422} and the high-mass star-forming region \object{Orion KL}, Table~\ref{abunv}: The methanol abundance $X$=$5\cdot10^{-6}$ in NGC~6334I is higher than in the Orion HC and IRAS~16293, which have the same ratio. As is shown by time-dependent chemical models of hot cores \citep{Nomura}, the parent molecule methanol (formed on grain surfaces) is stable for timescales of at least $10^4$ yr. The other organic molecules, dimethyl ether and methyl formate, are more abundant in NGC~6334I than in Orion, reflecting perhaps that NGC~6334I is more time evolved. These second generation molecules start to form in the gas phase between evaporated molecules and peak up in abundance at $10^{4-5}$~yr. After that, the parent molecules are eventually destroyed. For sulfur-bearing molecules, an anticorrelation between NGC~6334I and Orion is observable: while \ce{H2S} is two orders of magnitude less abundant in NGC~6334I, OCS is much more abundant than in Orion. The higher depletion of \ce{H2S} could be explained by the fact that this molecule is efficiently destroyed by \ce{H3O+} in hot cores and subsequently transformed into \ce{SO} and \ce{SO2} \citep{Wakelam}. Overall, the fractional abundance of the sulfur-bearing molecules and their sequence in NGC~6334I is more comparable to IRAS~16293 than to Orion (except for {CS}).
Some nitrogen-bearing molecules like nitriles have about the same fractional abundances in NGC~6334I as in Orion and IRAS~16293: \ce{CH3CN} with $X$=$1 \cdot 10^{-8}$, CN with $X$=$4 \cdot 10^{-10}$, \ce{HC3N} with $X$=$2 \cdot 10^{-9}$ and \ce{C2H5CN} with X=$1 \cdot 10^{-8}$. When comparing fractional abundances in the literature, it should be pointed out that possible uncertainties are the adopted \ce{H2} column density and the source size. It needs to be further investigated if all these deviations are due to different precursor abundances, varying gas temperatures and densities or are due to age effects solely.

Results for most molecules are discussed briefly below, see Fig.~\ref{HIFIfits} and \ref{HIFIfitsb} for their spectra. They are grouped by their moment of inertia, ranging from the lightest diatomic to the heaviest complex organic molecules. The electronic state for molecules is denoted by $^{2S+1}\Lambda_{\Omega}$, where $S$ is the total electronic spin, $\Lambda$
the projection of the total electronic orbital angular momenta along the internuclear axis and $\Omega=\Lambda+\Sigma$, where $\Sigma$ is the projection of $S$ onto the axis.

\subsection{Atomic and diatomic species}
C{\sc i} and C{\sc ii}, Fig.~\ref{fig:candc}. Two fine structure transitions of neutral carbon fall into the submillimeter range, $^3P_1$--$^3P_0$ at 492\,161~MHz and $^3P_2$--$^3P_1$ at 809\,342~MHz. Up to five velocity components can be observed, with the main emission  at --7.3 km\,s$^{-1}$. The nondetection of \ce{^{13}C} is used to set an upper limit for the abundance. The [C{\sc ii}] $^2P_{3/2}$--$^2P_{1/2}$ fine-structure transition at 1.9 THz is one of the most important cooling lines in the ISM and one of the most intense single lines ($\int T {\rm d} v$=262~K$\cdot$km\,s$^{-1}$) in the HIFI spectrum besides CO. Three components are found, corresponding to the ones in C{\sc i}. The large source sizes for all components indicate a broad distribution. The dust attenuation is neglected in the fit of the C{\sc ii} emission line because it is assumed that most of the emission originates from the diffuse outer layer. Therefore, the derived values for $N$ and $T$ should be taken as lower limits, and we notice that some of the spectral features could be affected or caused by contamination in the off-beam.

HF, Fig.~\ref{fig:hf}. The ground transition $J$=1-0 of hydrogen fluoride at 1\,232\,476~MHz is seen in absorption \citep{Emprechtinger}. Three components at $-5$, 0.3 and 7~km\,s$^{-1}$ are visible. The broad velocity width of the $-5$~km\,s$^{-1}$ component indicates that HF is widely distributed and extended over NGC~6334I. HF is considered to be a good tracer of \ce{H2} \citep{Neufeld}, since the formation is directly coupled through the exothermic reaction \ce{F + H2 -> HF + H}, and the ratio in the diffuse medium is $X=3.6\cdot 10^{-8}$. This would correspond to $N_{\rm H}=3\times10^{21}$ cm$^{-2}$ in this diffuse layer.

SO, Fig.~\ref{fig:so}. Sulfur monoxide is a radical with two unpaired electrons (ground state $^3\Sigma$) and a spin-multiplicity of 3, so that each total angular momentum level $\vec{J}=\vec{N}+\vec{S}$ is split into 3 levels according to the projection of $S$. The notation is $J,N \rightarrow J-1,N-1$, where $N$ is the rotational momenta. Beginning from $J$=12, the transitions are strong and can be observed up to $J$=24. Additionally, the isotopologue \ce{^{33}SO} is present for $J$=12 and \ce{^{34}SO} for $J$=12 up to $J$=16. Weaker SO transitions are present $4_3-1_2$ and $5_4-2_3$ in Band 1a and $7_7-6_7$ up to $12_{12}-11_{12}$. The emission comes predominantly from SMA2, as revealed by the velocity of $-7.4$~km\,s$^{-1}$ and the SMA map. The abundance is derived from the isotopologues, but it was necessary to implement a second component which is optically thin.

SiO, Fig.~\ref{fig:sio}. Silicon monoxide is viewed as a good tracer for outflows and the $J$=5--4 transition was observed by \cite{sio}. NGC~6334I shows a bipolar outflow with a red lobe in the north-east and a blue lobe in the south-west direction. With better resolution and additional transitions in the HIFI survey the abundance and spatial position can be given more securely. The $J$=12 and $J$=13 transitions in the HIFI data are quite weak. Therefore the parameters were only manually determined. Notable in the SMA data is the narrow $-17$~km\,s$^{-1}$ component of the blue lobe and the wide, more diffuse outflow in the red wing. Note that the massive outflows are better traced by CO maps of other groups. \citet{outflow} analyzed the ratios of several CO transitions and derived 1--3 $M_{\odot}$ per lobe (dependent on the inclination angle), a dynamical time scale (lobe length divided by $v_{\rm max}$) of $2-4 \times 10^3$ yr and a total mass entrainment rate of $\dot{M}_{\rm tot}=2-15 \times 10^{-3} M_{\odot}$/yr.

NH, Fig.~\ref{fig:nh}. The ground state transitions $J$=1--0 of the radical Imidogen with a spin multiplicity of 3 ($^3\Sigma$) and its hyperfine structure are observed. 
The coupling of the nuclear spin $I$=1 of $^{14}$N with $J$ leads to the splitting into $\vec{F}=\vec{J}+\vec{I}$ components.
To fit it properly, a second broad component had to be included. The $J=1_2-0_1$ transition is blended by the $J$=11--10 transition of HCN.

\ce{SH+}, Fig.~\ref{fig:sh}. Sulfoniumylidene has been detected only recently in Sgr~B2(M) \citep{Menten}. The $1_2$-$0_1$ transition and its hyperfine structure are observed tentatively above 3$\sigma$ in absorption and is heavily blended with methanol. The velocity matches that of SMA1. A comparative study of \ce{SH+} absorption lines towards other high-mass star-forming regions has been made by \citet{Godard}.

CH and \ce{CH+}, Fig.~\ref{fig:ch} and \ref{fig:ch+}. The hyperfine transitions of methylidine (CH) were analyzed by \citet{Wiel}. We included the blended lines from methanol and find three
components: one broad emission from NGC~6334I and two absorption lines from the foreground. The values are in very good agreement with their model I. The two lowest transitions of methylidynium (\ce{CH+}) are observed including the isotope \ce{^{13}C} (Lis et al. 2012, in prep.). The three components are approximately the same as in CH and show the same abundance.

NO, Fig.~\ref{fig:no}. The radical nitrogen monoxide with its doublet structure ($^2\Pi$) is detected in the transitions $^2\Pi_{1/2}, J=$5.5--4.5 (e and f) up to $^2\Pi_{1/2}, J=$8.5--7.5 (e and f). The hyperfine structure is not resolved.

NS, Fig.~\ref{fig:ns}. The electronic ground state of nitrogen sulfide is $^2\Pi$, which splits the rotation levels due to the unpaired electron. All transitions from from $J$=11 to 19 are detected. Noticeable is that the fit for the $\Pi_{1/2}$ states is good, but the $\Pi_{3/2}$ states are underexcited. The optical depth is small, so the source size was fixed at $\theta=3$\(^{\prime\prime}\).

CN, Fig.~\ref{fig:cn}. Due to the weak emission and small number of transitions, there is a large uncertainty of the abundance depending on the velocity width.
A width of 4~km\,s$^{-1}$ would result in an abundance higher by a factor of 2. Here the best fit is taken. The $J=2_{0,3} - 1_{0,2}$ transition at 226.8~GHz in the SMA data is seen in absorption and resembles well the dust emission. The assignment to SMA1 due to $v_{LSR}=-6.4$ km\,s$^{-1}$ is confirmed by the map.
             
CO and CS, Fig.~\ref{fig:co} and \ref{fig:cs}. Carbon monoxide is the most abundant molecule  after \ce{H2} in the ISM. In the HIFI spectrum, CO has the most intense lines, which are optically thick. The line profile shows strong outflows with widths of over 80 km\,s$^{-1}$ and self-absorption. The same is true for carbon monosulfide,
so no direct abundance determination is possible. Therefore only the four detected isotopologues of CS and CO are considered, with a focus on \ce{C^{18}O} and \ce{C^{17}O} by Ceccarelli et al. (2012), in prep. For CS, a multi-component approach was tried with two components from the hot cores and one extended emission. But because the first and third component have the same offset velocity, there might be a redundancy.
\subsection{Polyatomic linear molecules}
\ce{N2H+}, Fig.~\ref{fig:n2h+}. Four emission lines from $J$=6 to 10 of diazenylium are detected which exhibit an asymmetric line profile (Ceccarelli et al. 2012, in prep.). The intensities could not be fitted properly with a single component. The two component model shows a much better fit, but the values are very uncertain and a narrow line width had to be taken. \citet{Russeil} find two components also by observing the hyperfine structure of the $1_{01}-0_{12}$ line, at $-7.3$ and $-10$~km\,s$^{-1}$, but with large differences between them concerning $T$ (127~K to 5~K) and $\Delta v$ (4 to 19 km\,s$^{-1}$). Another possibility could be self-absorption.

HNC and HCN, Fig.~\ref{fig:hnc} and \ref{fig:hcn}. Hydrogen isocyanide is an isomer of hydrogen cyanide (HCN). Transitions from $J$=6 to 10 are present for HNC and $J$=6 to 12 for HCN. As for $\rm N_2H^+$, a two component model was taken. The line profile is changing and affected by self absorption. HCN has a high critical density ($n_{cr}$ $>10^9$ for $J$=9--8 at 797~GHz) and shows an outflow in the red wing of the asymmetric line profile with self-absorption, which indicates an infall. Also vibrationally excited lines ($\nu_2=1,2$) of HCN are observed. Both aspects, infall and hot HCN, are characteristic for evolved cluster-forming regions \citep{Rolffs}. A uniform fit with the isotopologues of HCN is included for the completeness of the survey. For HNC, no other isotopologues are found apart from the \ce{HN^{13}C} $J$=6--5 line.

\ce{HCO+}, Fig.~\ref{fig:hco}. The rotational transitions of the formyl ion are relatively intense and present up to band 5a (Ceccarelli et al. 2012, in prep.). At first, one component including H\textsuperscript{13}CO\textsuperscript{+} was tried, but the fitting analysis revealed that either the lower-$J$ or the higher-$J$ transitions are consistent but not all together. This is an indication that either LTE is not satisfied, or several components with different excitation conditions overlap. Assuming two components with different optical depths, the fit could be improved. At higher transitions the component with the higher optical depth dominates.

OCS, Fig.~\ref{fig:ocs}. Carbonyl sulfide is detected in the rotational transitions from $J$=40 to 53. Because at first the source size was unknown and no isotopologues are detected in the HIFI spectrum, the abundance was underestimated due to a low opacity ($\tau<0.3$). With the detection of three isotopologues in the SMA bands the column density was corrected upwards. The emission of O\textsuperscript{13}CS comes predominantly from SMA1.

\ce{HC3N}, Fig.~\ref{fig:hc3n}. Cyanoacetylene lines are weak in the HIFI data, but more intense in the SMA data. The fit for SMA1 gives a large line width of 9 km\,s$^{-1}$, which is
taken for the HIFI model.

\subsection{Symmetric top molecules} 
\ce{CH3CN}, Fig.~\ref{fig:mcn}. Methyl cyanide is considered to be a good tracer of hot cores \citep{olmi}. Rotational transitions from $J$=27 to 39 with $K$-components $K$=0--6 are detected in the HIFI spectrum. \citet{Thorwirth} fitted the $J$=12--11 transition with an unknown source size resulting in $T$=145~K and $N$=$5.2 \times 10^{15}$ cm$^{-2}$.
The temperature is in good agreement with our value of 154~K. Additionally, weak ($<3\sigma$) bending modes $v_8$=1 are observed in band 1a. The small source size and high temperature confirms that methyl cyanide probes the central part of hot cores.

\ce{NH3}, Fig.~\ref{fig:nh3}. The ground transitions of ammonia are detected mostly in absorption. The overlapping emission and absorption components could not be fitted consistently, and the column density for the emission component associated with NGC~6334I is therefore probably underestimated. 

\subsection{Inorganic asymmetric molecules} 
\ce{NH2}, Fig.~\ref{fig:nh2}. Azanyl is an asymmetric top with a hyperfine structure. Four transitions are detected including the ground state transition. The emission component can be associated with NGC~6334I. As in the case of NH, a broad component ($\Delta v$=30 km\,s$^{-1}$) is added to fit the absorption.

{\ce{H2Cl+} and \ce{HCl}, Fig.~\ref{fig:h2cl} and \ref{fig:hcl}. Hydrogen chloride and chloronium were analyzed by \citet{Lis}. The $J$=2--1 at 1251~GHz in band 5b of \ce{HCl} is additionally
detected, showing a second component in absorption from the colder and more extended envelope at $-10$~km\,s$^{-1}$ with a broad width comparable to \ce{H2Cl+}. The emission component can be assigned to the center of NGC~6334I. The abundance for this component differs by 50\% compared to \citep{Lis} because the source size is not known and the lines are optically thick and affected by the absorption feature. For \ce{H2Cl+}, the blended lines of dimethylether are included in the fit (blue line). Its velocity and line width leads to the conclusion that it originates from a diffuse foreground molecular cloud. 

\ce{H2O+}, Fig.~\ref{fig:h2o+}. Oxidaniumyl was analyzed by \citet{Ossenkopf}. Additionally, another weak fine structure transition was observed at 1140~GHz from the $J$=0.5 level. The derived abundance is above the lower limit given by \citet{Ossenkopf}. Its radial velocity and line width leads to the conclusion that it is not associated with NGC~6334I. 

\ce{H2O}, Fig.~\ref{fig:h2o}. Water was analyzed by \citet{water,Emprechtinger} . It has complex line shapes with many components like optically thick emission and broad outflows, therefore no fit was attempted. Instead, the optically thin isotopes were included to get an estimation of the abundance. Most fits including deuterated water can not reproduce the
line strength, so that a non-LTE calculation method is necessary to model them properly. 

\ce{H2S}, Fig.~\ref{fig:h2s}. Hydrogen sulfide shows 3 components. The emission one at $-6.1$~km\,s$^{-1}$ is affected by two absorption lines, of which one is from the outflow. 
While the emission peak could be constrained through the isotopologues, the absorptions features could not be reproduced well.
\subsection{Organic asymmetric molecules}
\ce{H2CO}, Fig.~\ref{fig:fa}. Formaldehyde is a nearly prolate top and the asymmetry splits every $K\neq 0$ level into doublets. For astrophysical applications, due to its structure this molecule is a good probe of the kinetic temperature and volume density and traces the inner and outer part of hot cores \citep{probe}. Interstellar formaldehyde is formed by hydrogenation of solid CO: \ce{H + CO -> HCO + H -> H2CO}. Another less efficient pathway in the gas phase is \ce{CH3 + O -> H2CO + H}.
Besides of the weeds, this organic molecule has the most intense lines throughout the HIFI spectrum from band 1a to 4b. The rotational transitions range from $J$=7 to 15 with $K$=0 to $K$=6, and the isotopologues $\rm H_2^{13}CO$ and $\rm H_2C^{18}O$ are observed. The  fits with XCLASS show two problems: one component is optically thick, and all the $K$ components could not be described by a single rotation temperature. The $K$=0,1 do not match with $K$=2--6 components consistently. From the SMA data the $3_0-2_0$ transition shows that the $K$=0 value is lower in intensity than the $K$=2 component. This hints to self absorption in SMA1 or optical thickness.

\ce{H2CCO} and \ce{H2CS}, Fig.~\ref{fig:h2cco} and \ref{fig:h2cs}. Ethenone and thioformaldehyde are nearly prolate tops with a spectrum similar to \ce{H2CO}. For \ce{H2CCO}, only the $K$=1 components from $J$=24 to 27 are detected, the other $K$=0 and $K$=3 components  are included to set an upper limit. Since the transitions are weak and the optical depth is small, $\rm \theta$ and $\Delta v$ needed to be fixed. For \ce{H2CS} ranging from $J$=14 to 23 with $K$=0 to $K$=5 transitions, a good fit is obtained where all $K$ components can be described consistently with one excitation temperature.

\ce{CH3OH}, Fig.~\ref{fig:m1} and \ref{fig:m2}. Methanol is the strongest weed in the HIFI spectrum and accounts for about half of all detected lines. One component is optically thick, which can be derived from the fit of the \ce{^{13}CH3OH} isotopologue, and the other component accounts for the broad emission. The model fits the observation quite well, except of the most intense lines. As can be seen in the SMA spectra, the XCLASS LTE model cannot reproduce the methanol lines due to a high optical depth.
Furthermore, high energy lines $>500$~K are underestimated, so that a 3rd component might be necessary.
      
\ce{NH2CHO}, Fig.~\ref{fig:form}. Formamide is only detected weakly in band 1a. The optical depth is small, and the source size is determined from the SMA map which shows that
the emission comes mainly from SMA2.

\ce{C2H4O}, Fig.~\ref{fig:eto}. Ethylene oxide is surprisingly only detected in SMA2. As for formamide, many weak lines are observed. 
The source velocity $v_{\rm LSR}$ in the HIFI data is consistent with the general source velocity of SMA2.

HNCO, Fig.~\ref{fig:hnco}. Isocyanic acid is almost linear, but asymmetric and shows relatively weak emission. The very high temperature and small source size may trace the hot core SMA2.

\ce{CH3OCH3}, Fig.~\ref{fig:dm}. Dimethyl ether has two methyl groups, and similarly to methanol, a splitting into 4 sublevels (AA,EE,AE,EA). It is, after methanol, the organic molecule with the most dense and intense lines in the entire HIFI spectrum. The laboratory measured spectrum from the \ce{^{13}CH3OCH3} isotopologue are taken from \citep{koerber}, but are not yet publicly available. The SMA spectra show that the dominant emission arises from SMA1 and that they are weak associations with the outflows.

\ce{CH3OCHO}, Fig.~\ref{fig:mf}. Methyl formate has the densest spectrum of all detected molecules in the lowest HIFI band 1a (Brouillet et al. 2012, in prep.) and the highest number of lines in both SMA bands. The SMA maps show an association with the outflows. 

\section{Conclusions}
In this article, observational data from the HIFI instrument aboard the Herschel Space Observatory  and from the Submillimeter Array are analyzed for the high-mass star-forming region NGC~6334I. We employ a LTE radiative transfer model in order to get insights into the physical structure of the source. The molecular spectral line survey gives an overview of the chemical inventory of the molecular species and their excitation, whereas the high-resolution spectrometer allows tracing the kinematics of the different velocity components. The interferometer maps reveal the source  morphology and help to distinguish the embedded cores and resolve their sizes. The results fit well with the overall picture of a high-mass young stellar object and hot molecular cores described by \citet{Dishoeck}.

Two main software tools, XCLASS and MAGIX, are used to demonstrate that modeling and handling of the vast HIFI data sets is possible. These tools allow to decompose the spectrum into contributions of individual molecules and their isotopologues. By assuming LTE condition and employing the LM fitting algorithm, a fast extraction of information is possible. In order to take line blends into account, a model for every molecule is produced, which is consistent over multiple HIFI bands. It should be clear that for other excitation conditions than LTE and for optically thick lines like CO this method reaches its limitations. Sophisticated 3-d radiative transfer codes such as RADMC-3D\footnote{\url{http://www.ita.uni-heidelberg.de/~dullemond/software/radmc-3d/}} and LIME \citep{LIME} with the possibility of non-LTE calculations will be used in the future in order to model NGC~6334I into more detail, taking into account the geometry and non-homogeneous distribution of temperature and abundances. With the improved capabilities of ALMA (Atacama Large Millimeter Array), it will be possible to study the molecular chemistry and kinematics on sub-arcsecond scales ($<$~1000~AU) and to follow the fragmentation of protostellar cores with much improved mass sensitivity.

The molecular survey revealed a variety of chemical species, including saturated organic molecules. The astrochemical question is: how can these abundances be explained and how are they related to each other? The next step is therefore a careful comparison of the column densities presented here with predictions of astrochemical models and detailed comparisons with the molecular abundances and abundance ratios in other CHESS targets.

\begin{acknowledgements}
This work was carried out within the Collaborative Research Council 956, funded by the Deutsche Forschungsgemeinschaft (DFG).
C. Ceccarelli acknowledges the financial support from the French spatial agency CNES and the Agence Nationale pour la Recherche (ANR), France (project FORCOMS, contracts ANR-08-BLAN-022). Support for this work was provided by NASA through an award issued by JPL/Caltech.

HIFI has been designed and built by a consortium of institutes and university departments from across Europe, Canada and the United States under the leadership of SRON Netherlands Institute for Space Research, Groningen, The Netherlands and with major contributions from Germany, France and the US. Consortium members are: Canada: CSA, U.Waterloo; France: CESR, LAB, LERMA, IRAM; Germany: KOSMA, MPIfR, MPS; Ireland, NUI Maynooth; Italy: ASI, IFSI-INAF, Osservatorio Astrofisico di Arcetri-INAF; Netherlands: SRON, TUD; Poland: CAMK, CBK; Spain: Observatorio Astronómico Nacional (IGN), Centro de Astrobiología (CSIC-INTA). Sweden: Chalmers University of Technology - MC2, RSS \& GARD; Onsala Space Observatory; Swedish National Space Board, Stockholm University - Stockholm Observatory; Switzerland: ETH Zurich, FHNW; USA: Caltech, JPL, NHSC. 
\end{acknowledgements}
\bibliographystyle{aa} 
\bibliography{ngc}

\Online
\begin{appendix} 
\onecolumn
\section{SMA spectra and tables}
\begin{table*}[htbp]
  \centering
  \caption{XCLASS fit results from SMA for NGC~6334I}
    \begin{tabular}{llrrrrrr}
\hline \hline
   Source & Species & $\Theta$   & $T_{ex}$ & $N$         & $\Delta v$ & $v_{LSR}$ & Notes  \\
          &         & (\arcsec)  &  (K)     & (cm$^{-2}$) & (km/s)     & (km/s)    &        \\
\hline
   SMA1& CN    & 2  & 3    & $0.2 \times 10^{15}$ & 2.2     & 1.3  &  \\
   SMA1& CN      & 2  & 3    & $0.8 \times 10^{15}$ & 6.2     & -4.6  & \\
   SMA2& CN      & 2  & 3    & $0.7 \times 10^{15}$ & 6.9     & -0.6  &  \\
   SMA2& CN     & 2  & 3    & $1.5 \times 10^{15}$ & 3.2     & -4.5  & \\
   SMA1& DCN   & 3.5   & 359    & $1.6 \times 10^{15}$ & 5.8     & -4.9  &   2     \\
   SMA2& DCN       & 5.0   & 146    & $0.6 \times 10^{15}$ & 6     & -7.8  & 2       \\
   SMA1& SiO   & 1.7  & 40    & $1.0 \times 10^{15}$ & 8     & -3.2 &     2  \\
   SMA2& SiO       & 2.0  & 40    & $1.0 \times 10^{15}$ & 6     & -5.2  &  2     \\
   SMA1& \ce{H2S}       & 3       & 62    & $1.4 \times 10^{17}$ & 5       & -4  & 2,3 \\
   SMA2& \ce{H2S}       & 3       & 147   & $0.4 \times 10^{17}$ & 6.5     & -6  &  2,3       \\
   SMA1& CS & 2.4     & 40    & $1.6 \times 10^{18}$ & 5.1     & -3.2    &  2,3,5 \\
   SMA2& CS & 2.2     & 32    & $6.6 \times 10^{17}$ & 2.7     & -6.4   &   2,3,5  \\
   SMA1& OCS       & 2.5     & 137   & $2.3 \times 10^{18}$ & 4       & -5.6  & 1    \\
   SMA2& OCS      & 2.5     & 126   & $1.1 \times 10^{18}$ & 4       & -7  & 1    \\
   SMA1& \ce{CH3OH} & 4   & 78   & $2.4 \times 10^{19}$ & 3.6     & -5.2  &     \\
   SMA2& \ce{CH3OH} & 4   & 104    & $2.0 \times 10^{19}$ & 3.0     & -7.5  &       \\
   SMA1&  \ce{CH3OCH3} & 3.0   & 78    & $2.3 \times 10^{18}$ & 4.9   & -5.6  &    \\
   SMA2&  \ce{CH3OCH3}  & 3.0     & 98    & $0.7 \times 10^{18}$ & 5.4   & -7.7    &        \\
   SMA1& \ce{CH3OCHO} & 2.3     & 104   & $7.7 \times 10^{17}$ & 4.6   & -6.1  &       \\
   SMA2& \ce{CH3OCHO}         & 3     & 115   & $9.6 \times 10^{17}$ & 3.8   & -7.6  &       \\
   SMA1& \ce{H2CO}  & 3.0   & 104    & $1.6 \times 10^{17}$ & 5.7   & -5.5  &  \\
   SMA2& \ce{H2CO}  & 3.0   & 138    & $1.3 \times 10^{17}$ & 4.6   & -7.5 &        \\
   SMA1& \ce{NH2CHO} & 2     & 25   & $4.7 \times 10^{15}$ & 5.2     & -4.7 &   2,3    \\
   SMA2& \ce{NH2CHO} & 2     & 45   & $5.0 \times 10^{15}$ & 4.0     & -6.9 &   2,3    \\
   SMA1& \ce{HC3N}  & 2.8     & 208   & $4.2 \times 10^{15}$ & 9.4     & -5.7  &       \\
   SMA2& \ce{HC3N}  & 2.4     & 223   & $1.4 \times 10^{15}$ & 5.8     & -6.6  &       \\
   SMA1& \ce{C2H3CN}    & 2.0   & 147   & $5.2 \times 10^{15}$ & 6.0   & -6.3    &       \\
   SMA1& \ce{C2H5CN}    & 2.2   & 280   & $4.2 \times 10^{16}$ & 6.6   & -5.5    &       \\
   SMA2& \ce{C2H5CN}    & 1.0   & 123   & $1.6 \times 10^{16}$ & 4.1   & -6.7    &        \\
   SMA1& \ce{CH3CH3CO}  & 2.6   & 73   & $2.1 \times 10^{16}$ & 5.6   & -5.4    &        \\
   SMA2& \ce{CH3CH3CO}  & 2.6   & 159   & $2.9 \times 10^{16}$ & 4.8   & -7.4    &        \\
    SMA2& \ce{HCOOH}    & 2.0   & 258   & $5.5 \times 10^{16}$ & 4.3   & -9   &    2    \\          
    SMA2& \ce{(CH2OH)_2} & 2.2   & 217   & $2.6 \times 10^{16}$ & 3.4   & -8.2    &       \\
    SMA2& \ce{C2H5OH}  & 2.5     & 77   & $1.4 \times 10^{17}$ & 4.9     & -7.9    &      \\
    SMA2& \ce{C2H4O}  & 2.1   & 335    & $6.7 \times 10^{16}$ & 5     & -6.8  &       \\
    SMA2& \ce{SO}    & 2.5   & 70    & $1.7 \times 10^{17}$ & 4.2     & -8.7  &  2,6 \\
\hline
    \end{tabular}%
  \label{tsmaf}%
\tablefoot{(1) including only isotopologues \ce{^{18}OCS}, \ce{O^{13}CS}, \ce{OC^{33}S}. (2) 1 or few lines. (3) blended.
            (4) including \ce{^{13}CH3OH}. (5) including only \ce{^{13}C^{34}S}. (6) including only \ce{^{33}SO}.}
\end{table*}%

\begin{landscape}
\begin{figure}
  \centering 
  \subfloat{\includegraphics[angle={270},width=12cm]{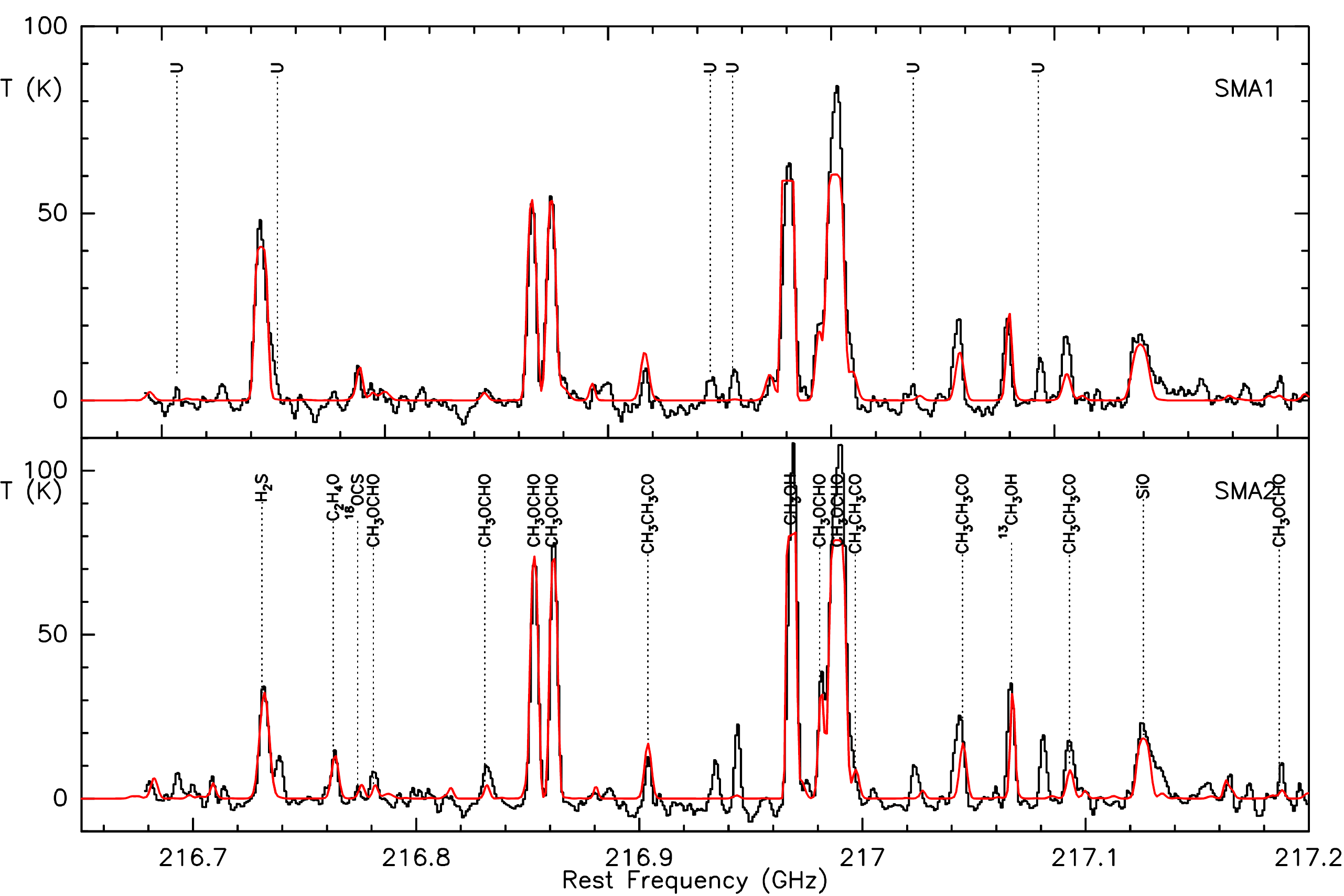}}
  \subfloat{\includegraphics[angle={270},width=12cm]{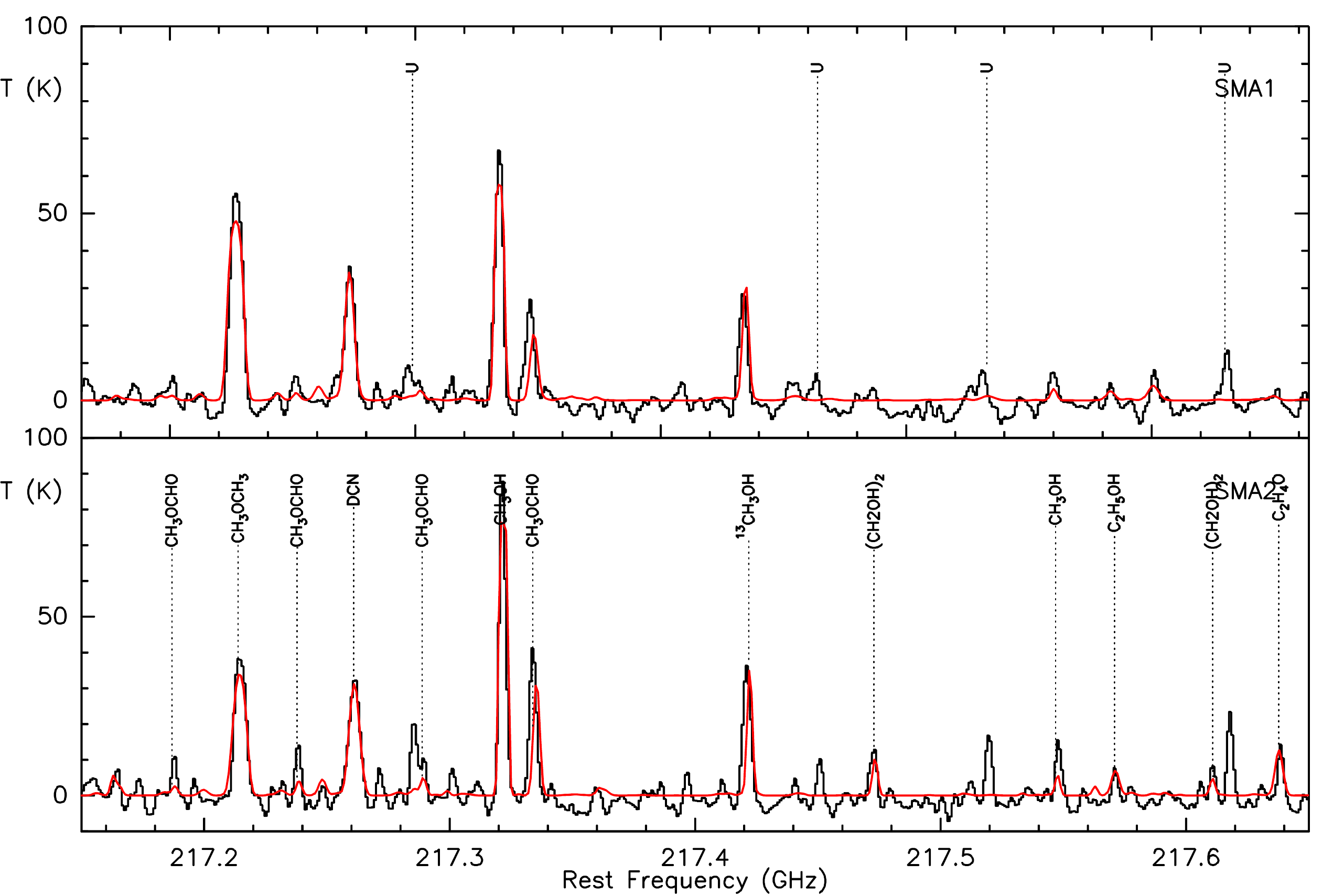}}\\
  \subfloat{\includegraphics[angle={270},width=12cm]{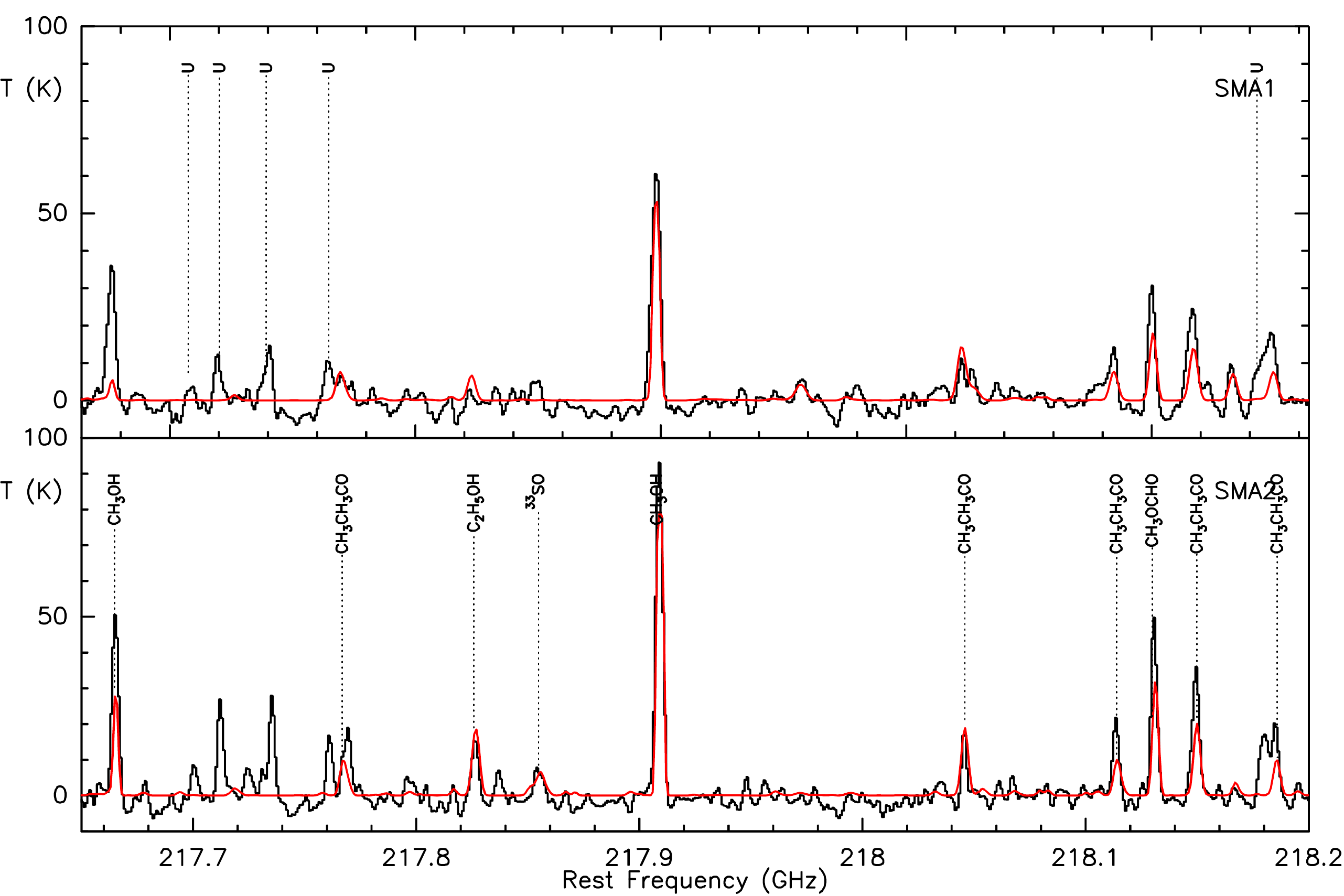}}
  \subfloat{\includegraphics[angle={270},width=12cm]{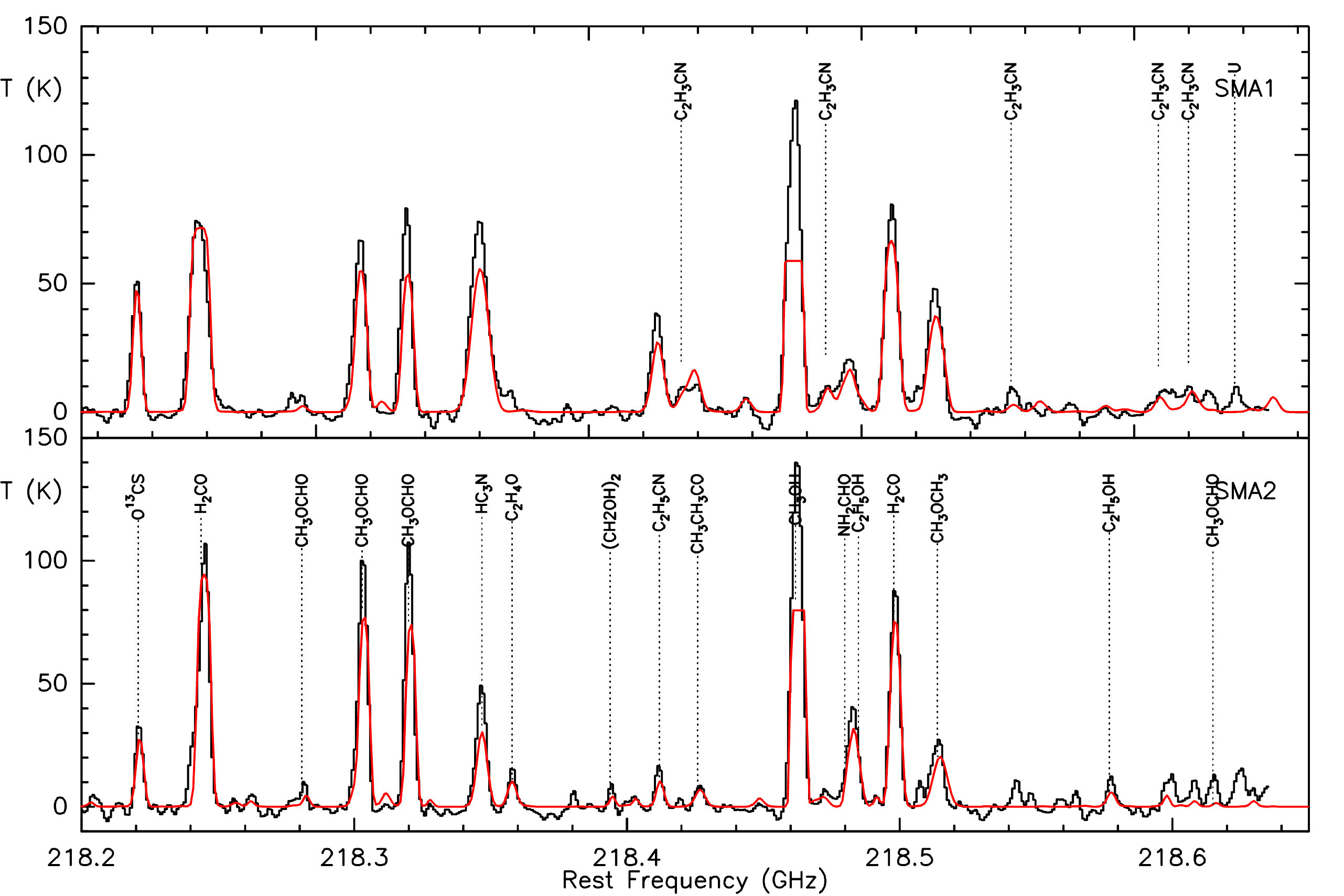}}
  \caption{LSB Spectra (in steps of 500 MHz) of the two hot cores SMA1 (upper) and SMA2 (lower) of NGC 6334I, observed with the Submillimeter Array. Overlaid in red are the XCLASS fits. For more detailed information about each identified spectral line, see Table \ref{smail}.}
  \label{smafits}
\end{figure}
\end{landscape}

\begin{landscape}
\begin{figure}
 \ContinuedFloat 
  \centering  
  \subfloat{\includegraphics[angle={270},width=12cm]{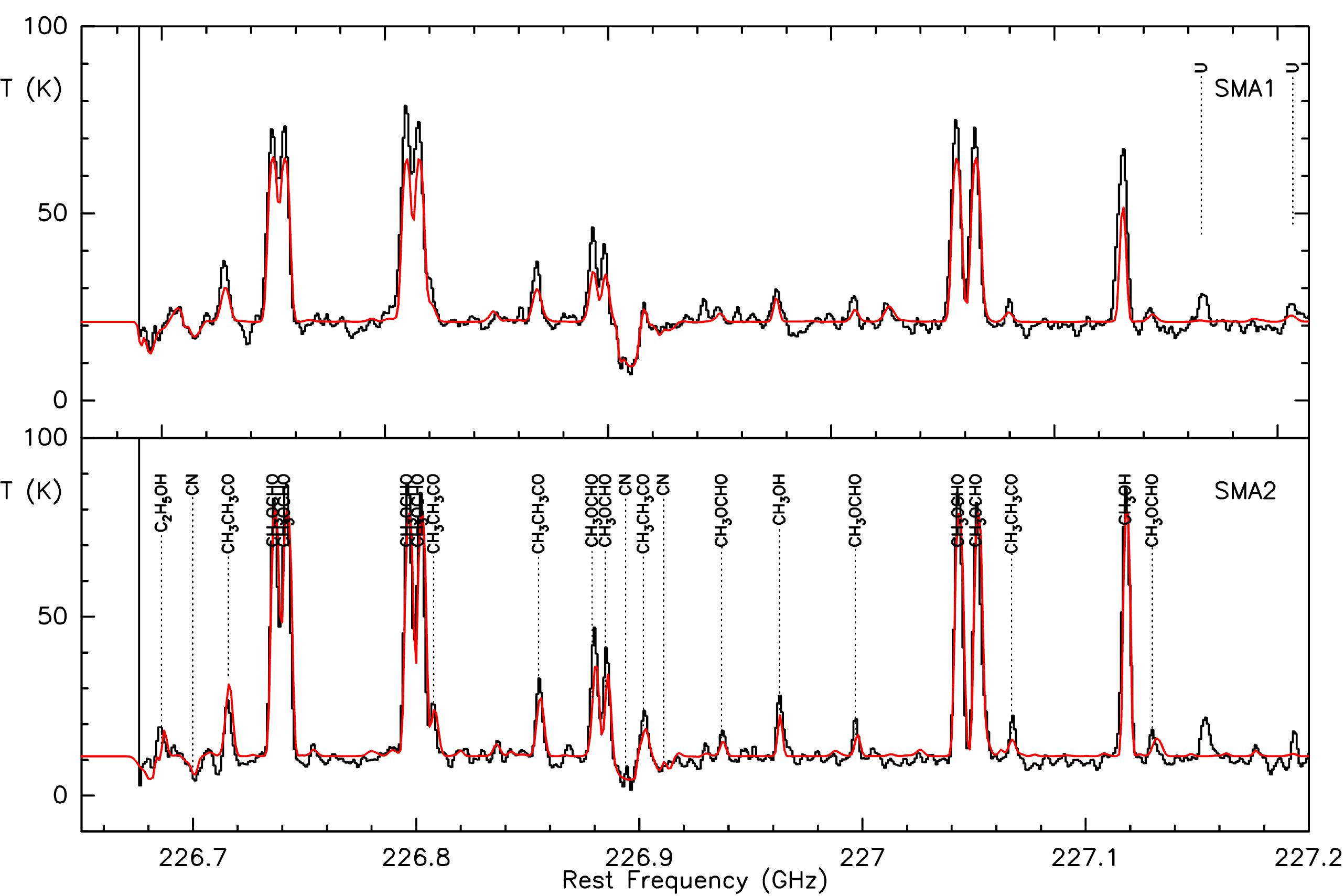}}
  \subfloat{\includegraphics[angle={270},width=12cm]{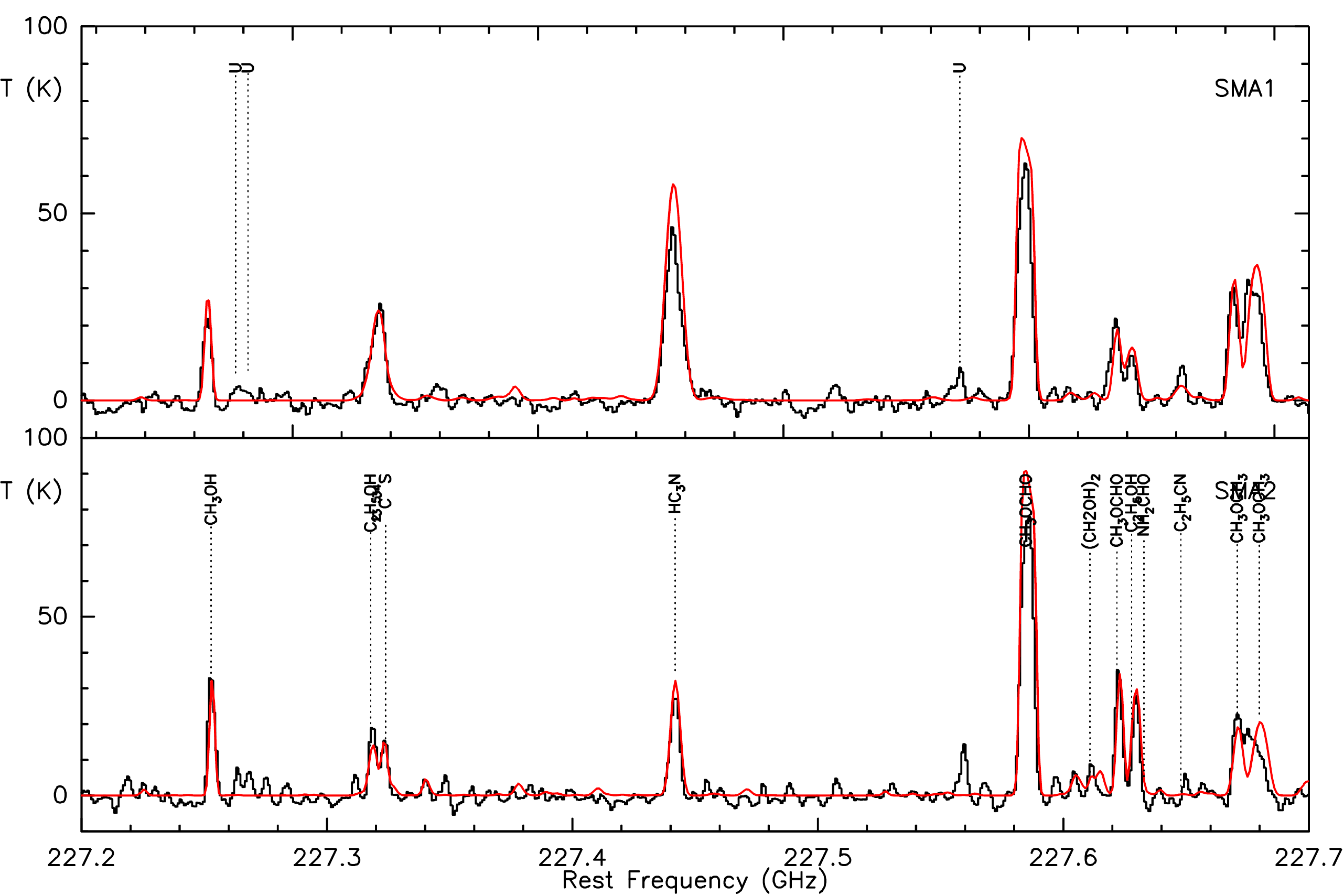}}\\
  \subfloat{\includegraphics[angle={270},width=12cm]{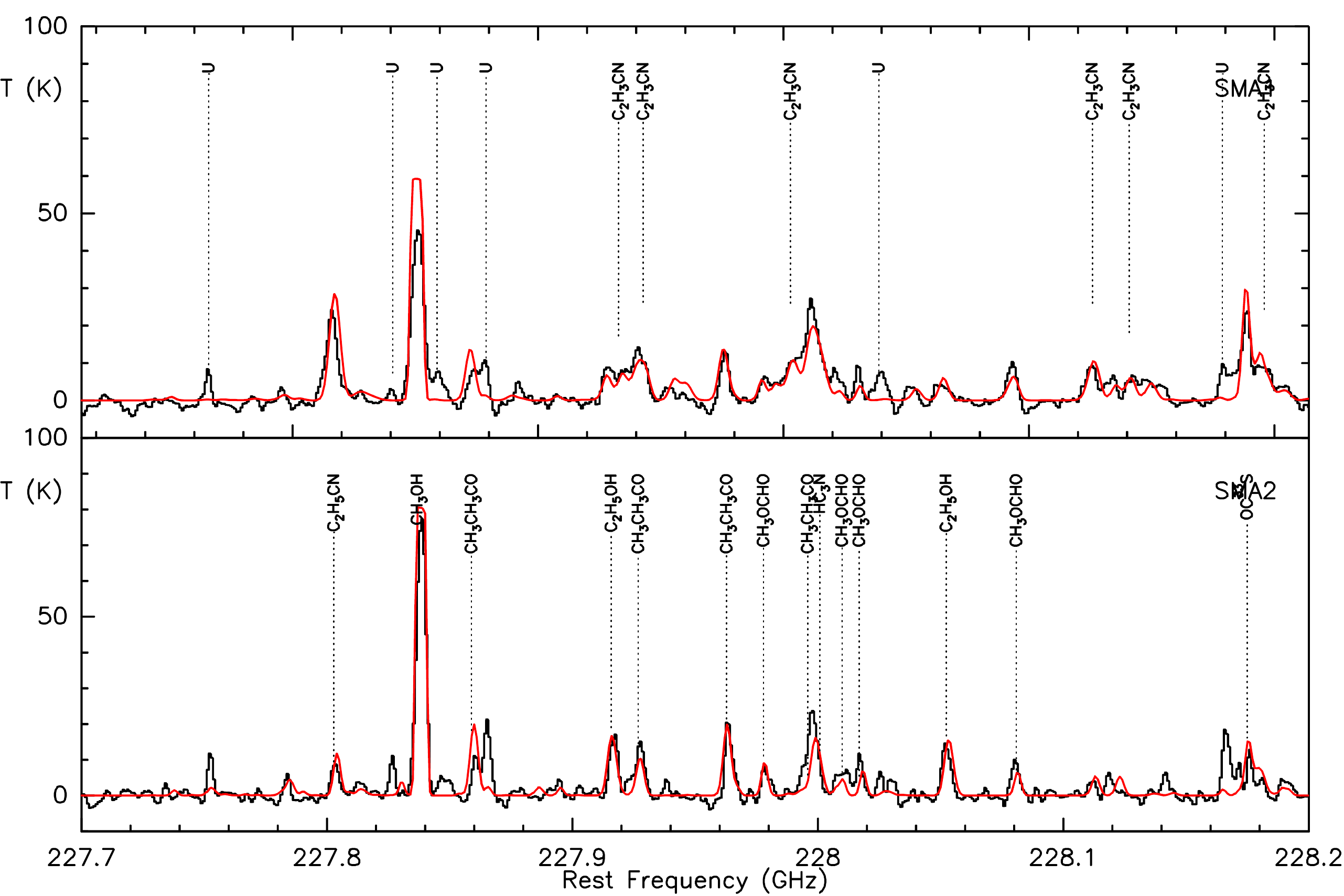}}
  \subfloat{\includegraphics[angle={270},width=12cm]{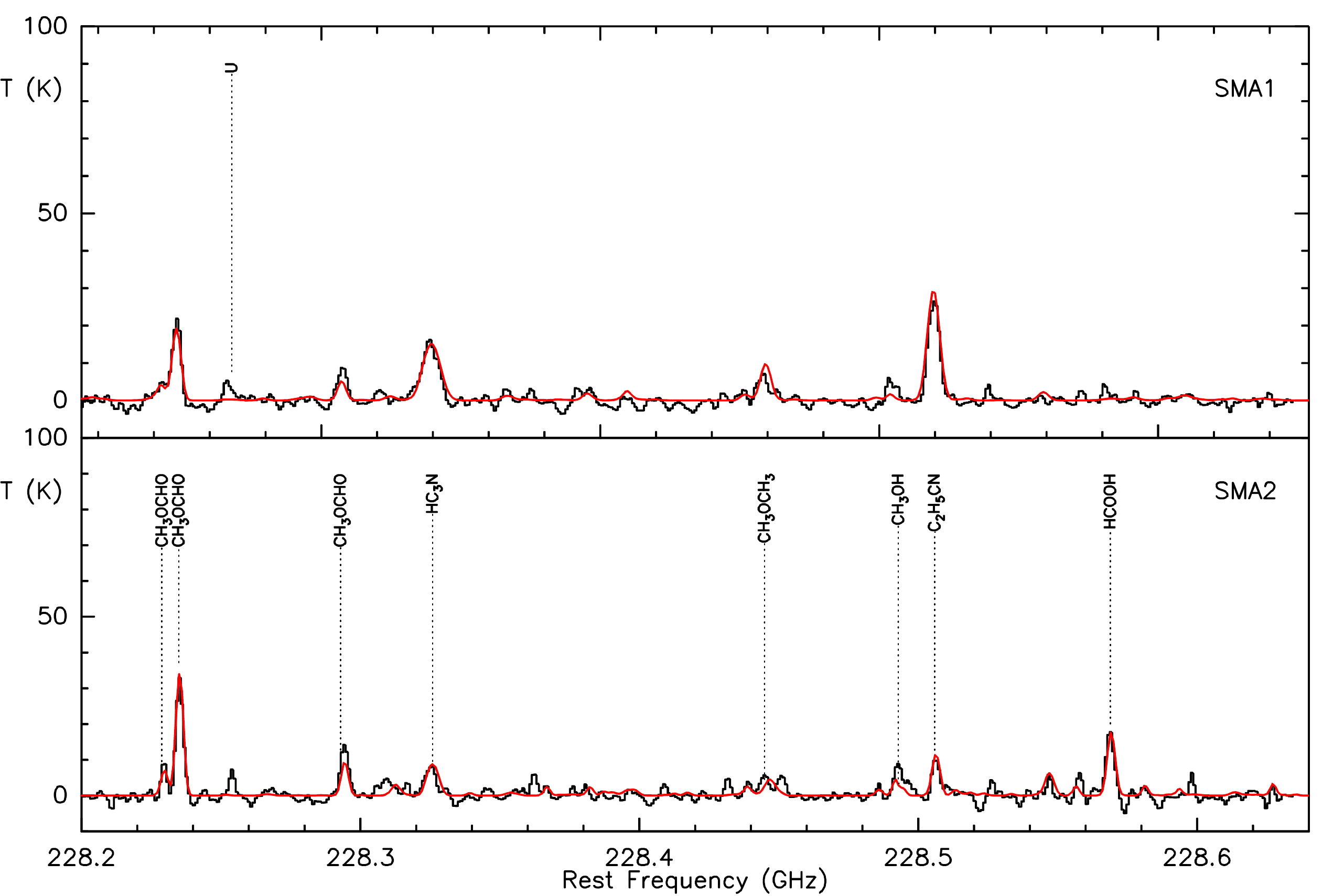}}
 \caption{Continued. SMA USB spectra.}
\end{figure}
\end{landscape}

\renewcommand{\baselinestretch}{1}
\begin{table*}\scriptsize
\centering 
\caption{List of identified lines in the SMA bands of NGC 6334I in order of increasing rest frequency}
\begin{tabular}{lll|lll}
\hline \hline
         &                         &     LSB            &           &                        &        USB        \\
Molecule & Transition & Frequency (MHz) & Molecule & Transition & Frequency (MHz)\\
\hline
\ce{H2S} &  2 2 0 - 2 1 1  & 216710.437        	        & \ce{C2H5OH} &  10 2 9 2      - 9 1 8 2 & 226661.701 \\ 
\ce{C2H4O} &  16 5 11       - 16 4 12 & 216741.468 	        & CN &  2 0 2 2       - 1 0 1 1 & 226679.311 \\ 
\ce{C2H4O}&  16 6 11       - 16 5 12 & 216741.712 	        & \ce{CH3CH3CO},v=1 &  20 3 17 0     - 19 4 16 1 & 226692.48 \\ 
\ce{^{18}OCS} &  19            - 18 & 216753.488        	& \ce{CH3CH3CO},v=1  &  20 4 17 0     - 19 3 16 1 & 226692.48 \\ 
\ce{CH3OCHO} &  32 6 27 1     - 32 5 28 1 & 216758.969 	        & \ce{CH3OCHO} &  20 2 19 1     - 19 2 18 1 & 226713.06 \\ 
\ce{CH3OCHO} &  32 6 27 0     - 32 5 28 0 & 216809.014 	        & \ce{CH3OCHO} &  20 2 19 0     - 19 2 18 0 & 226718.688 \\ 
\ce{CH3OCHO} &  18 2 16 2     - 17 2 15 2 & 216830.197 	        & \ce{^{13}CH3OH} &  12 5 8 0      - 13 4 9 0 & 226716.392 \\ 
\ce{CH3OCHO} &  18 2 16 0     - 17 2 15 0 & 216838.889 	        & \ce{CH3CH3CO},v=0 &  20 4 17 1     - 19 3 16 1 & 226784.639 \\ 
\ce{CH3CH3CO},v=1 &  19 4 16 0     - 18 4 15 1 & 216881.337  	& \ce{CH3CH3CO},v=0 &  20 3 17 1     - 19 4 16 1 & 226784.639 \\ 
\ce{CH3CH3CO},v=1 &  19 3 16 0     - 18 4 15 1 & 216881.337 	& \ce{CH3CH3CO},v=0 &  20 4 17 1     - 19 3 16 2 & 226784.67 \\ 
\ce{CH3CH3CO},v=1 &  19 4 16 0     - 18 3 15 1 & 216881.337	& \ce{CH3CH3CO},v=0 &  20 3 17 1     - 19 4 16 2 & 226784.67 \\ 
\ce{CH3CH3CO},v=1 &  19 3 16 0     - 18 3 15 1 & 216881.337 	& \ce{CH3OCHO} &  20 1 19 2     - 19 1 18 2 & 226773.13 \\ 
\ce{CH3OH} &  5 1 0 0       - 4 2 0 0 & 216945.521           	& \ce{CH3OCHO} &  20 1 19 0     - 19 1 18 0 & 226778.786 \\ 
\ce{CH3OCHO} &  17 3 14 5     - 16 3 13 5 & 216958.834 	        & \ce{CH3CH3CO},v=0 &  20 4 17 0     - 19 3 16 1 & 226832.055 \\ 
\ce{CH3OCHO} &  20 0 20 2     - 19 1 19 1 & 216962.989 	        & \ce{CH3CH3CO},v=0 &  20 3 17 0     - 19 4 16 1 & 226832.055 \\ 
\ce{CH3CH3CO},v=0 &  19 4 16 1     - 18 3 15 1 & 216974.426 	& \ce{CH3OCHO} &  20 2 19 1     - 19 1 18 2 & 226856.872 \\ 
\ce{CH3CH3CO},v=0 &  19 3 16 1     - 18 4 15 1 & 216974.426 	& \ce{CH3OCHO} &  20 2 19 0     - 19 1 18 0 & 226862.257 \\ 
\ce{CH3CH3CO},v=0 &  19 4 16 1     - 18 3 15 2 & 216974.464 	& CN &  2 0 3 3       - 1 0 2 2 & 226874.191 \\ 
\ce{CH3CH3CO},v=0 &  19 3 16 1     - 18 4 15 2 & 216974.464	& CN &  2 0 3 4       - 1 0 2 3 & 226874.781 \\ 
\ce{CH3CH3CO},v=0 &  19 4 16 0     - 18 3 15 1 & 217022.509	& \ce{CH3CH3CO},v=0 &  20 4 17 0     - 19 3 16 0 & 226879.393 \\ 
\ce{CH3CH3CO},v=0 &  19 3 16 0     - 18 4 15 1 & 217022.509 	& \ce{CH3CH3CO},v=0 &  20 3 17 0     - 19 4 16 0 & 226879.393 \\ 
\ce{^{13}CH3OH} &  14 1 13 0     - 13 2 12 0 & 217044.616       & CN &  2 0 3 2       - 1 0 2 2 & 226887.42 \\ 
\ce{CH3CH3CO},v=0 &  19 4 16 0     - 18 3 15 0 & 217070.504 	& CN &  2 0 3 3       - 1 0 2 3 & 226892.128 \\ 
\ce{CH3CH3CO},v=0 &  19 3 16 0     - 18 4 15 0 & 217070.504 	& \ce{CH3OCHO} &  6 6 1 4       - 5 5 1 4 & 226913.736 \\ 
\ce{CH3OCHO} &  30 4 26 0     - 30 4 27 0 & 217077.079 		& \ce{CH3OH} &  14 -2 0 0     - 13 3 0 0 & 226939.41 \\ 
SiO & 5 - 4 & 217104.98 					& \ce{CH3OCHO} &  25 9 17 1     - 25 8 18 1 & 226974.039 \\ 
\ce{CH3OCHO} &  35 9 27 1     - 35 8 28 1 & 217165.366 		& \ce{CH3OCHO} &  19 2 17 2     - 18 2 16 2 & 227019.55 \\ 
\ce{CH3OCH3} &  22 4 19 3     - 22 3 20 3 & 217189.668	        & \ce{CH3OCHO} &  19 2 17 0     - 18 2 16 0 & 227028.121 \\ 
\ce{CH3OCHO} &  32 9 24 0     - 32 8 25 0 & 217215.847 		& \ce{CH3CH3CO},v=1  &  20 3 17 1     - 19 4 16 1 & 227032.073 \\ 
DCN &  3 2           - 2 2 & 217238.631				 & \ce{CH3CH3CO},v=1  &  20 4 17 1     - 19 3 16 1 & 227032.073 \\ 
\ce{CH3OCHO} &  30 4 26 0     - 30 3 27 0 & 217266.549 		& \ce{CH3CH3CO},v=1  &  20 4 17 1     - 19 3 16 2 & 227043.496 \\ 
\ce{CH3OH} &  6 1 -1 1      - 7 2 -1 1 & 217299.205 		& \ce{CH3CH3CO},v=1  &  20 3 17 1     - 19 4 16 2 & 227043.496 \\ 
\ce{CH3OCHO} &  17 4 13 3     - 16 4 12 3 & 217312.626 		& \ce{CH3OH} &  21 1 0 0      - 21 0 0 0 & 227094.747 \\ 
\ce{^{13}CH3OH} &  10 2 8 0      - 9 3 7 0 & 217399.55 		& \ce{CH3OCHO} &  25 9 16 2     - 25 8 18 1 & 227106.67 \\ 
\ce{(CH2OH)2} &  24 1 24 0     - 23 1 23 1 & 217449.995 	& \ce{CH3OH} &  12 -1 0 0     - 11 2 0 0 & 227229.511 \\ 
\ce{(CH2OH)2} &  24 0 24 0     - 23 0 23 1 & 217450.27 		& \ce{C2H5OH} &  13 3 10 1     - 12 3 9 1 & 227294.752 \\ 
\ce{CH3OH} &  16 1 0 0      - 15 3 0 0 & 217525.002 		& \ce{^{13}C^{34}S} & 5 - 4 & 227300.506 \\ 
\ce{C2H5OH} &  5 1 4 1       - 4 0 4 0 & 217548.152 		& \ce{HC3N},v=0 &  25            - 24 & 227418.905 \\ 
\ce{C2H5OH} &  25 3 22 0     - 24 4 20 1 & 217549.328 		& \ce{CH3CH3CO},v=1 &  21 3 19 0     - 20 3 18 0 & 227560.862 \\ 
\ce{(CH2OH)2} &  21 2 19 1     - 20 2 18 0 & 217587.548 	& \ce{CH3CH3CO},v=1 &  21 2 19 0     - 20 2 18 0 & 227560.862 \\ 
\ce{C2H4O} &  15 4 11       - 15 3 12 & 217615.673		 & \ce{CH3OCHO} &  21 0 21 2     - 20 0 20 2 & 227561.741 \\ 
\ce{C2H4O} &  15 5 11       - 15 4 12 & 217615.72 		& \ce{(CH2OH)2} &  22 4 18 0     - 21 4 17 1 & 227587.074 \\ 
\ce{CH3OH} &  15 6 1 1      - 16 5 1 1 & 217642.678 		& \ce{CH3OCHO} &  18 3 15 3     - 17 3 14 3 & 227599.261 \\ 
\ce{CH3CH3CO},v=1  &  20 2 18 0     - 19 2 17 0 & 217744.747     & \ce{NH2CHO} &  11 0 11       - 10 0 10 & 227606.176 \\ 
\ce{CH3CH3CO},v=1  &  20 3 18 0     - 19 3 17 0 & 217744.747     & \ce{C2H5OH} &  18 5 13 2     - 18 4 14 2 & 227606.079 \\ 
\ce{C2H5OH} &  5 3 3 2       - 4 2 2 2 & 217803.689		 & \ce{C2H5CN},v=1 &  26 1 26 1     - 25 1 25 1 & 227624.735 \\ 
\ce{^{33}SO} &  5 6 8         - 4 5 7 & 217832.642		 & \ce{CH3OCH3} &  47 11 36 1    - 46 12 35 1 & 227668.331 \\ 
\ce{CH3OH} &  20 1 0 0      - 20 0 0 0 & 217886.504 		& \ce{CH3OCH3} &  47 11 37 3    - 46 12 34 3 & 227675.733 \\ 
\ce{CH3CH3CO},v=1  &  20 2 18 0     - 19 3 17 1 & 218023.182     & \ce{C2H5CN} &  25 3 22       - 24 3 21 & 227780.972 \\ 
\ce{CH3CH3CO},v=1  &  20 3 18 0     - 19 2 17 1 & 218023.182 	& \ce{CH3OH} &  16 1 1 0      - 15 2 1 0 & 227814.528 \\ 
\ce{CH3CH3CO},v=0 &  20 3 18 1     - 19 2 17 2 & 218091.448	 &\ce{CH3CH3CO},v=1  &  21 3 19 0     - 20 3 18 1 & 227836.273 \\ 
\ce{CH3CH3CO},v=0 &  20 2 18 1     - 19 3 17 2 & 218091.448 	& \ce{CH3CH3CO},v=1  &  21 2 19 0     - 20 2 18 1 & 227836.273 \\ 
\ce{CH3OCHO} &  17 4 13 5     - 16 4 12 5 & 218108.438		& \ce{C2H3CN} &  24 7 18       - 23 7 17 & 227897.607 \\ 
\ce{CH3CH3CO},v=0 &  20 2 18 0     - 19 3 17 1 & 218127.207 	& \ce{C2H3CN} &  24 7 17       - 23 7 16 & 227897.608 \\ 
\ce{CH3CH3CO},v=0&  20 3 18 0     - 19 2 17 1 & 218127.207 	& \ce{CH3CH3CO},v=0 &  21 3 19 1     - 20 3 18 1 & 227903.922 \\ 
\ce{CH3CH3CO},v=0 &  20 3 18 0     - 19 2 17 0 & 218162.929 	& \ce{CH3CH3CO},v=0&  21 2 19 1     - 20 2 18 1 & 227903.922 \\ 
\ce{CH3CH3CO},v=0 &  20 2 18 0     - 19 3 17 0 & 218162.929 	& \ce{CH3CH3CO},v=0 &  21 3 19 1     - 20 3 18 2 & 227903.953 \\ 
\ce{O^{13}CS} &  18            - 17 & 218198.998 		& \ce{CH3CH3CO},v=0 &  21 2 19 1     - 20 2 18 2 & 227903.953 \\ 
\ce{H2CO} &  3 0 3         - 2 0 2 & 218222.192 		& \ce{C2H3CN} &  24 6 19       - 23 6 18 & 227906.683 \\ 
\ce{CH3OCHO} &  31 9 23 0     - 31 8 24 0 & 218259.594 		& \ce{C2H3CN} &  24 6 18       - 23 6 17 & 227906.709 \\ 
\ce{CH3OCHO} &  17 3 14 2     - 16 3 13 2 & 218280.9		&\ce{CH3CH3CO},v=0 &  21 2 19 0     - 20 2 18 1 & 227939.374 \\ 
\ce{CH3OCHO} &  17 3 14 0     - 16 3 13 0 & 218297.89 		& \ce{CH3CH3CO},v=0&  21 2 19 0     - 20 3 18 1 & 227939.374 \\ 
\ce{HC3N},v=0 &  24            - 23 & 218324.723 		& \ce{CH3CH3CO},v=0&  21 3 19 0     - 20 2 18 1 & 227939.374 \\ 
\ce{C2H4O} &  14 3 11       - 14 2 12 & 218335.626             & \ce{CH3CH3CO},v=0 &  21 3 19 0     - 20 3 18 1 & 227939.374 \\ 
\ce{C2H4O} &  14 4 11       - 14 3 12 & 218335.634             & \ce{CH3OCHO} &  24 9 15 0     - 24 8 16 0 & 227954.452 \\ 
\ce{(CH2OH)2} &  22 4 19 0     - 21 4 18 1 & 218371.495 	& \ce{C2H3CN} &  24 5 20       - 23 5 19 & 227966.032 \\ 
\ce{C2H5CN} &  24 3 21       - 23 3 20 & 218389.97 	      & \ce{C2H3CN} &  24 5 19       - 23 5 18 & 227967.589 \\ 
\ce{C2H3CN} &  23 7 17       - 22 7 16 & 218398.555 		& \ce{CH3CH3CO},v=0 &  21 3 19 0     - 20 3 18 0 & 227974.758 \\ 
\ce{C2H3CN} &  23 7 17       - 22 7 16 & 218398.555 		& \ce{CH3CH3CO},v=0 &  21 2 19 0     - 20 2 18 0 & 227974.758 \\ 
\ce{C2H3CN} &  23 7 16       - 22 7 15 & 218398.555 		& \ce{HC3N},v$_7$=1 &  25 1          - 24 -1 & 227977.277 \\ 
\ce{C2H3CN} &  23 6 18       - 22 6 17 & 218402.435 		& \ce{CH3OCHO} &  31 4 27 2     - 31 3 28 2 & 227986.255 \\ 
\ce{C2H3CN} &  23 6 17       - 22 6 16 & 218402.451 		& \ce{CH3OCHO} &  24 9 15 2     - 24 8 16 2 & 227994.545 \\ 
\ce{CH3CH3CO},v=1  &  11 11 1 0     - 10 10 0 0 & 218402.853 	& \ce{C2H5OH} &  13 1 12 1     - 12 1 11 1 & 227891.911 \\ 
\ce{CH3CH3CO},v=1  &  11 11 0 0     - 10 10 1 0 & 218404.816 	& \ce{C2H5OH} &  13 3 10 0     - 12 3 9 0 & 228029.05 \\ 
\ce{CH3OH} &  4 2 0 0       - 3 1 0 0 & 218440.063 		& \ce{CH3OCHO} &  31 4 27 0     - 31 3 28 0 & 228057.899 \\ 
\ce{C2H3CN} &  23 5 19       - 22 5 18 & 218451.297 		& \ce{C2H3CN} &  24 11 13      - 23 11 12 & 228087.245 \\ 
\ce{C2H3CN} &  23 5 18       - 22 5 17 & 218452.357 		& \ce{C2H3CN} &  24 11 14      - 23 11 13 & 228087.245 \\ 
\ce{NH2CHO} &  10 1 9        - 9 1 8 & 218459.653		 & \ce{C2H3CN} &  24 4 21       - 23 4 20 & 228104.614 \\ 
\ce{C2H5OH} &  5 3 2 2       - 4 2 3 2 & 218461.226 		& \ce{OC^{33}S} &  19            - 18 & 228151.943 \\ 
\ce{H2CO} &  3 2 2         - 2 2 1 & 218475.632		         & \ce{C2H3CN} &  24 4 20       - 23 4 19 & 228160.305 \\ 
\ce{CH3OCH3} &  23 3 21 1     - 23 2 22 1 & 218491.914 		& \ce{CH3OCHO} &  24 9 16 1     - 24 8 17 1 & 228205.832 \\ 
\ce{CH3OCH3} &  23 3 21 0     - 23 2 22 0 & 218494.39 		& \ce{CH3OCHO} &  18 3 15 5     - 17 3 14 5 & 228211.291 \\ 
\ce{C2H3CN} &  23 10 13      - 22 10 12 & 218519.997 		& \ce{CH3OCHO} &  24 9 16 0     - 24 8 17 0 & 228270.535 \\ 
\ce{C2H3CN} &  23 10 14      - 22 10 13 & 218519.997		 & \ce{HC3N},v$_7$=1 &  25 -1         - 24 1 & 228303.174 \\ 
\ce{C2H5OH} &  21 5 16 2     - 21 4 17 2 & 218554.499   	& \ce{CH3OCH3} &  26 3 24 0     - 25 4 21 0 & 228421.737 \\ 
\ce{C2H3CN} &  23 4 20       - 22 4 19 & 218573.646 		& \ce{CH3OCH3} &  26 3 24 1     - 25 4 21 1 & 228422.878 \\ 
\ce{C2H3CN} &  23 3 21       - 22 3 20 & 218585.072 		& \ce{CH3OH} &  21 0 0 0      - 20 3 0 0 & 228467.902 \\ 
\ce{CH3OCHO}    &   27 7 21 0  - 27 5 22 0 &218593.341  	& \ce{C2H5CN} &  25 2 23       - 24 2 22 & 228483.136 \\ 
             &                          &            		& \ce{HCOOH} &  10 2 8      - 9 2 7 & 228544.168 \\ 
\hline
\end{tabular}
\label{smail}
\end{table*}
\twocolumn
\clearpage
\section{HIFI fitted curves of all molecules}
Examples of selected, fitted spectra for each molecule. In red is the model, in blue are mostly annotations or blended lines.
If only few transitions of a molecule are present, they are all shown in the plot. For weeds or molecules with many lines, no quantum numbers are given and
only selected frequency ranges over different bands are presented. Heavy blended lines are excluded for simplicity.
All intensities are given in units of Kelvin. The quantum numbers refer to upper state - lower state. For absorption lines, this sequence should be read reversed.

The entire observed and simulated spectrum for each HIFI band will be made available online in the CLASS format on the CHESS website (\url{http://www-laog.obs.ujf-grenoble.fr/heberges/hs3f/}), as soon as the HIPE 8 reprocessed data are available. Additionally, the modeled spectrum for each molecule (including isotopologues and vibrationally excited states) can be loaded.
It should be noticed that for the simulated spectrum of only one molecule, the blending by other molecules is not taken into account. This results in a deviation for the fits of absorption lines when a nearby emission line contributes to the continuum. In XCLASS, at first all the emission components are calculated and then the absorption components.
   
\begin{landscape}
\begin{figure}
\centering
\subfloat[][C,\ce{C+}]{\label{fig:candc}\includegraphics[angle=270,width=0.75\textwidth]{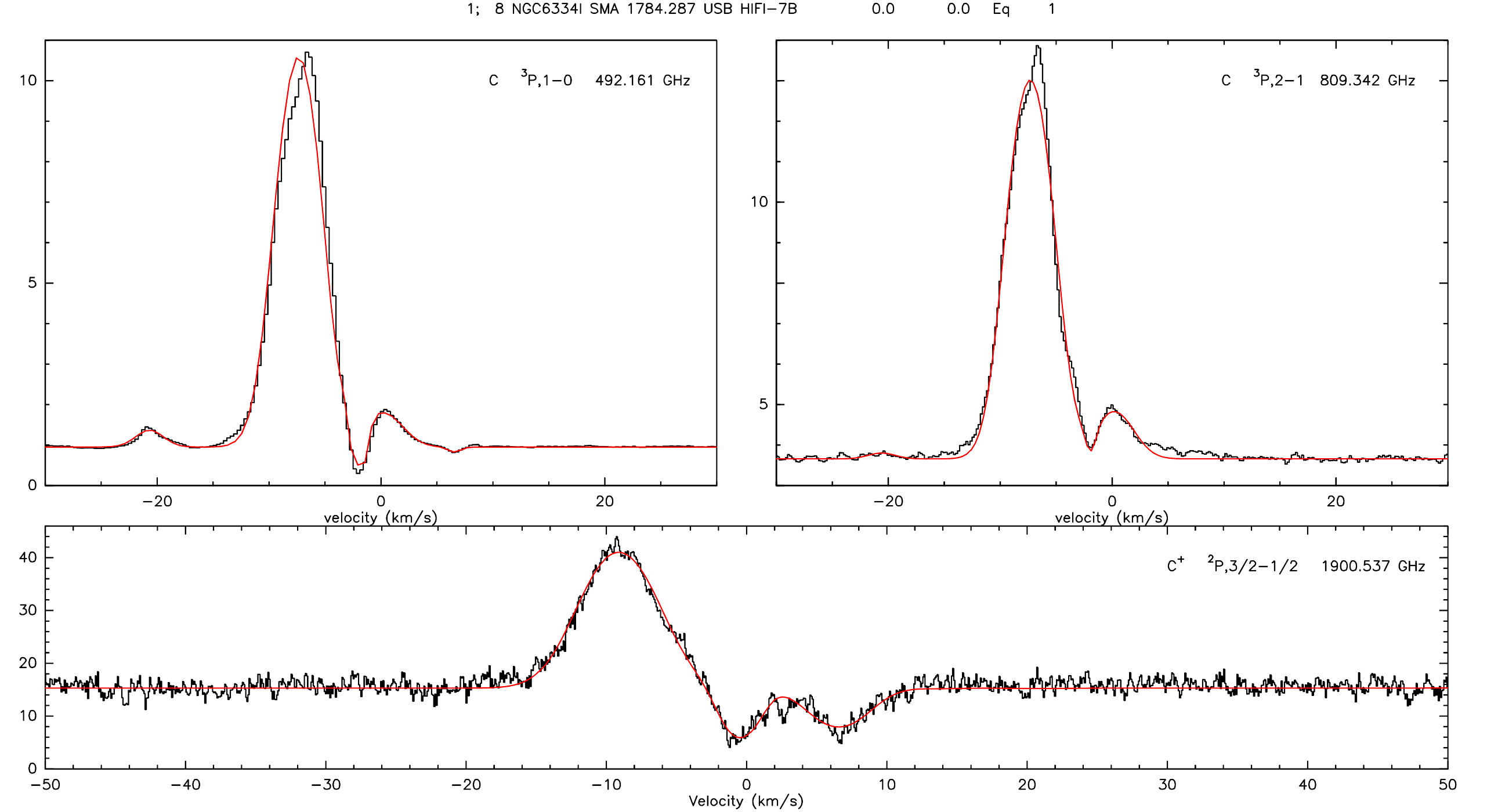}}\\
\subfloat[][\ce{CH+}]{\label{fig:ch+}\includegraphics[angle=270,width=0.70\textwidth]{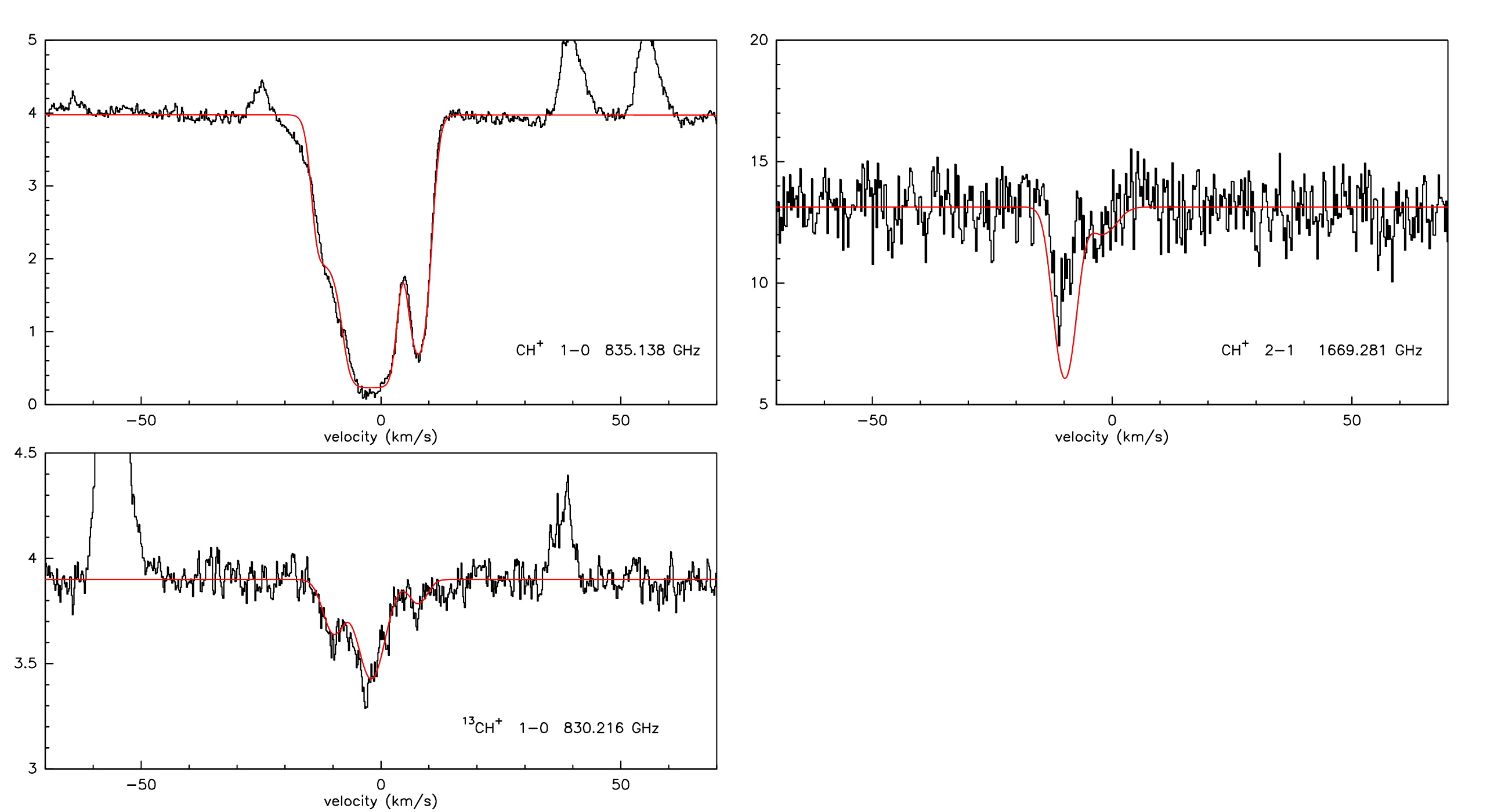}}
\subfloat[][CH]{\label{fig:ch}\includegraphics[angle=270,width=0.65\textwidth]{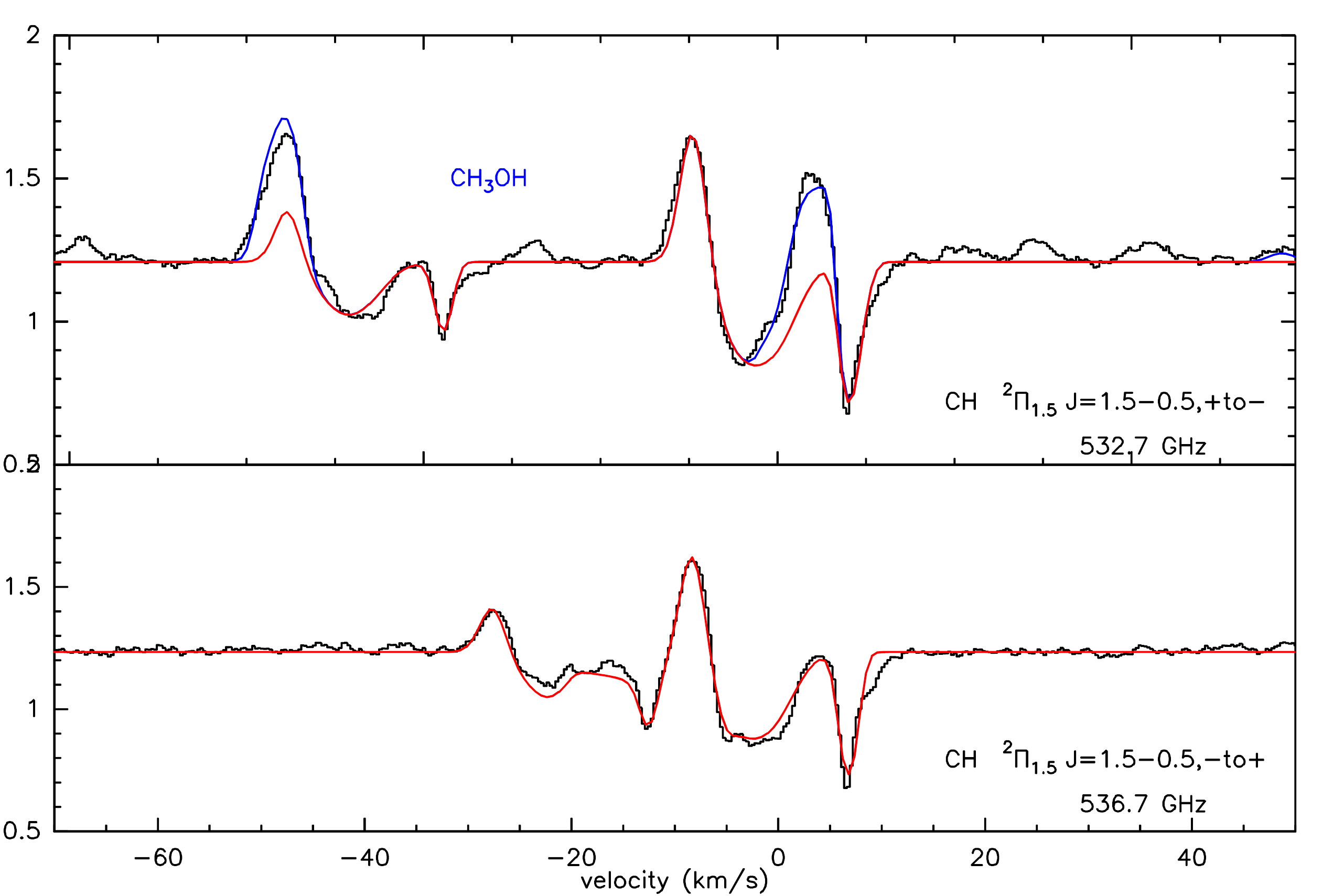}}
\caption{XCLASS fits of chemical species in the HIFI spectrum of NGC 6334I.}
\label{HIFIfits}
\end{figure}
\end{landscape}

\begin{landscape}
\begin{figure}
 \ContinuedFloat
\centering
\subfloat[][NH]{\label{fig:nh}\includegraphics[angle=270,width=0.65\textwidth]{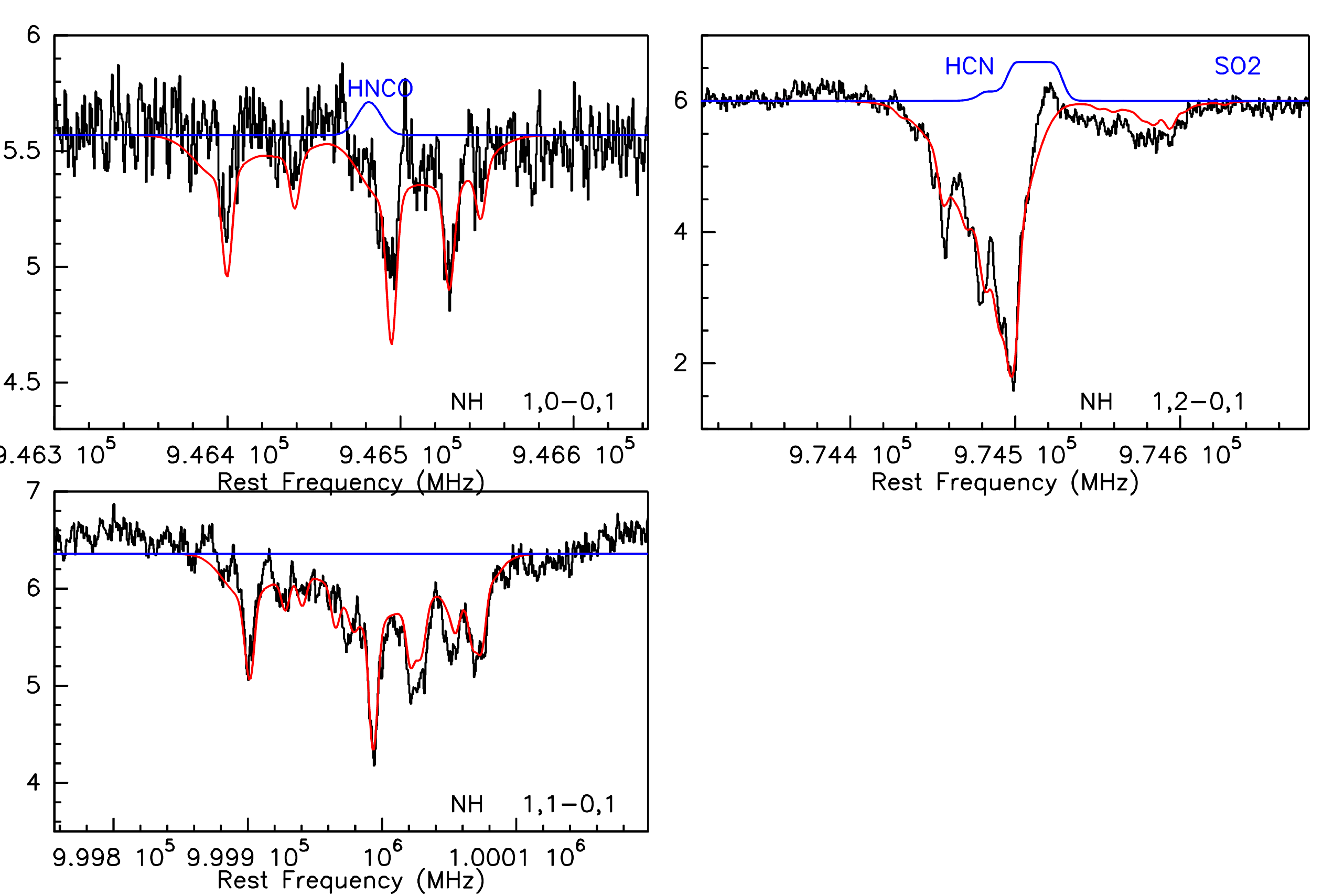}}
\subfloat[][\ce{NH2}]{\label{fig:nh2}\includegraphics[angle=270,width=0.65\textwidth]{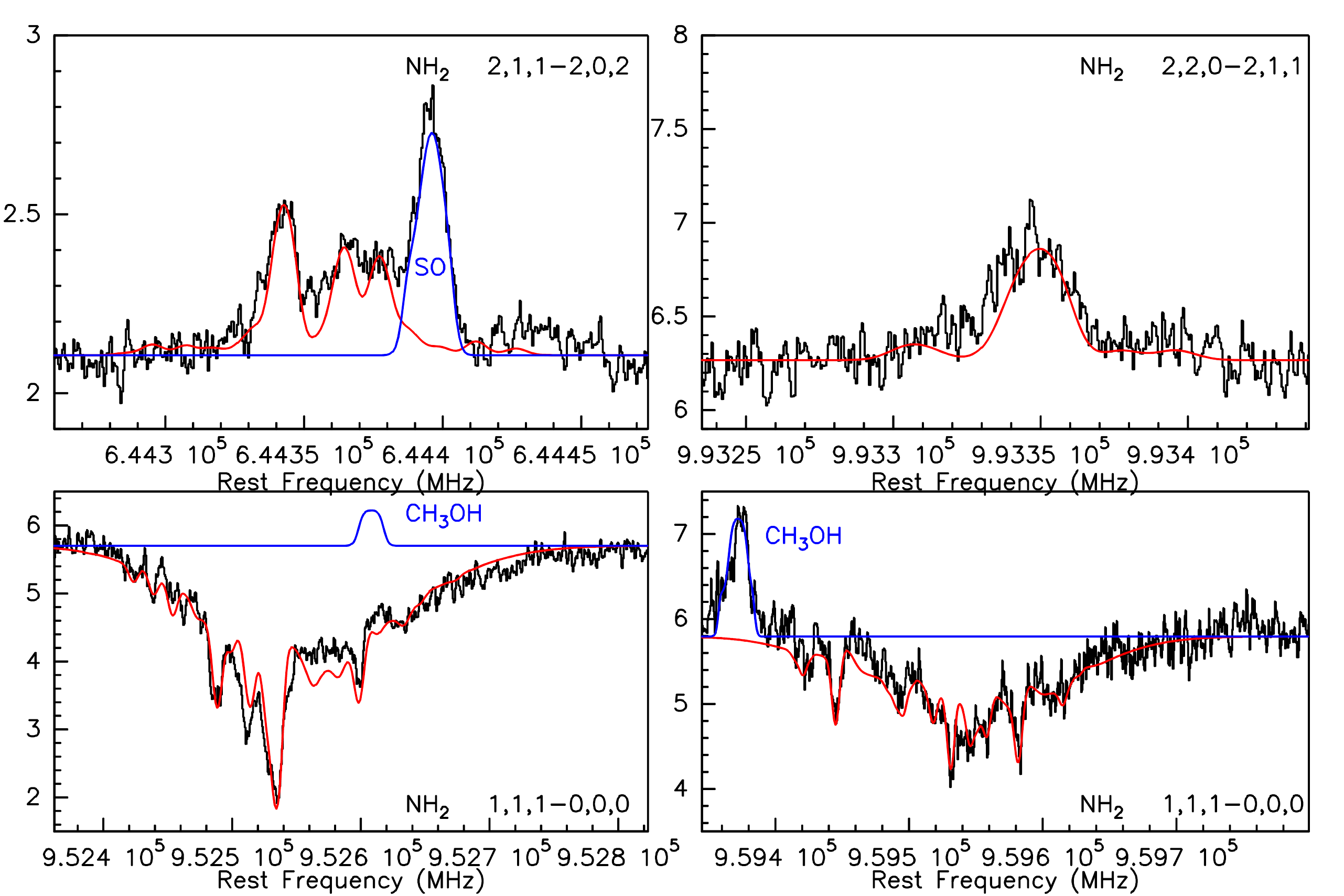}}\\
\subfloat[][\ce{NH3}]{\label{fig:nh3}\includegraphics[angle=270,width=0.65\textwidth]{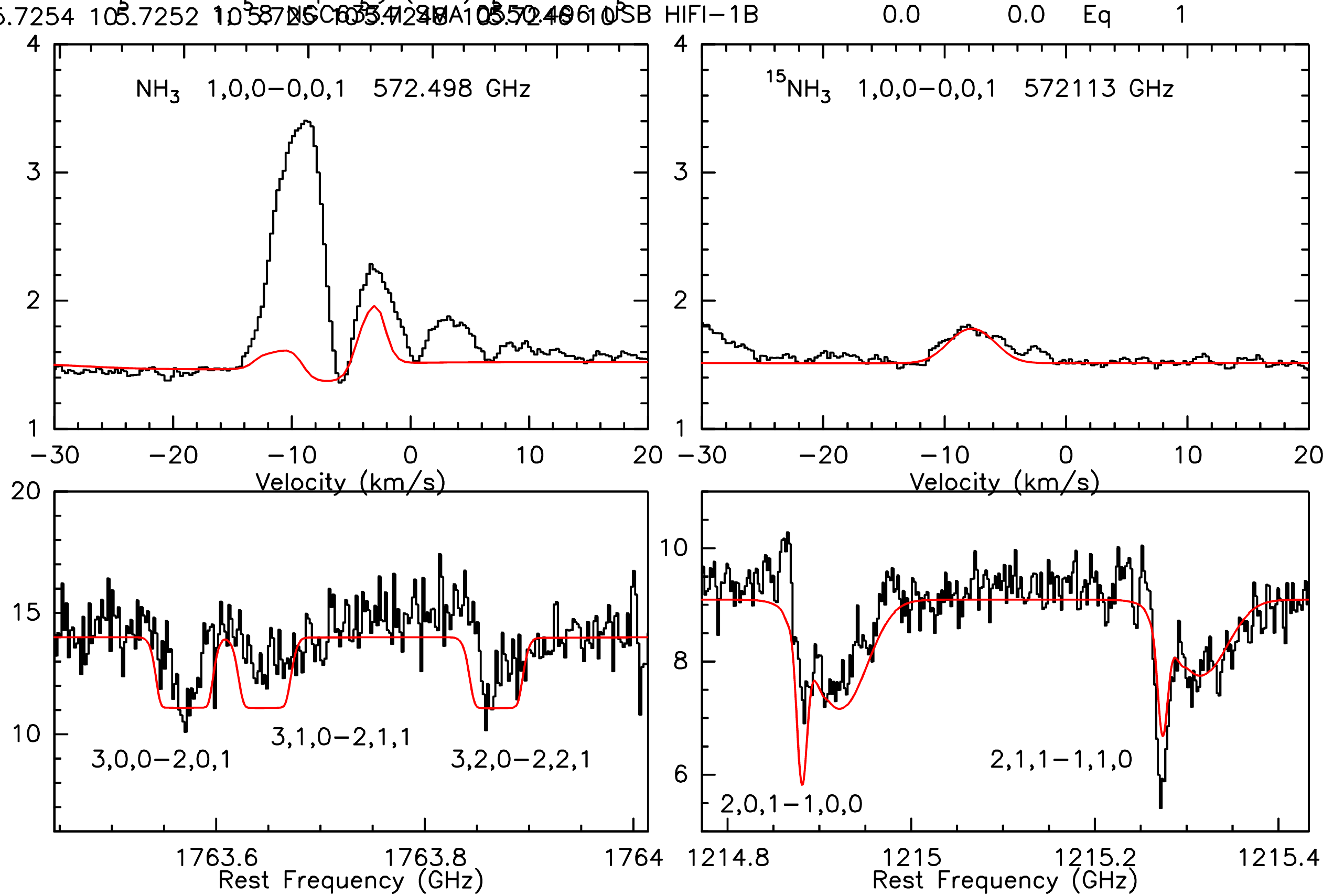}}
\subfloat[][\ce{N2H+}]{\label{fig:n2h+}\includegraphics[angle=270,width=0.65\textwidth]{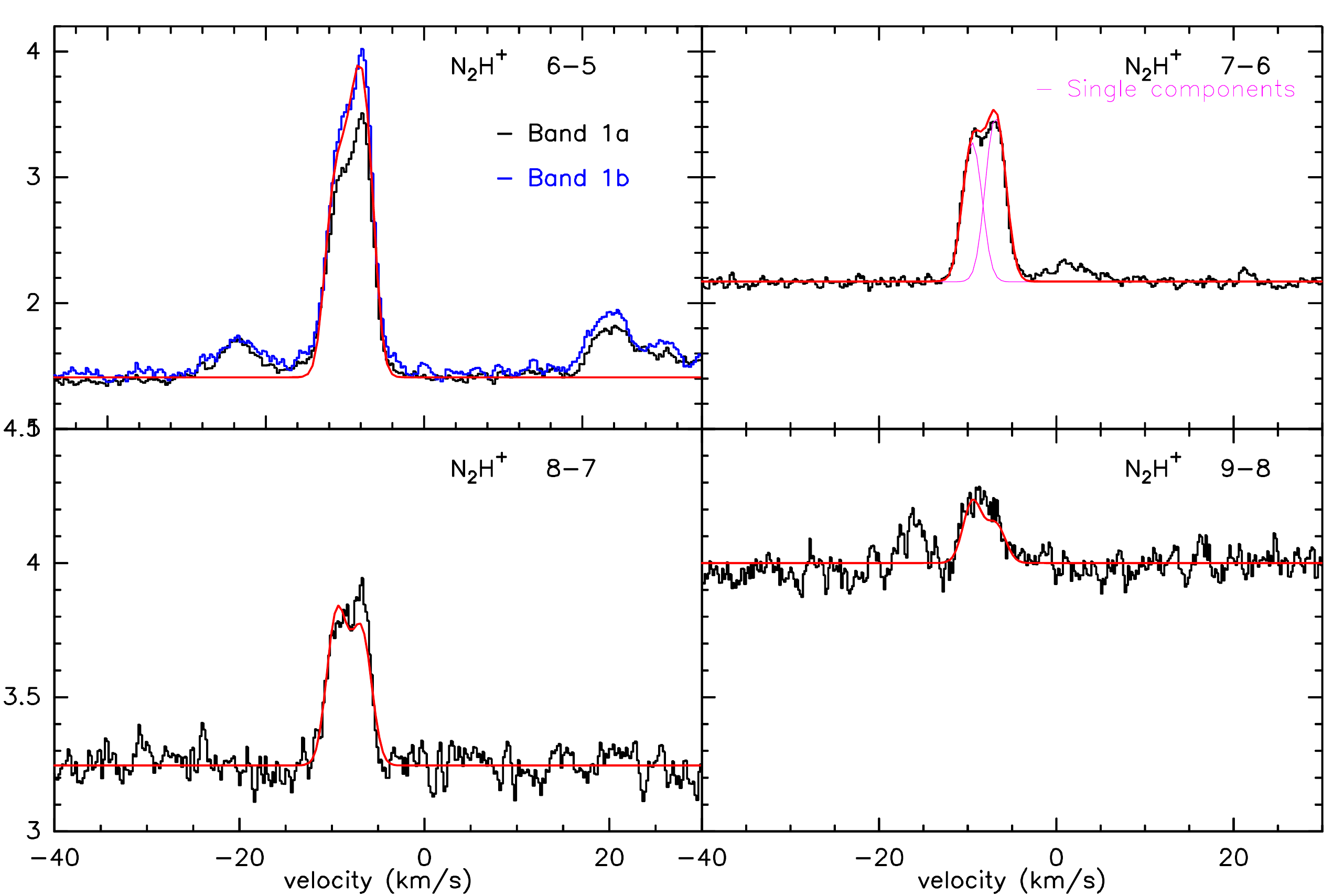}}
\caption{Continued}
\end{figure}
\end{landscape}

\begin{landscape}
\begin{figure}
 \ContinuedFloat
\centering
\subfloat[][\ce{OH+}]{\label{fig:oh+}\includegraphics[angle=270,width=0.65\textwidth]{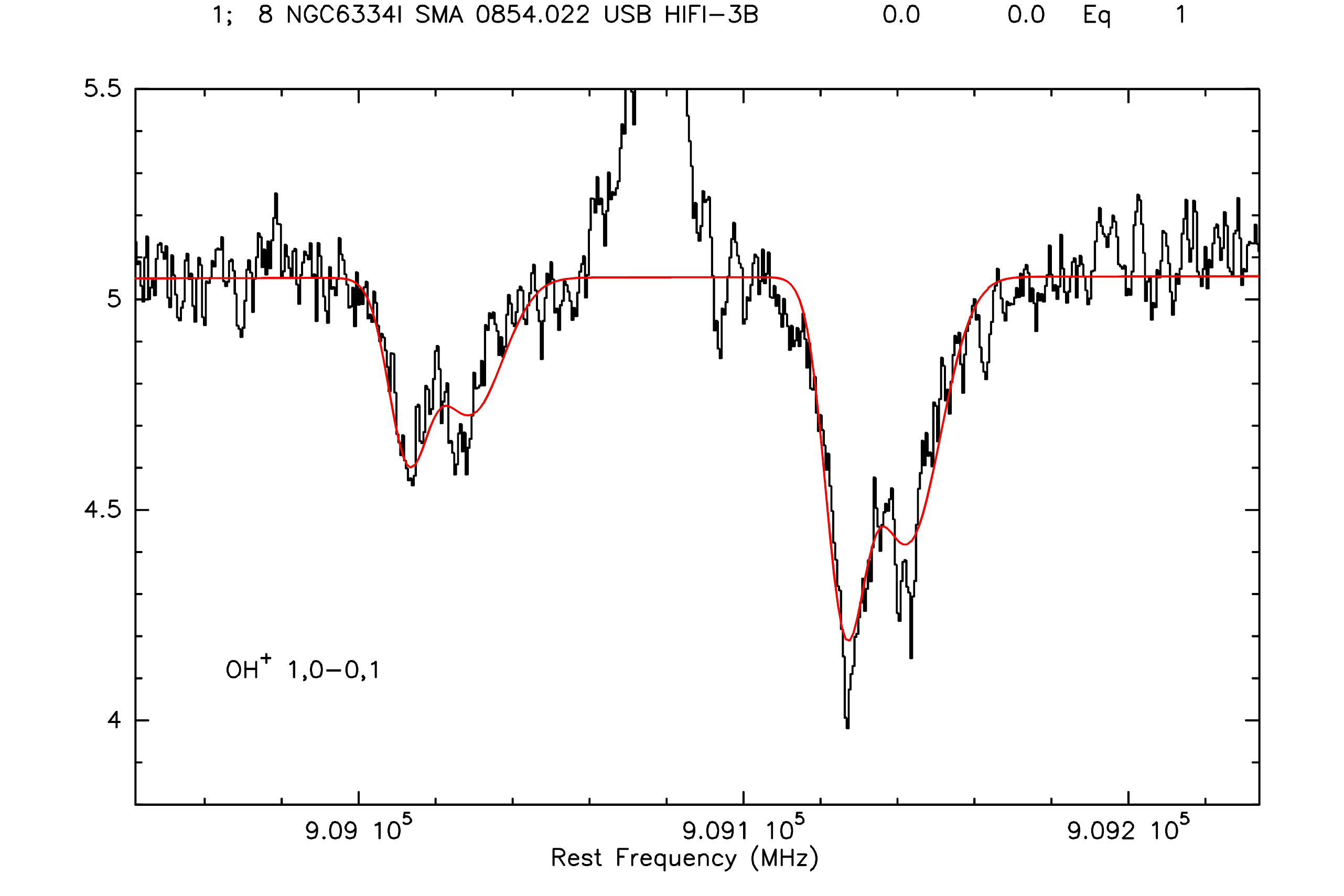}}
\subfloat[][\ce{H2O}]{\label{fig:h2o}\includegraphics[angle=270,width=0.65\textwidth]{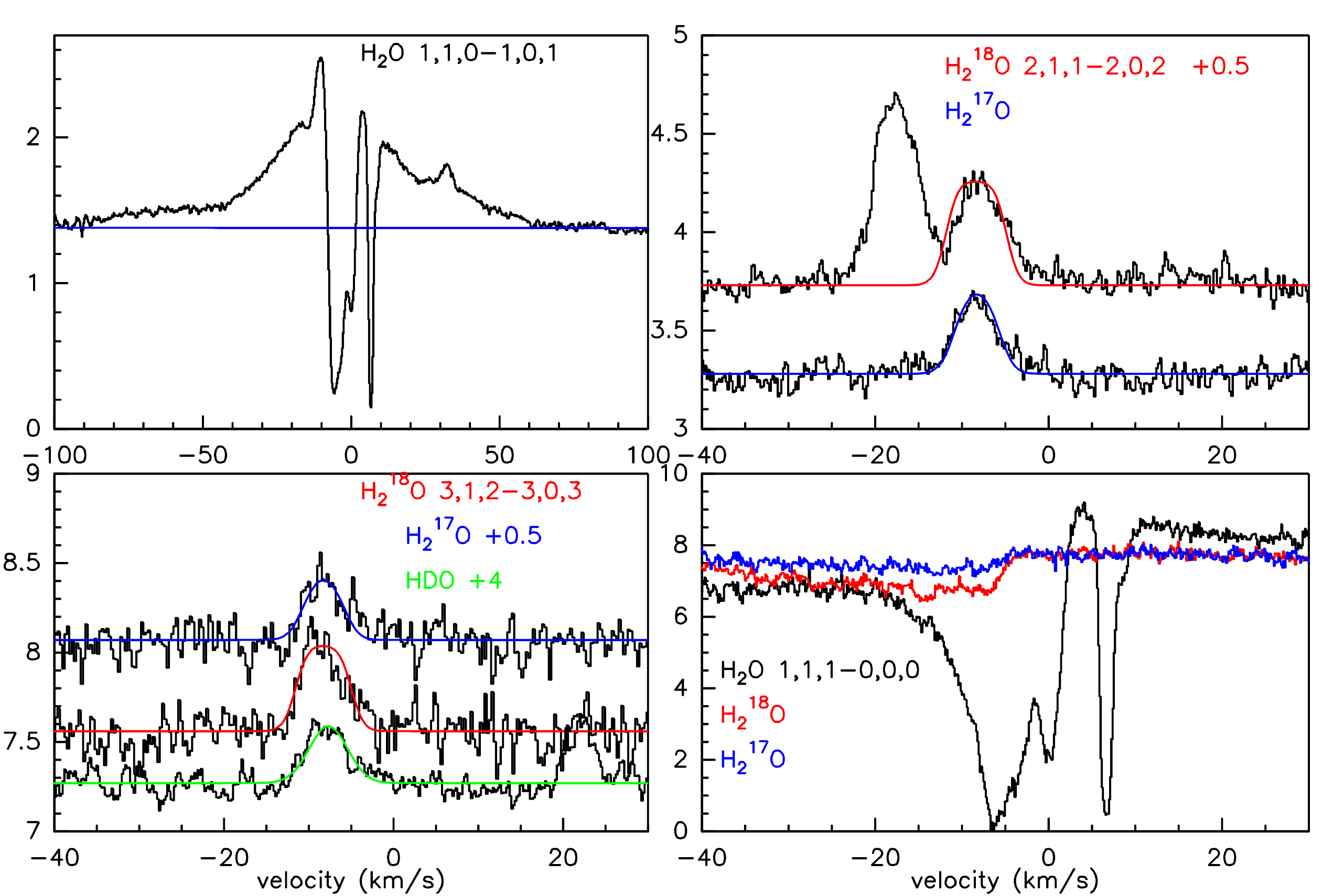}}\\
\subfloat[][\ce{H2O+}]{\label{fig:h2o+}\includegraphics[angle=270,width=0.65\textwidth]{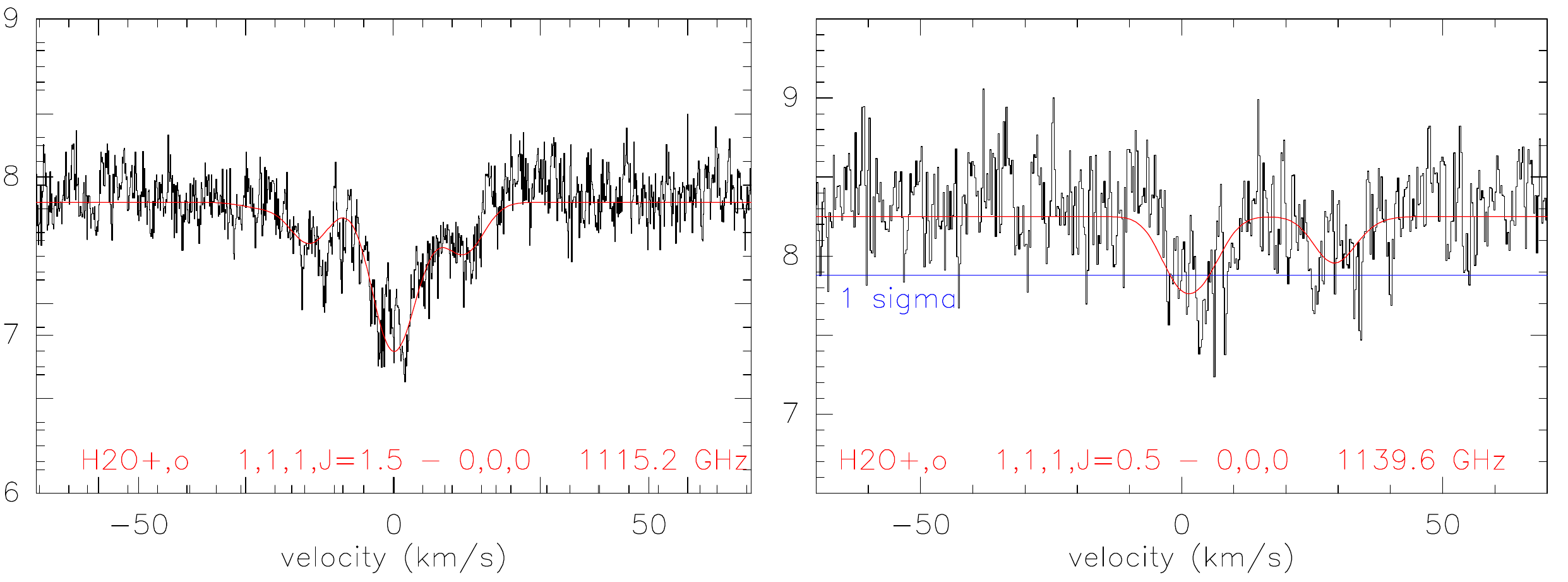}}
\subfloat[][\ce{C2H}]{\label{fig:c2h}\includegraphics[angle=270,width=0.65\textwidth]{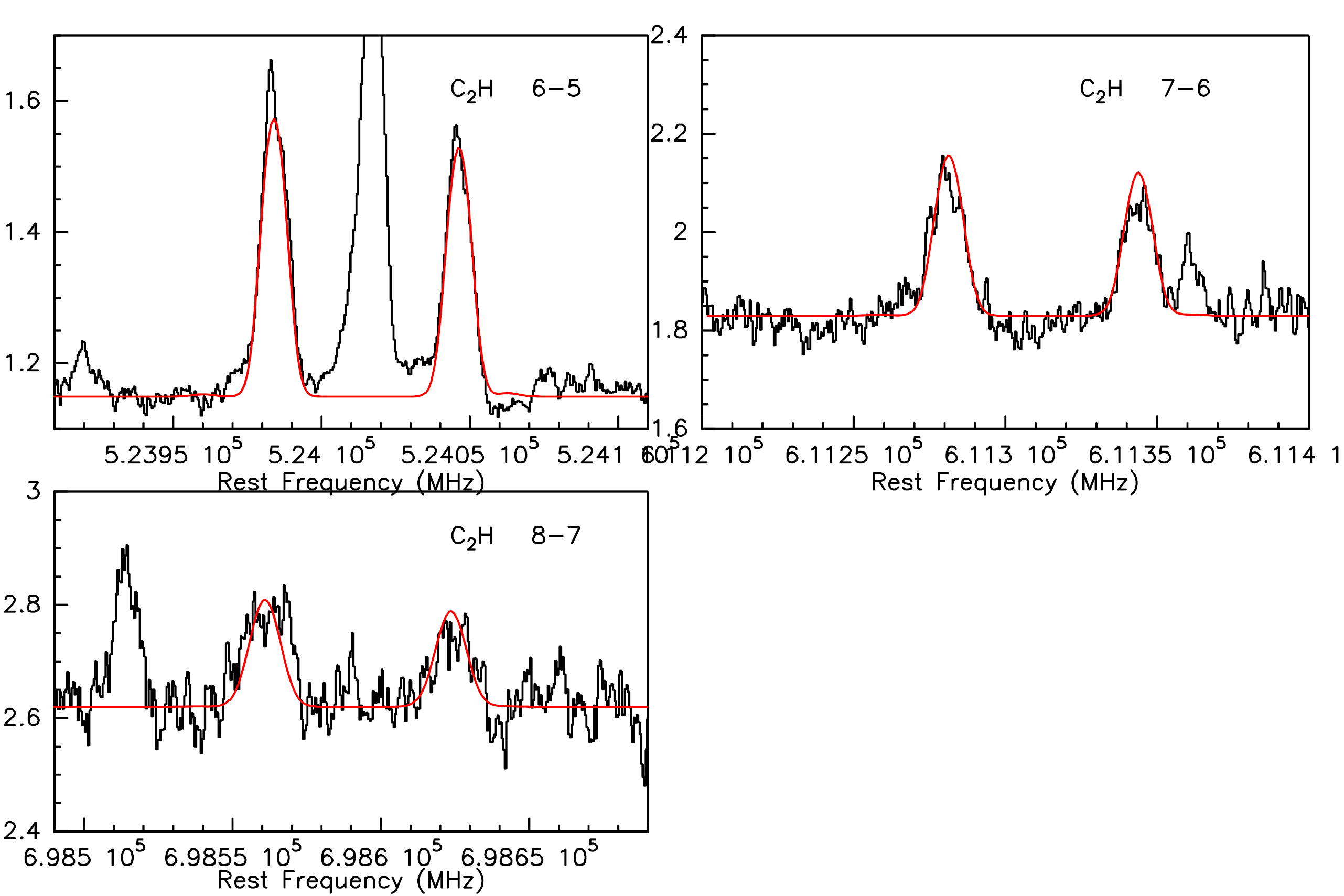}}
\caption{Continued}
\end{figure}
\end{landscape}

\begin{landscape}
\begin{figure}
 \ContinuedFloat
\centering
\subfloat[][CN]{\label{fig:cn}\includegraphics[angle=270,width=0.65\textwidth]{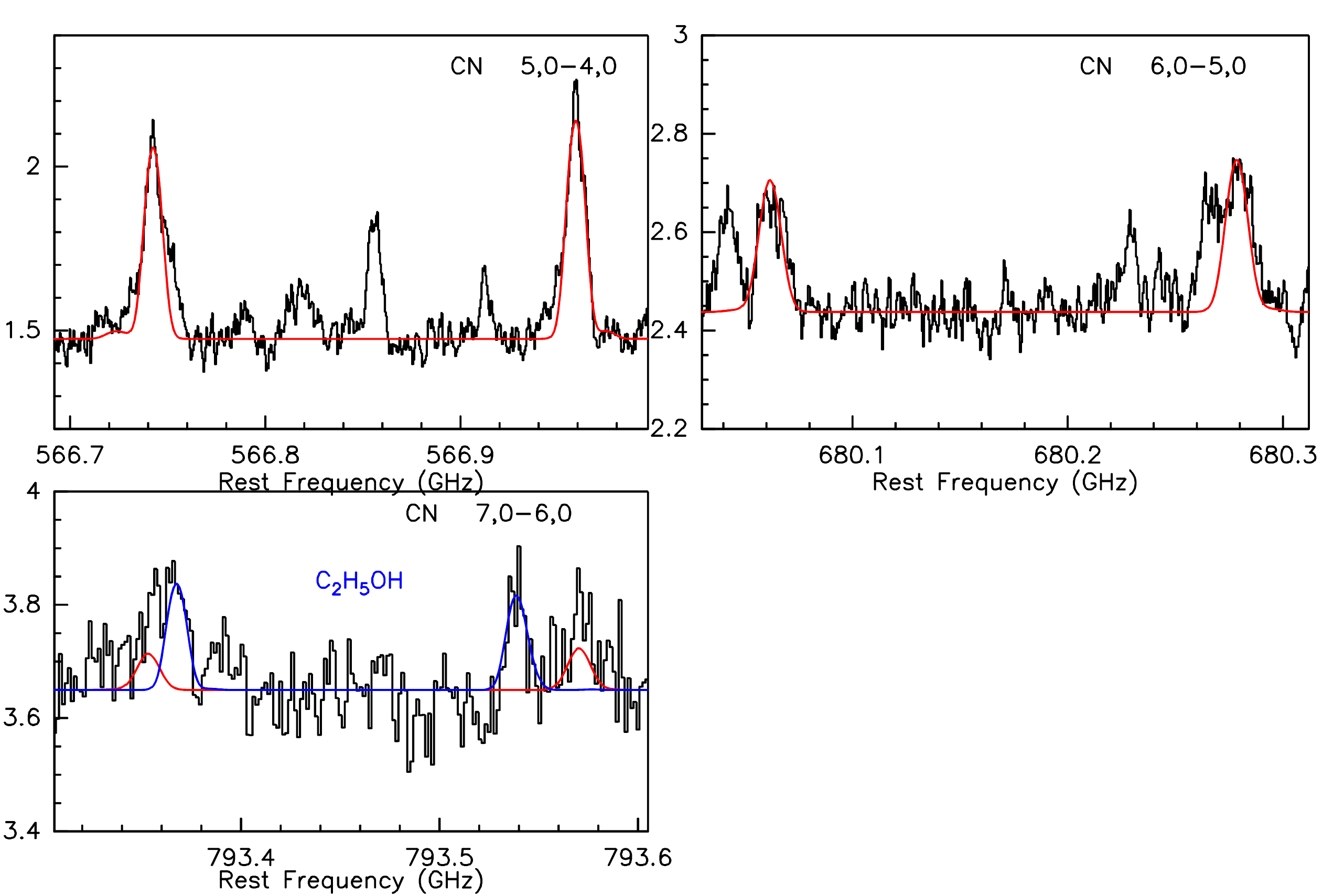}}
\subfloat[][CO]{\label{fig:co}\includegraphics[angle=270,width=0.65\textwidth]{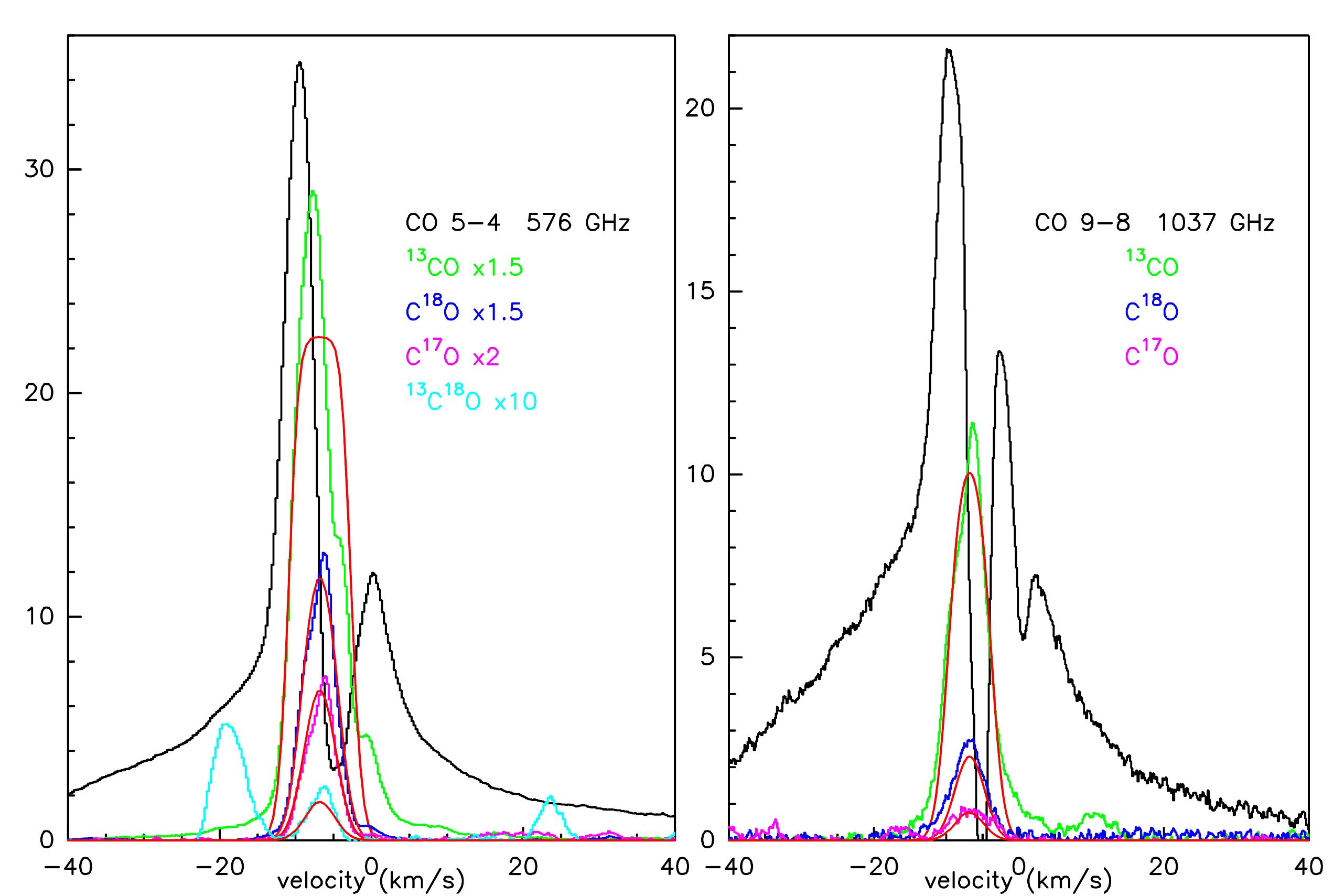}}\\
\subfloat[][HCN]{\label{fig:hcn}\includegraphics[angle=270,width=0.65\textwidth]{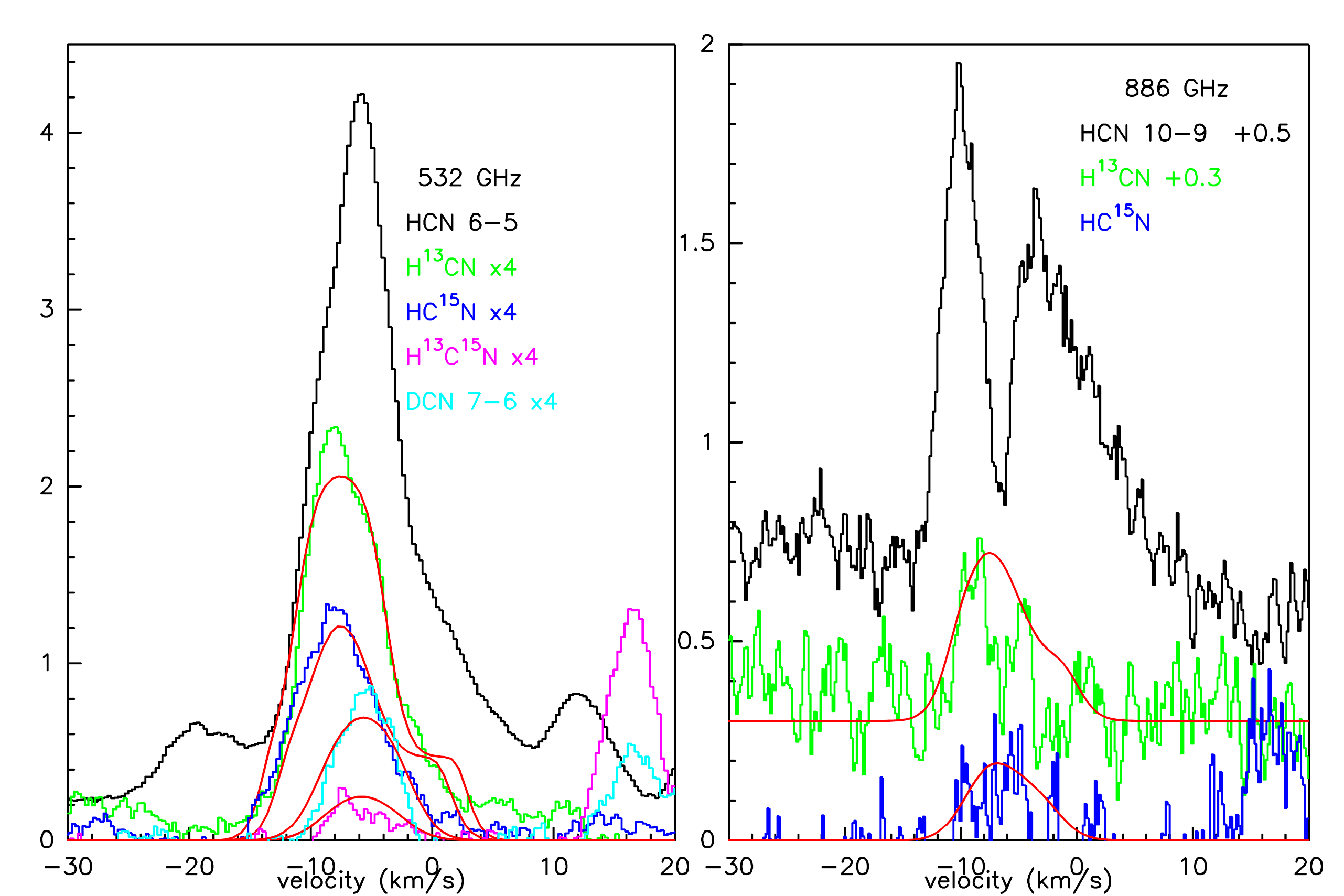}}
\subfloat[][HNC]{\label{fig:hnc}\includegraphics[angle=270,width=0.65\textwidth]{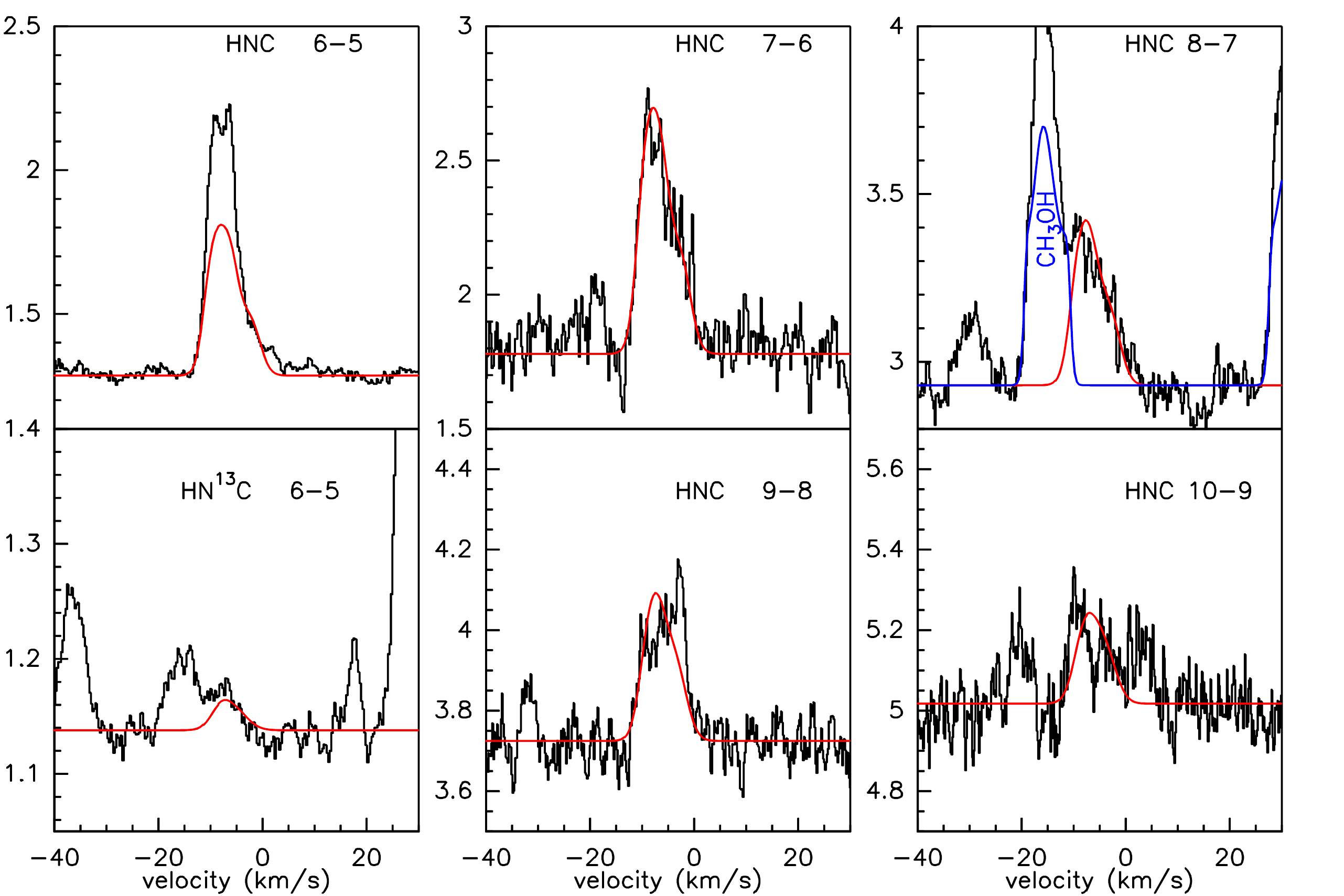}}
\caption{Continued}
\end{figure}
\end{landscape}

\begin{landscape}
\begin{figure}
 \ContinuedFloat
\centering
\subfloat[][\ce{HCS+}]{\label{fig:hcs}\includegraphics[angle=270,width=0.65\textwidth]{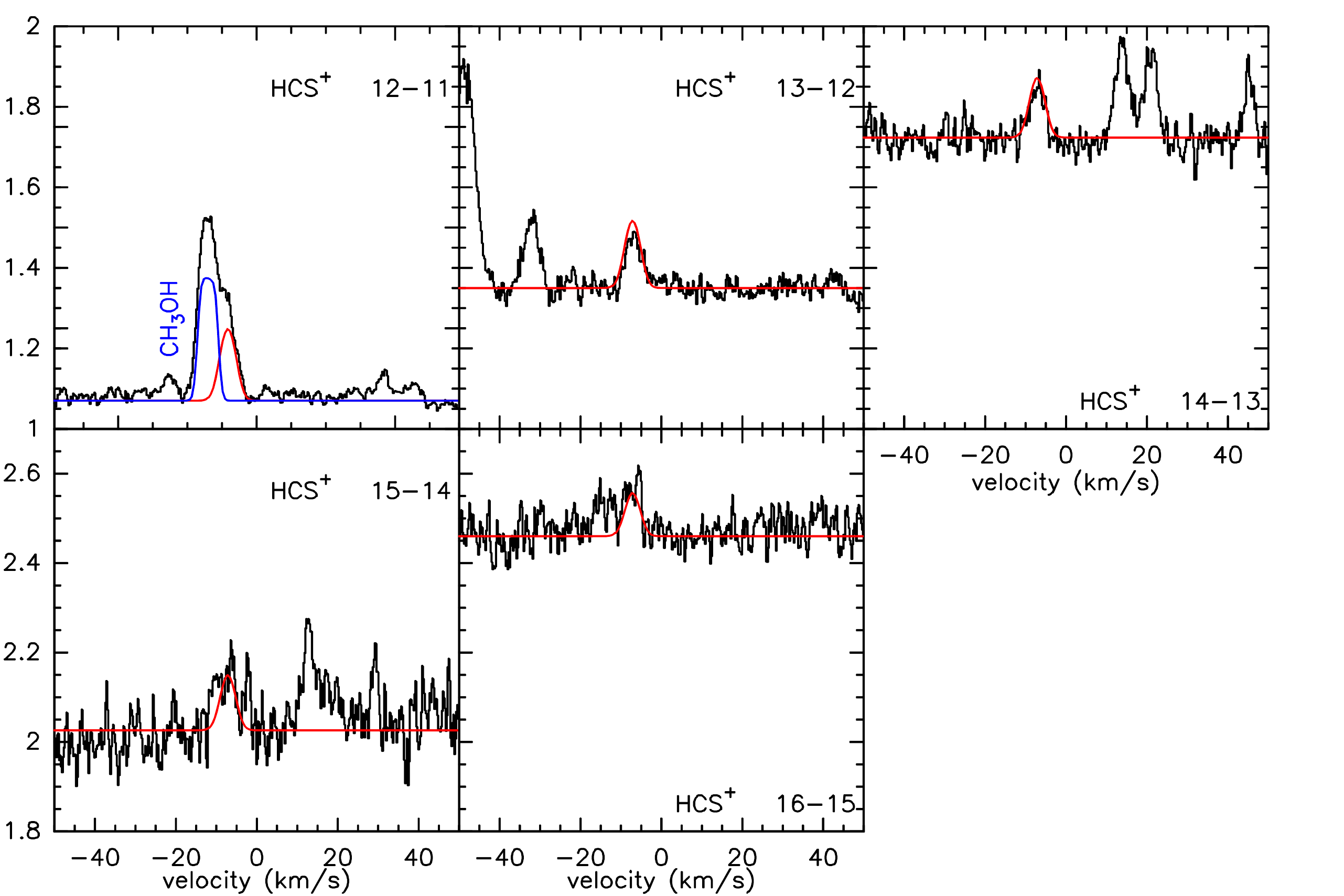}}
\subfloat[][NO]{\label{fig:no}\includegraphics[angle=270,width=0.65\textwidth]{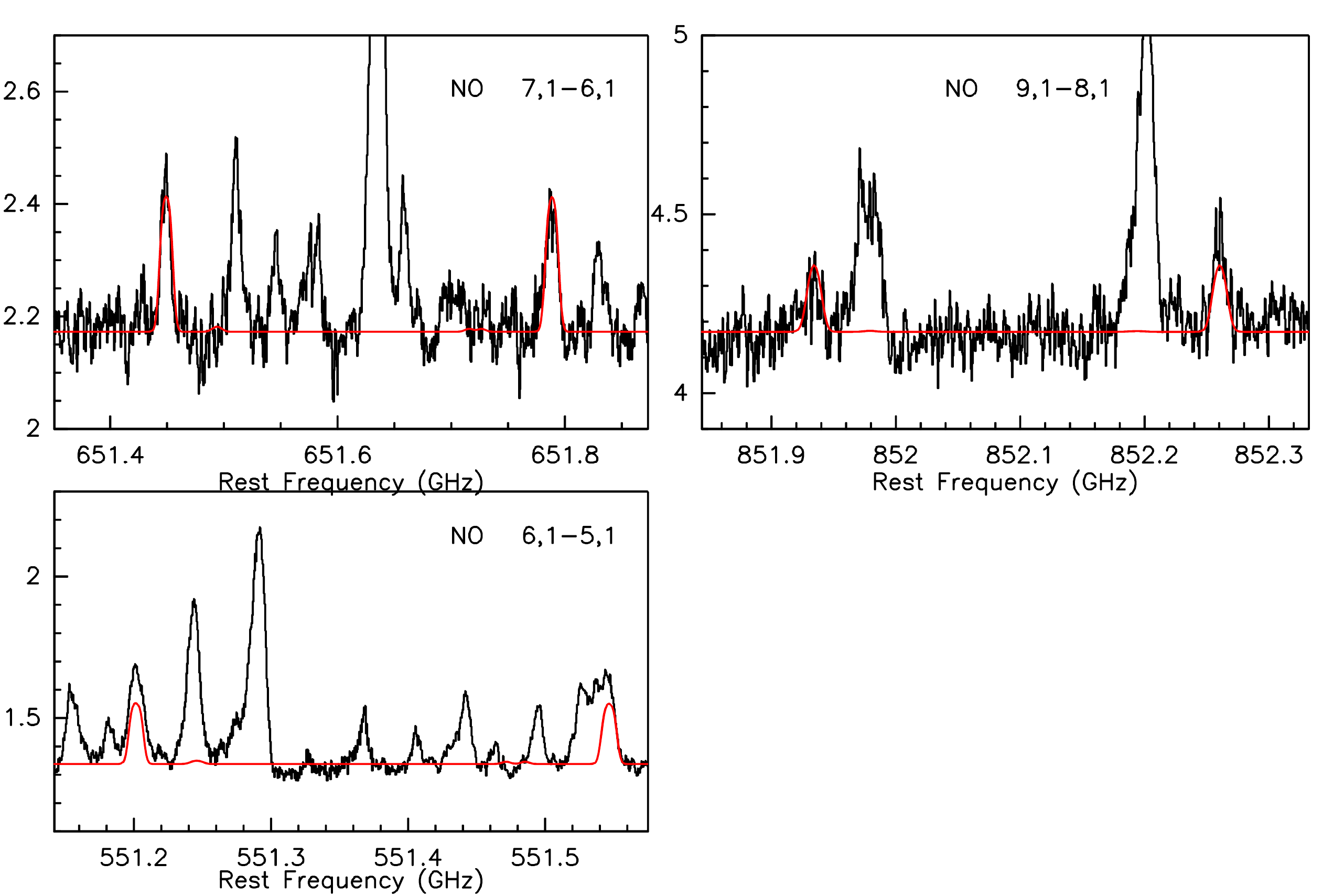}}\\
\subfloat[][SiO]{\label{fig:sio}\includegraphics[angle=270,width=0.65\textwidth]{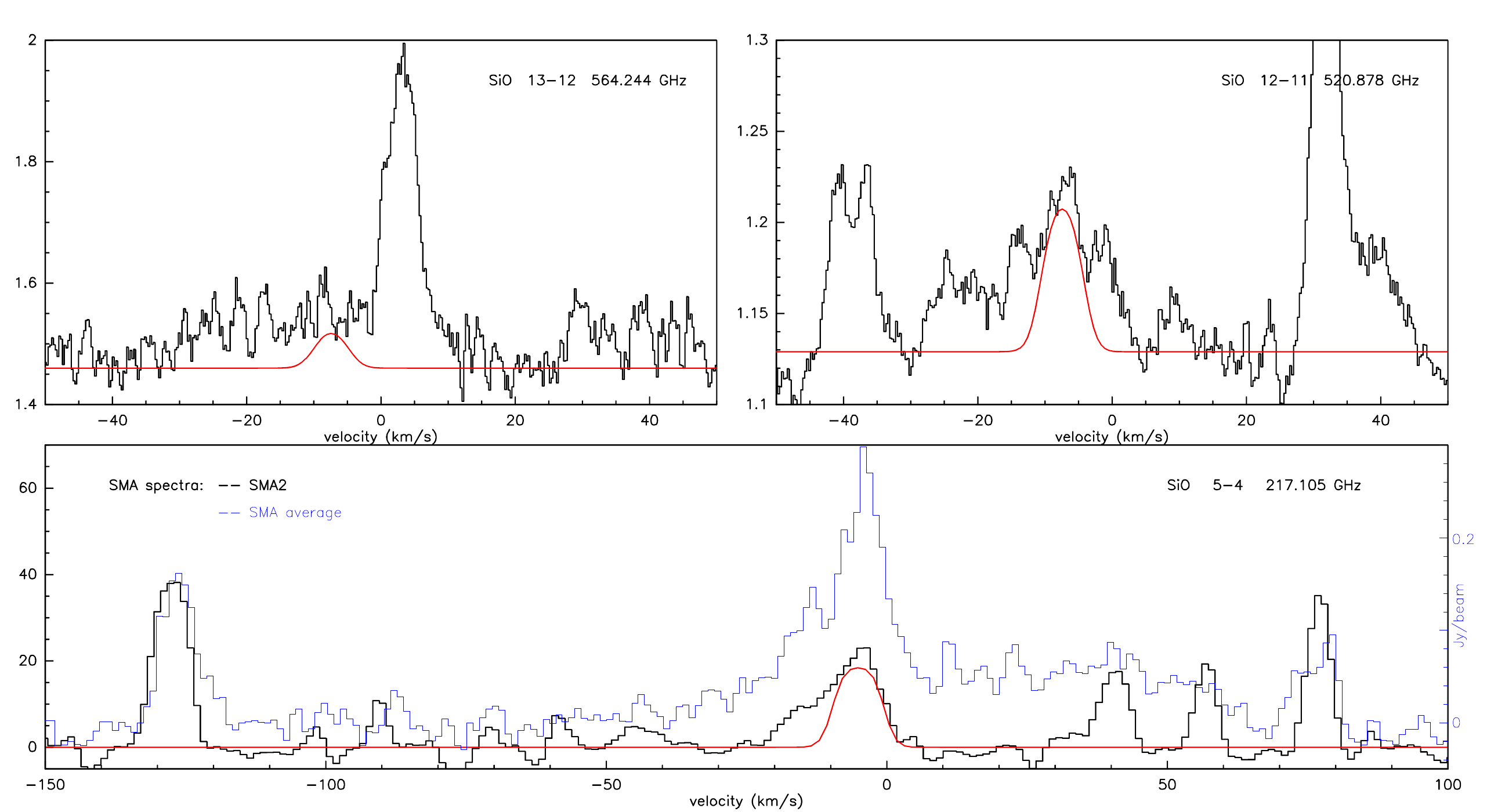}}
\subfloat[][HF]{\label{fig:hf}\includegraphics[angle=270,width=0.65\textwidth]{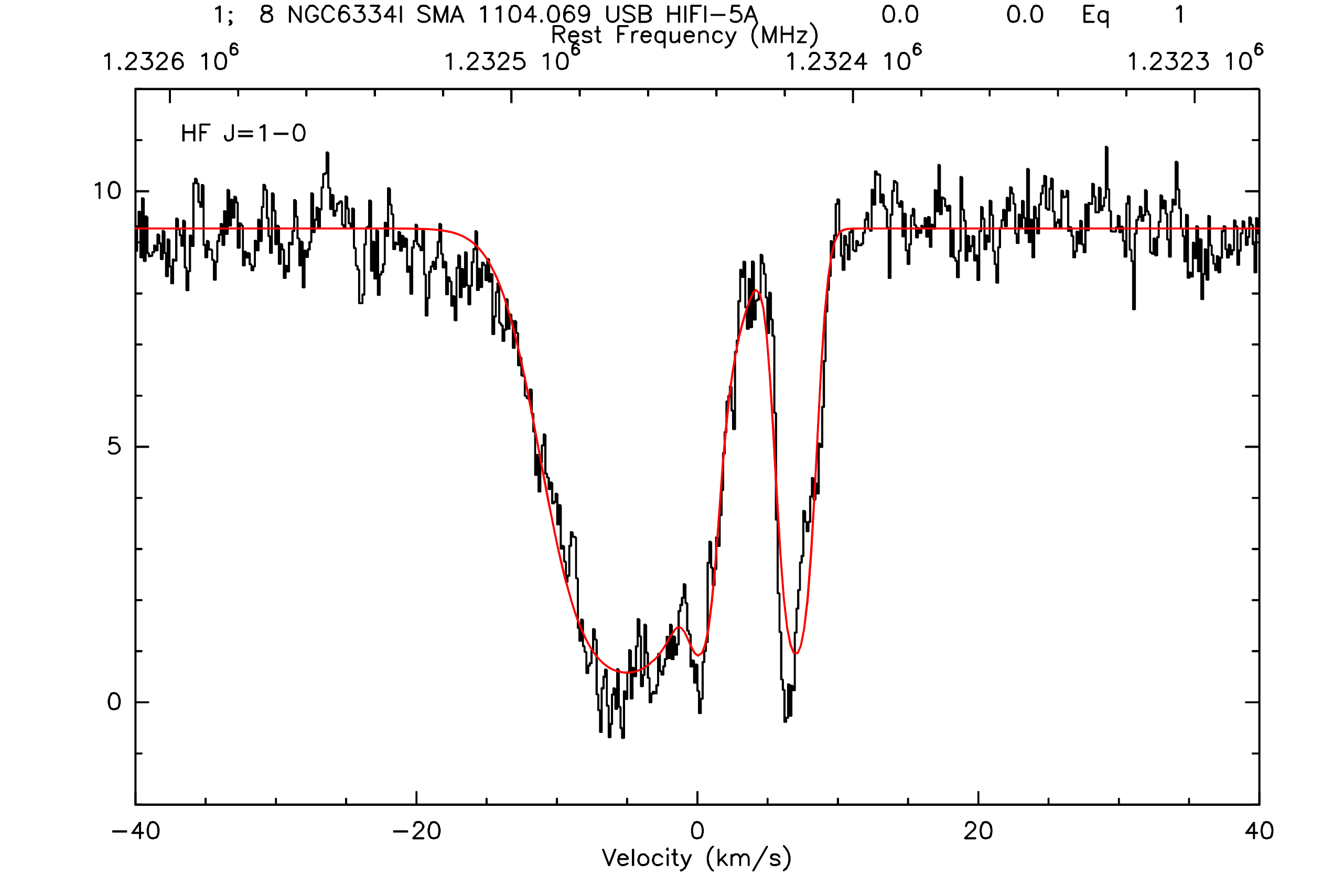}}
\caption{Continued}
\end{figure}
\end{landscape}

\begin{landscape}
\begin{figure}
 \ContinuedFloat
\centering
\subfloat[][HCl]{\label{fig:hcl}\includegraphics[angle=270,width=0.65\textwidth]{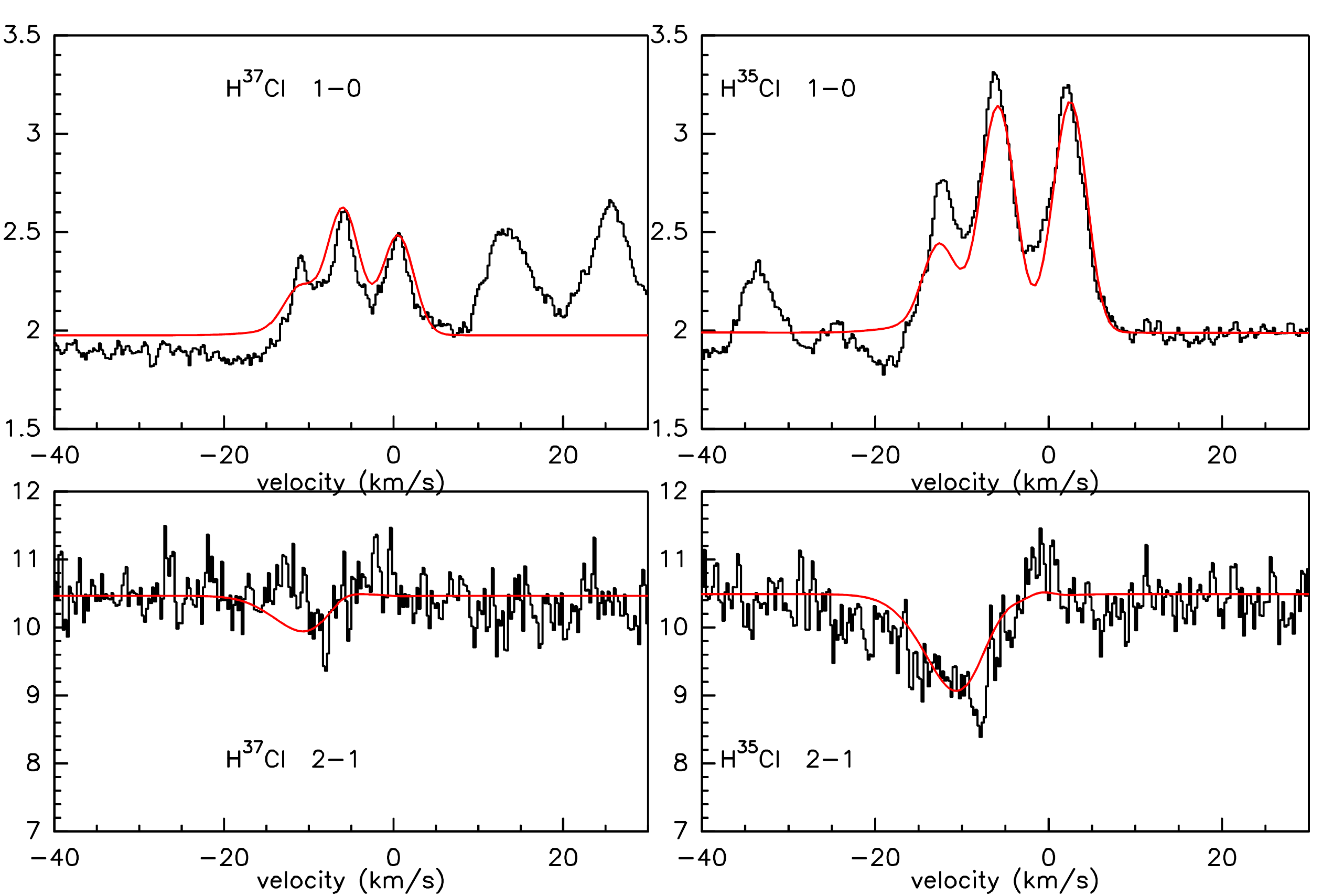}}
\subfloat[][\ce{H2Cl+}]{\label{fig:h2cl}\includegraphics[angle=270,width=0.65\textwidth]{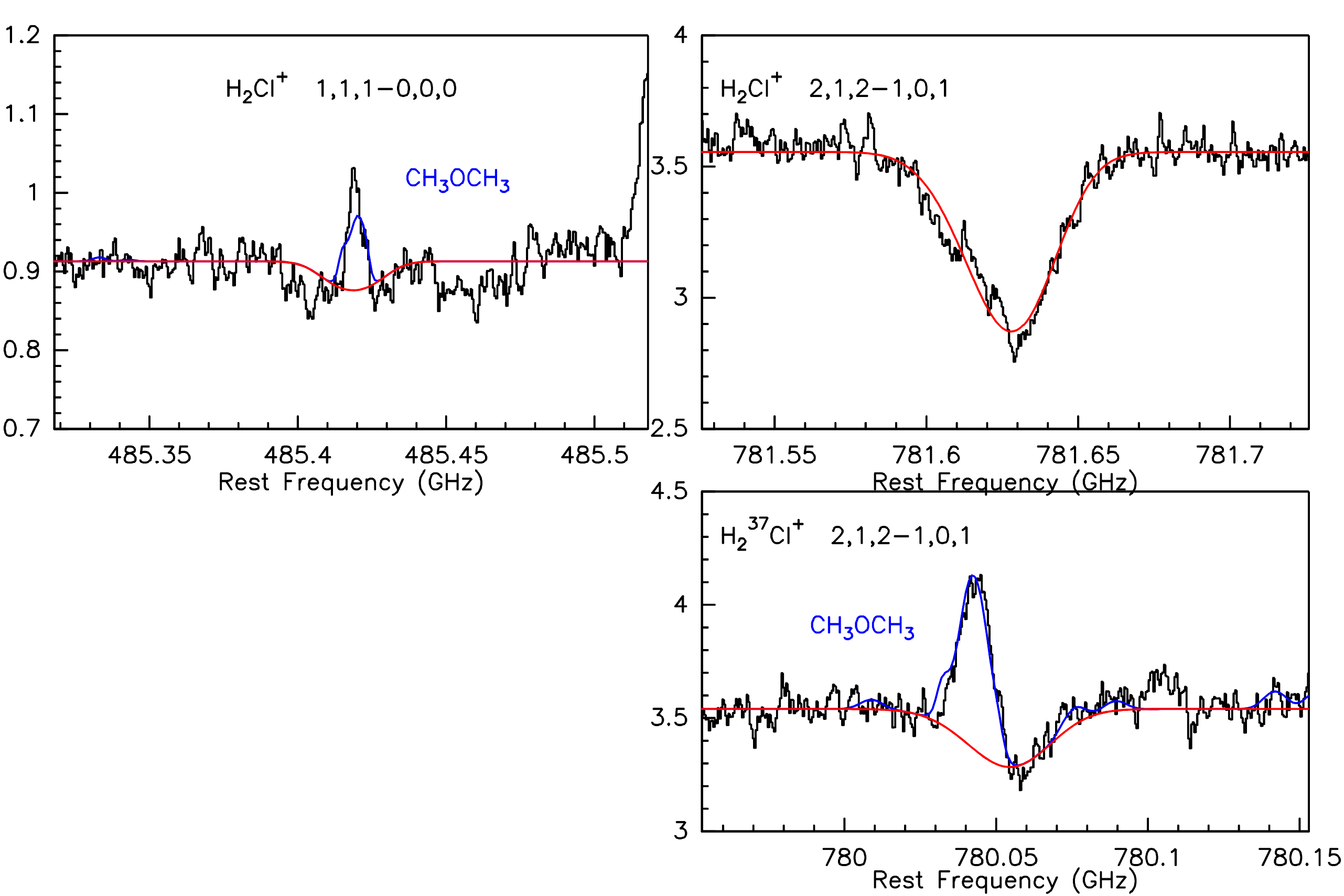}}\\
\subfloat[][\ce{SH+}]{\label{fig:sh}\includegraphics[angle=270,width=0.65\textwidth]{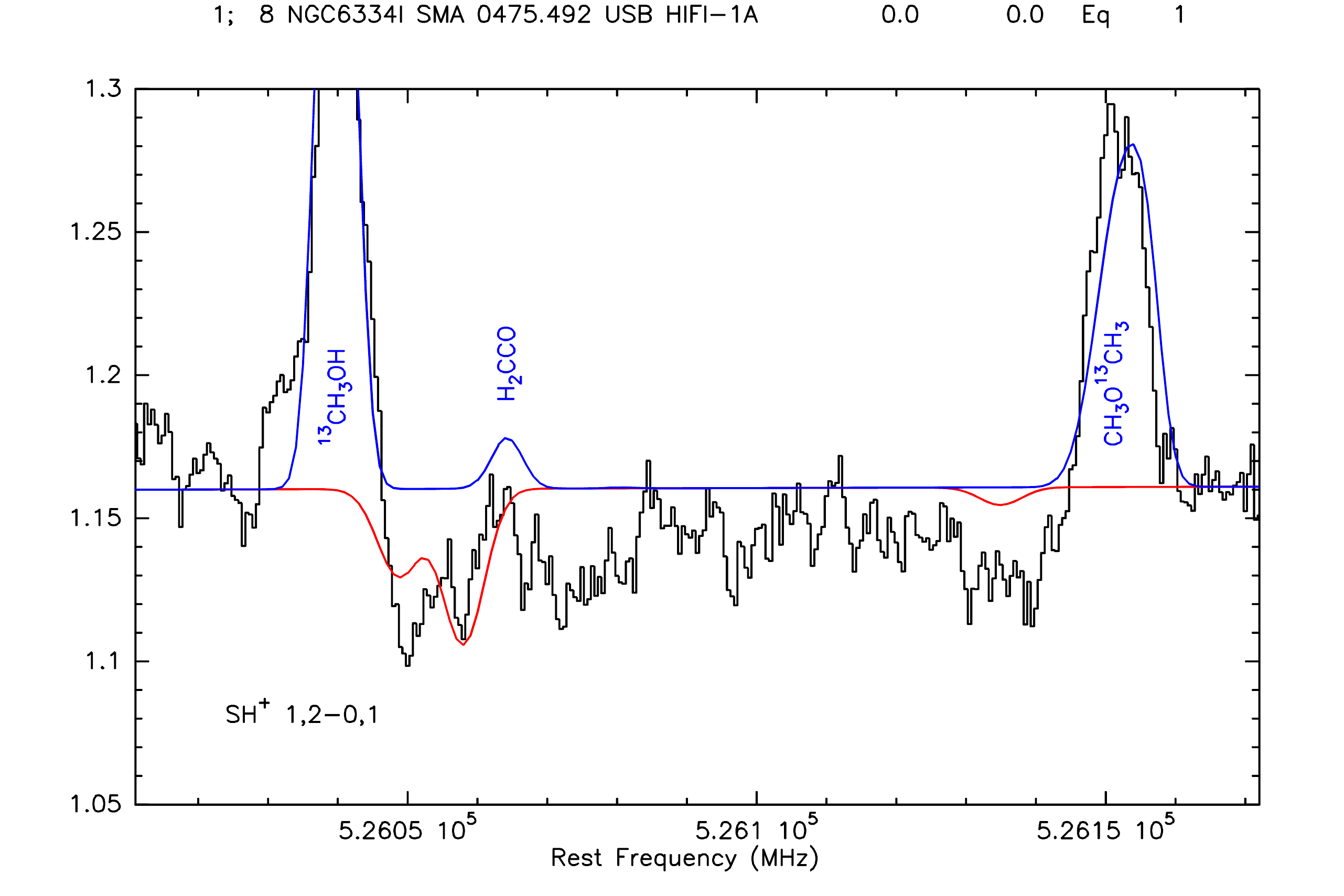}}
\subfloat[][\ce{H2S}]{\label{fig:h2s}\includegraphics[angle=270,width=0.65\textwidth]{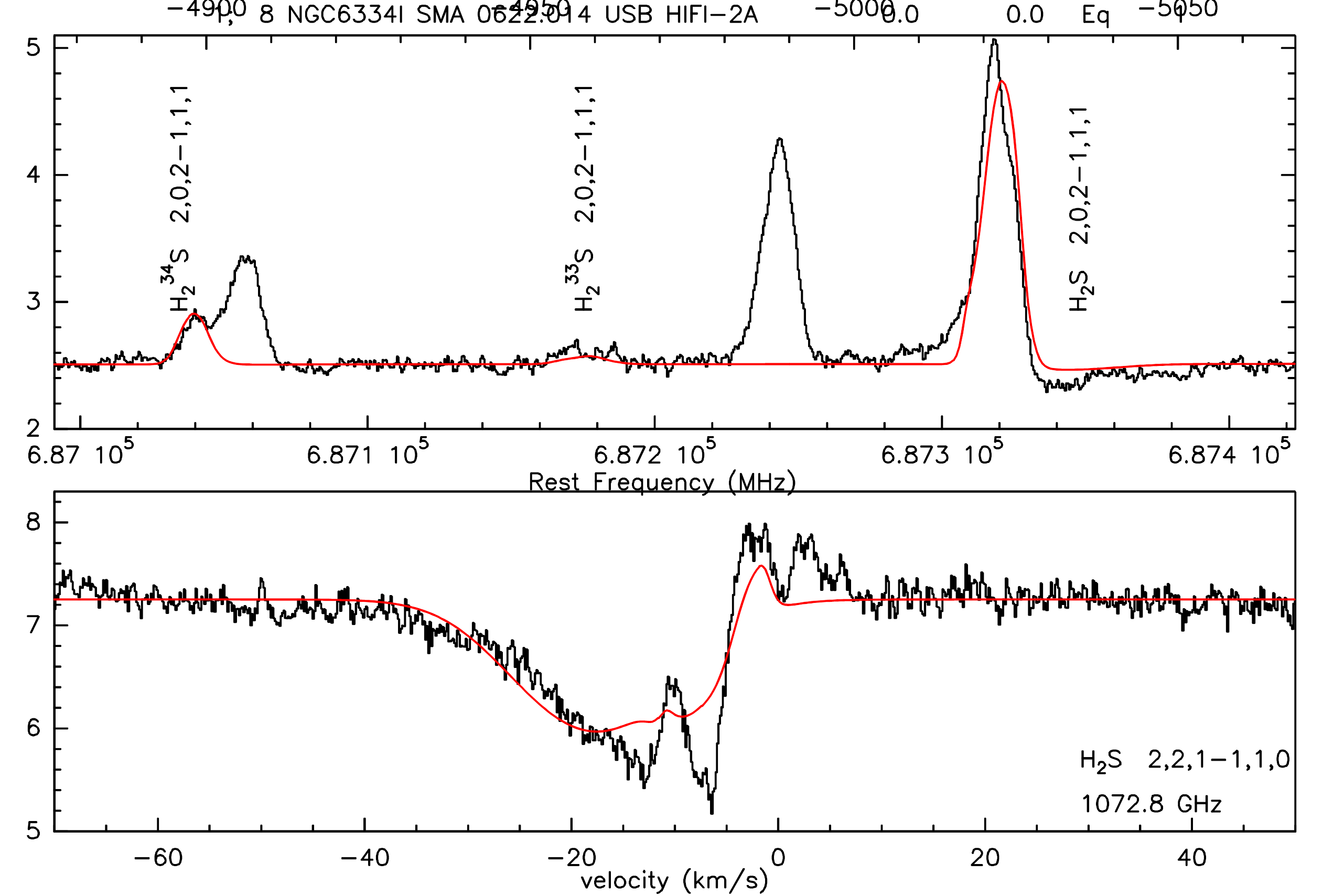}}
\caption{Continued}
\end{figure}
\end{landscape}

\begin{landscape}
\begin{figure}
\centering
\subfloat[][CS]{\label{fig:cs}\includegraphics[angle=270,width=0.65\textwidth]{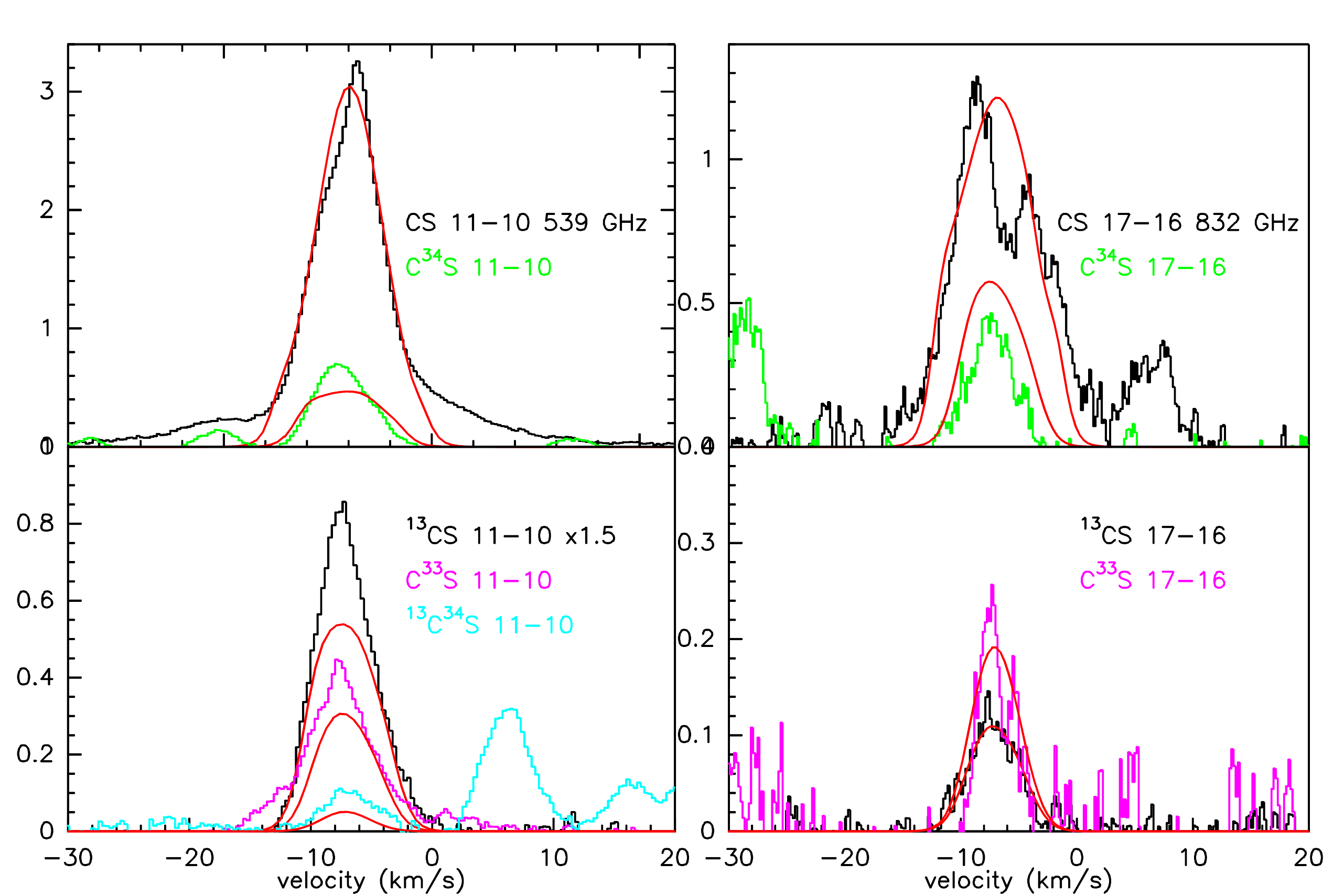}}
\subfloat[][NS]{\label{fig:ns}\includegraphics[angle=270,width=0.65\textwidth]{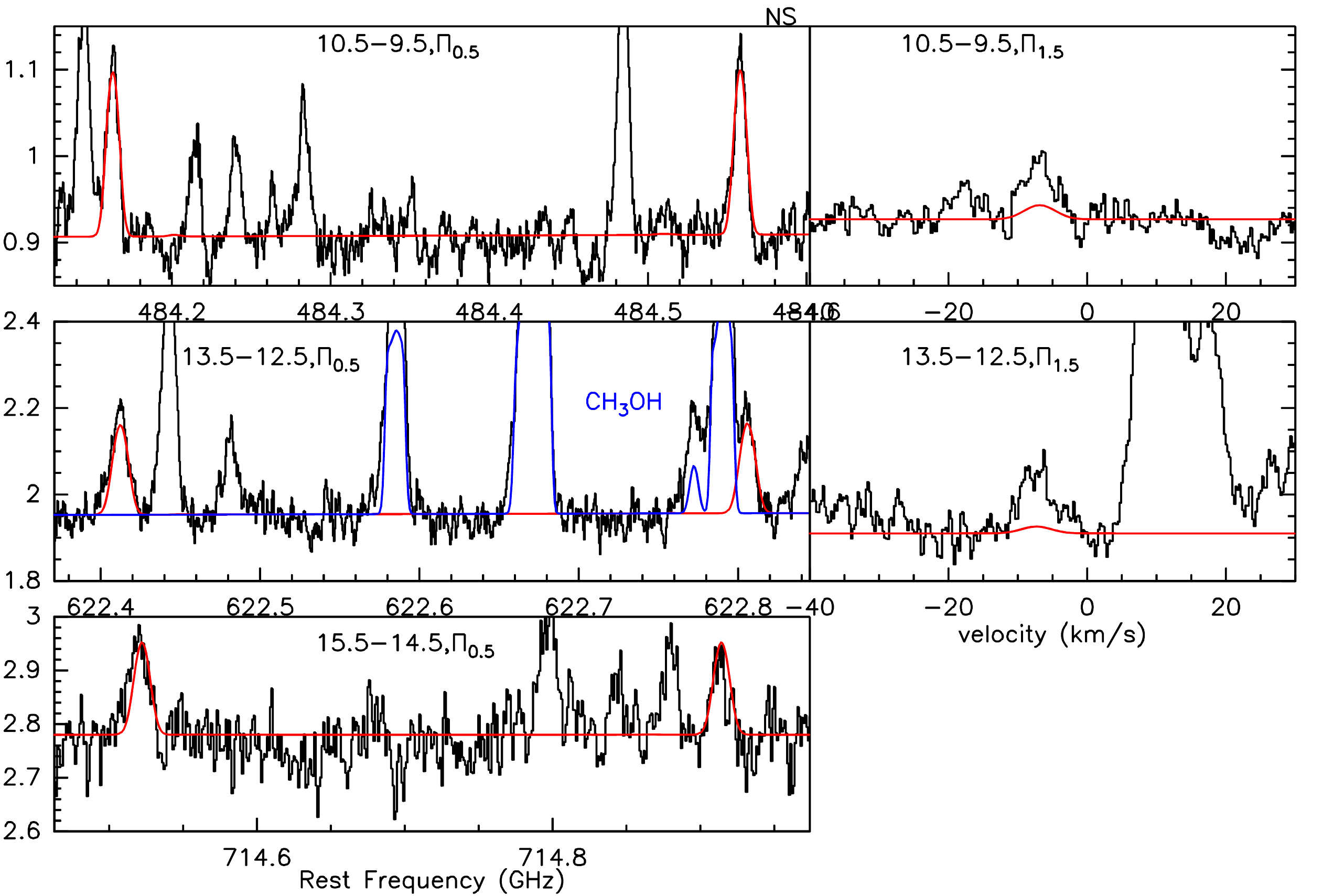}}\\
\subfloat[][OCS]{\label{fig:ocs}\includegraphics[angle=270,width=0.65\textwidth]{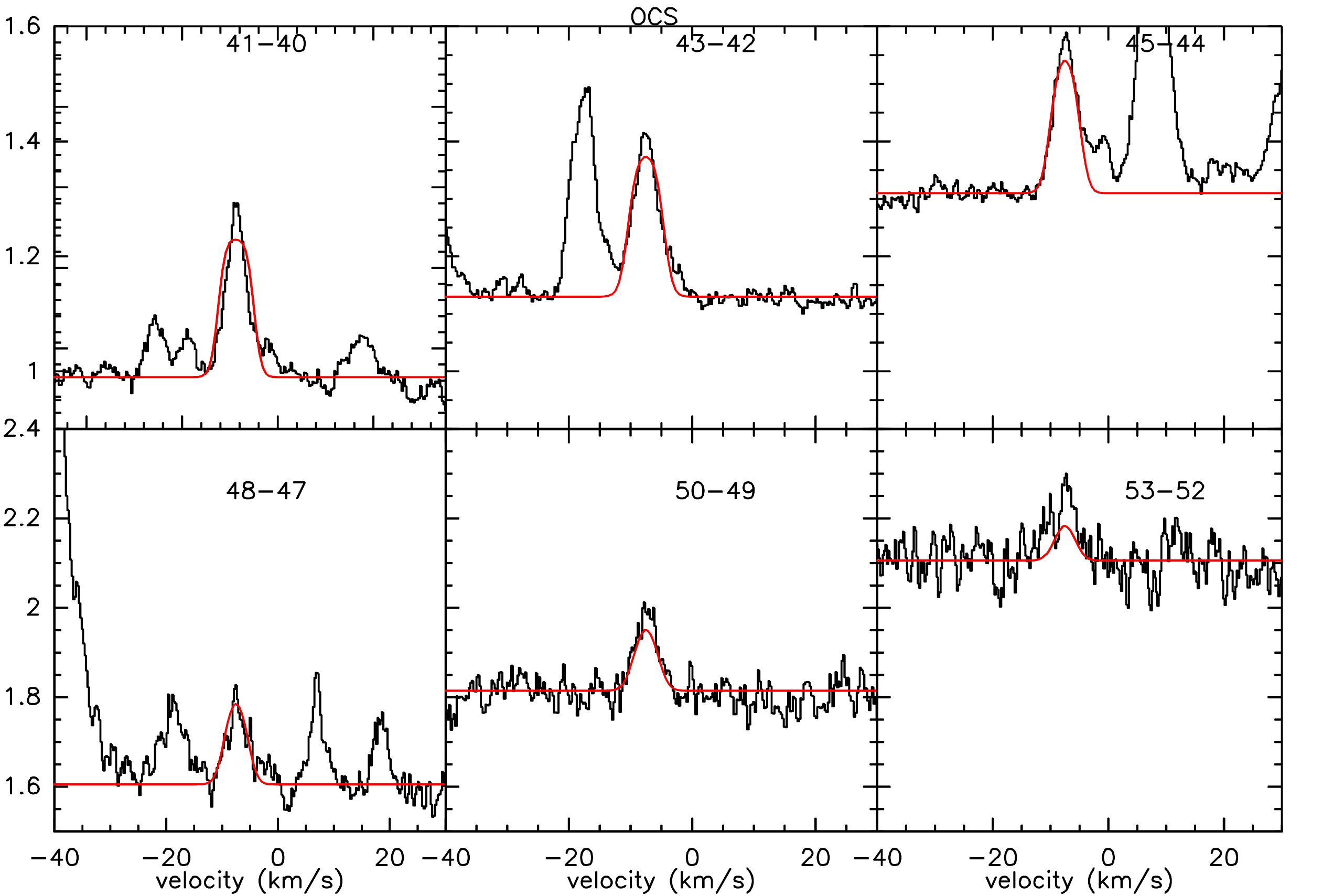}}
\subfloat[][SO]{\label{fig:so}\includegraphics[angle=270,width=0.65\textwidth]{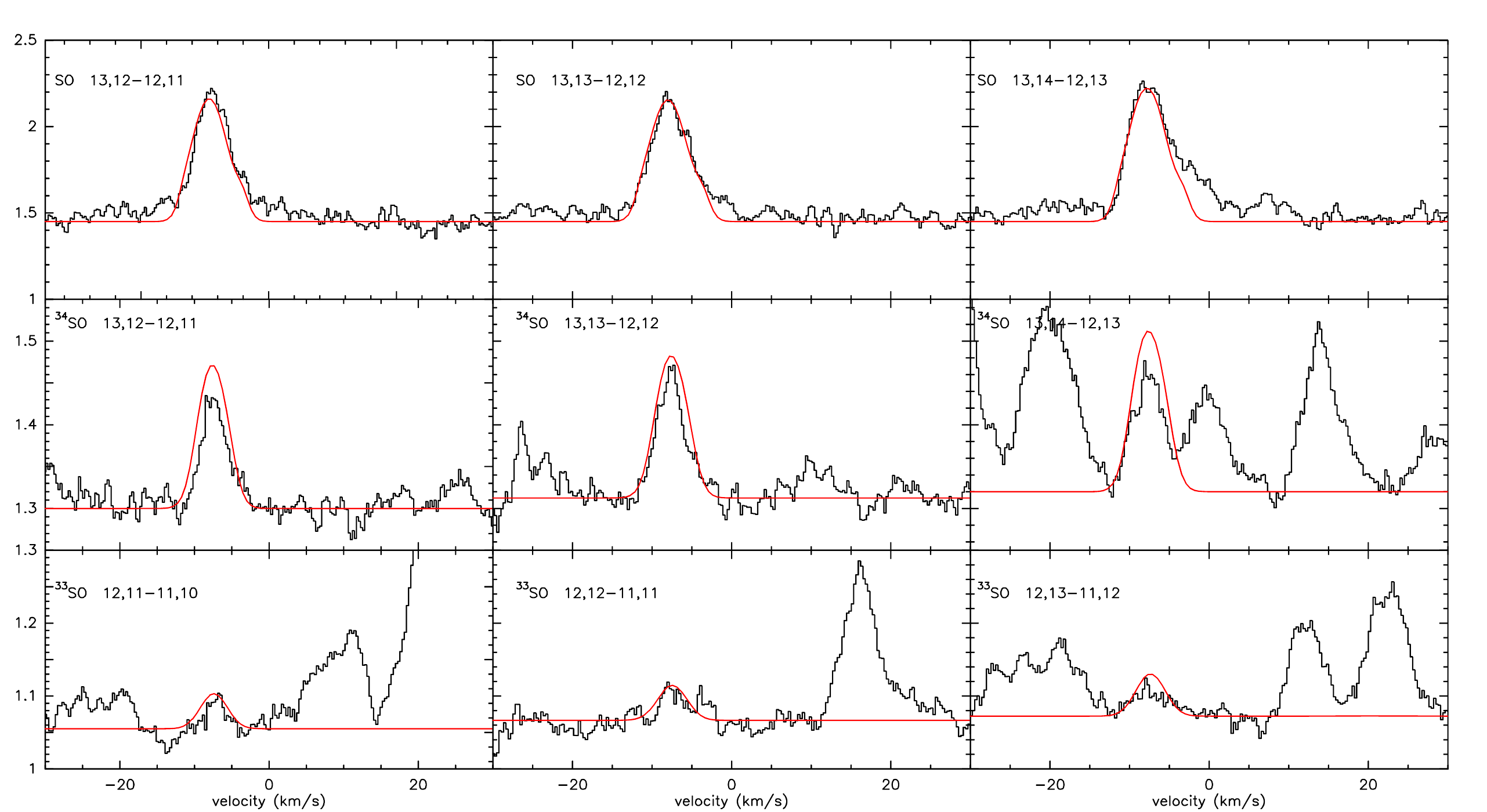}}
\caption{XCLASS fits of chemical species in the HIFI spectrum of NGC 6334I. Continued}
\label{HIFIfitsb}
\end{figure}
\end{landscape}

\begin{landscape}
\begin{figure}
 \ContinuedFloat
\centering
\subfloat[][\ce{SO2}]{\label{fig:so2}\includegraphics[angle=270,width=0.65\textwidth]{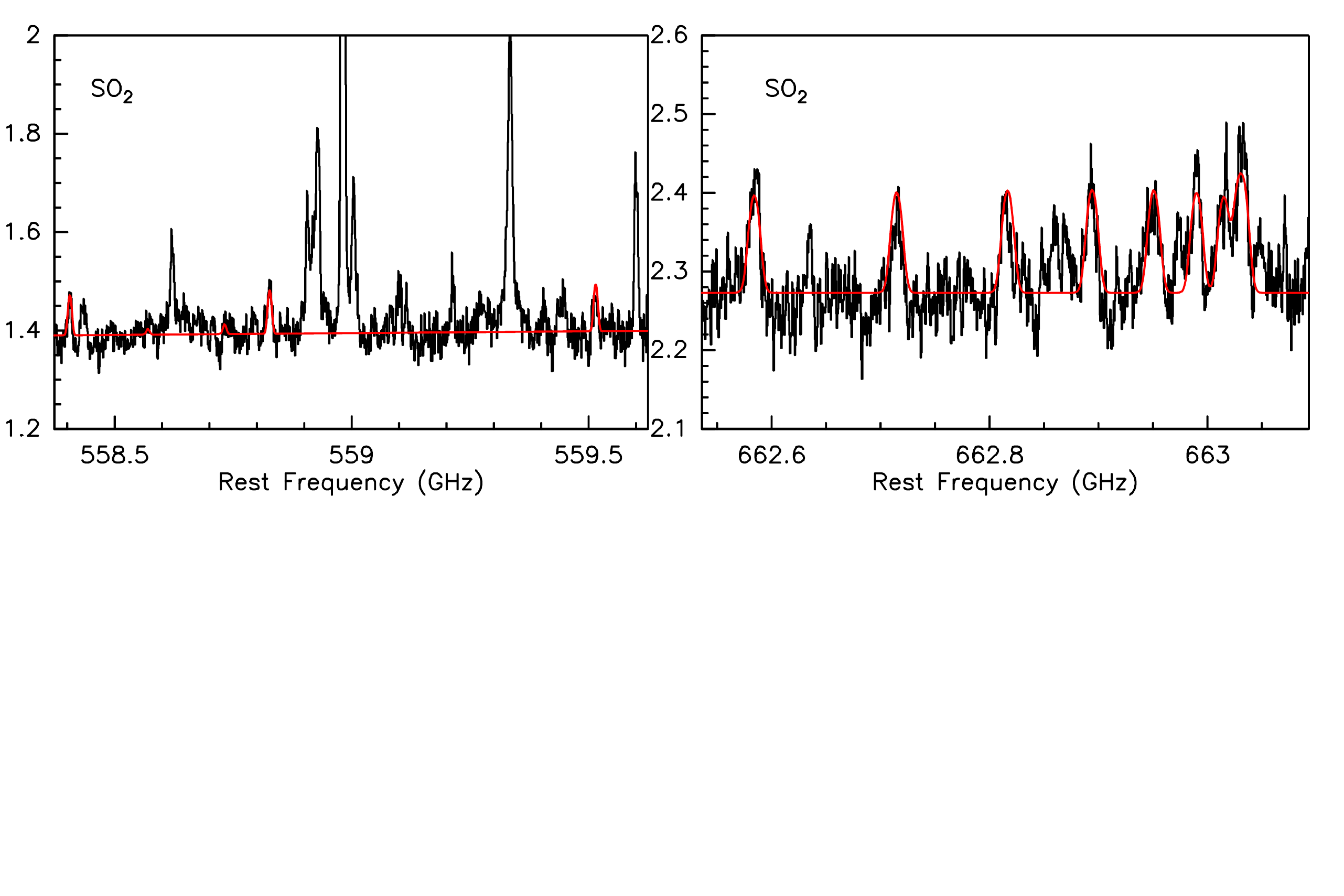}}
\subfloat[][\ce{H2CS}]{\label{fig:h2cs}\includegraphics[angle=270,width=0.65\textwidth]{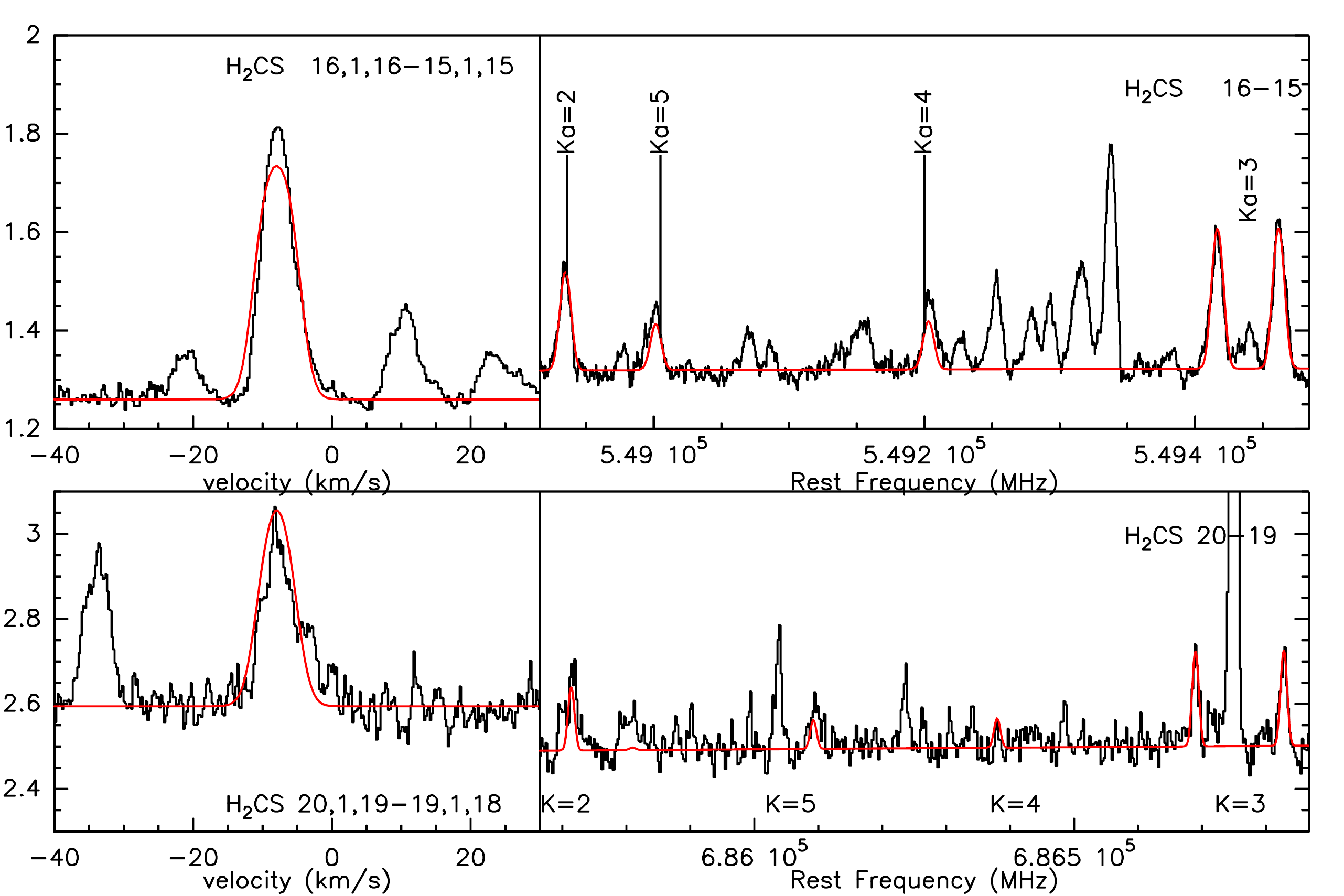}}\\
\subfloat[][\ce{CH3OH}]{\label{fig:m1}\includegraphics[angle=270,width=0.65\textwidth]{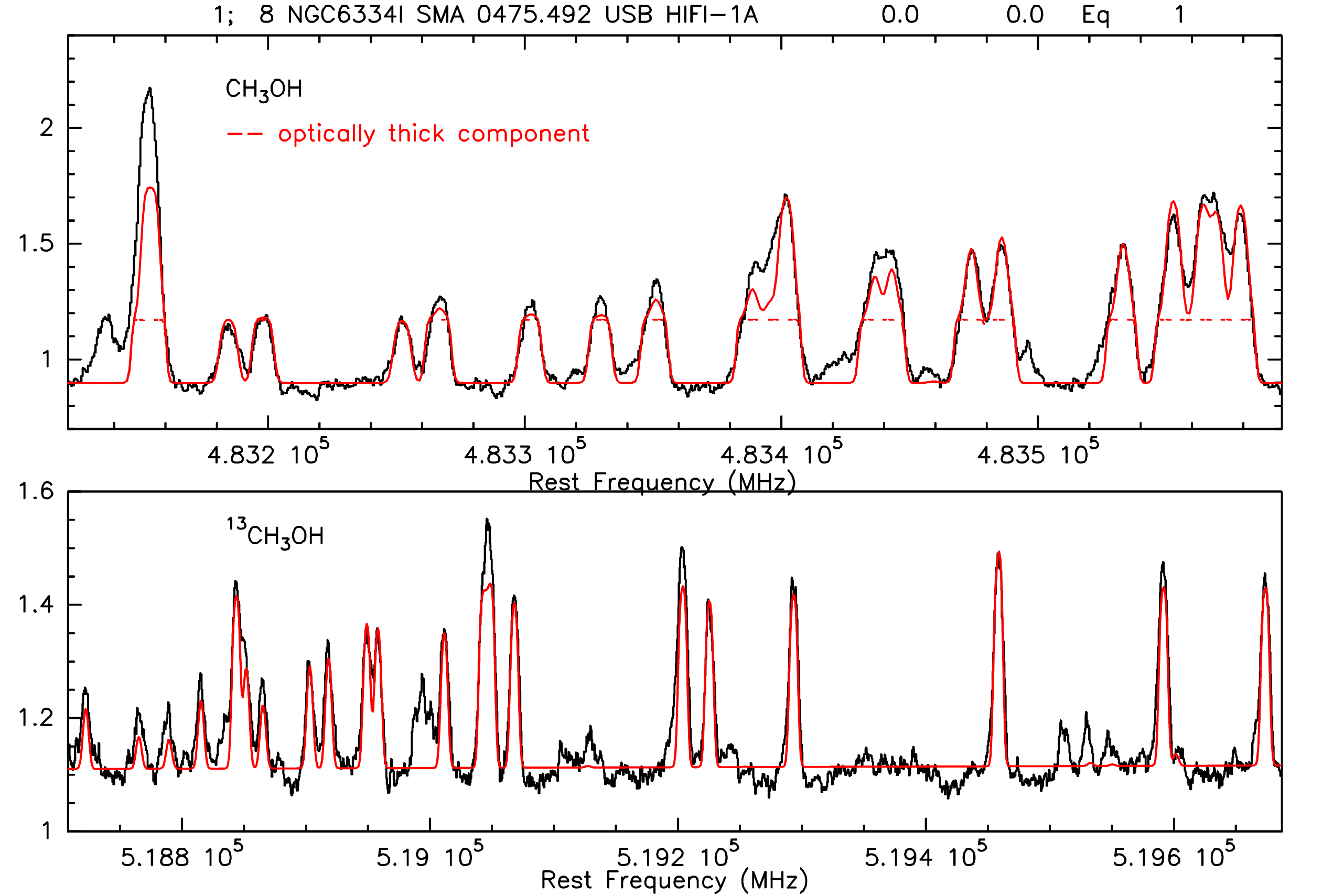}}
\subfloat[][\ce{CH3OH}]{\label{fig:m2}\includegraphics[angle=270,width=0.65\textwidth]{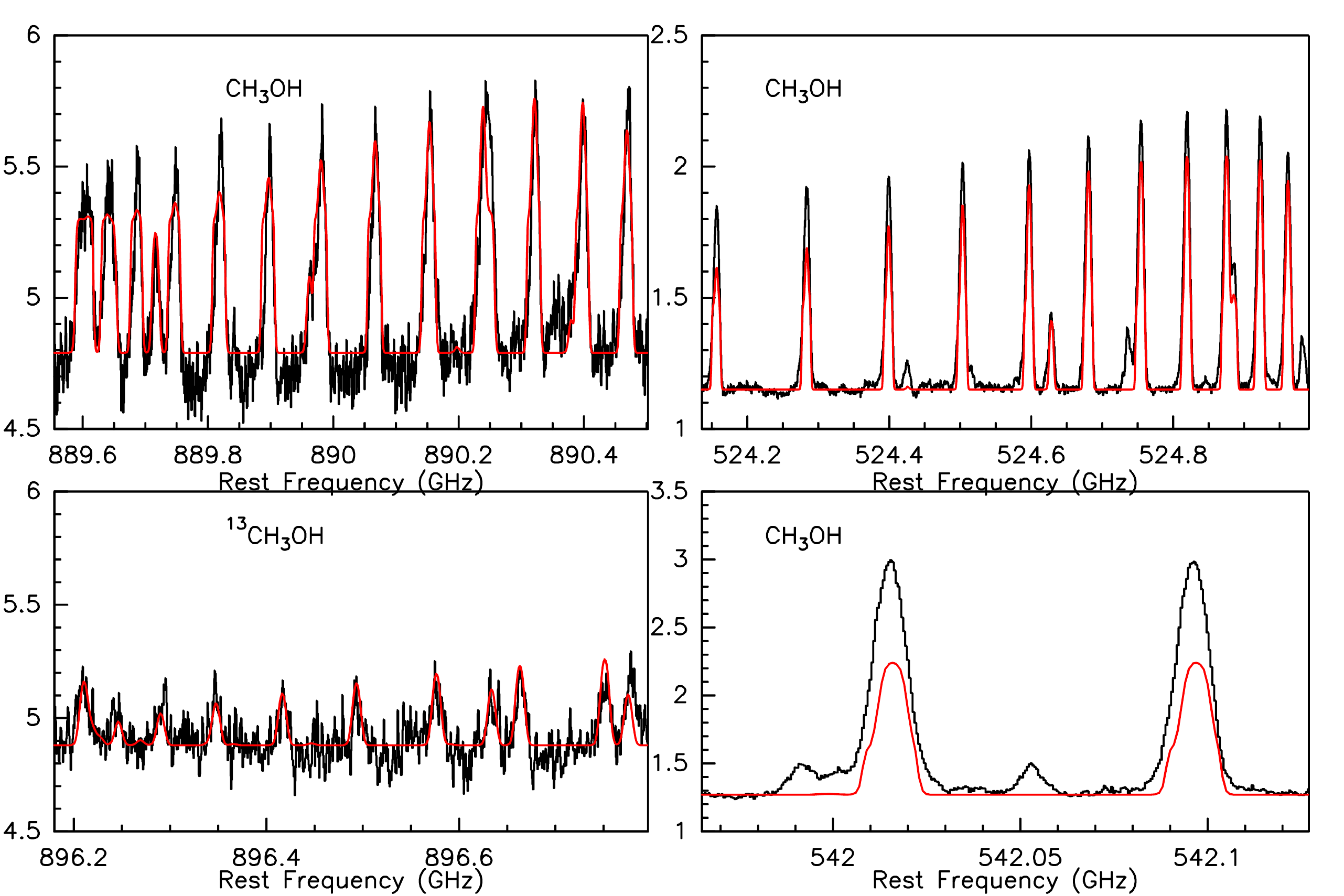}}
\caption{Continued}
\end{figure}
\end{landscape}

\begin{landscape}
\begin{figure}
 \ContinuedFloat
\centering
\subfloat[][\ce{CH3OCH3}]{\label{fig:dm}\includegraphics[angle=270,width=0.65\textwidth]{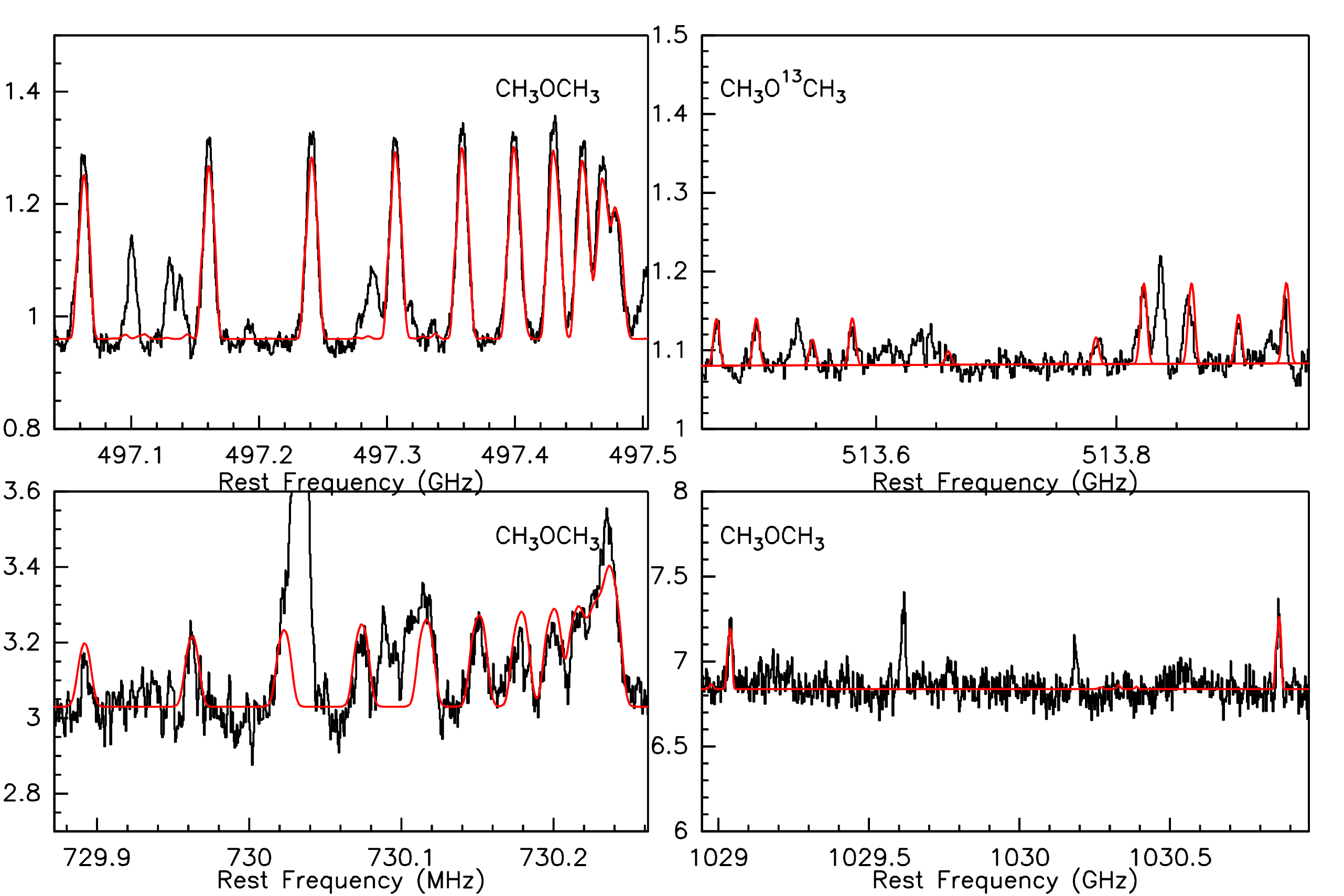}}
\subfloat[][\ce{CH3OCHO}]{\label{fig:mf}\includegraphics[angle=270,width=0.65\textwidth]{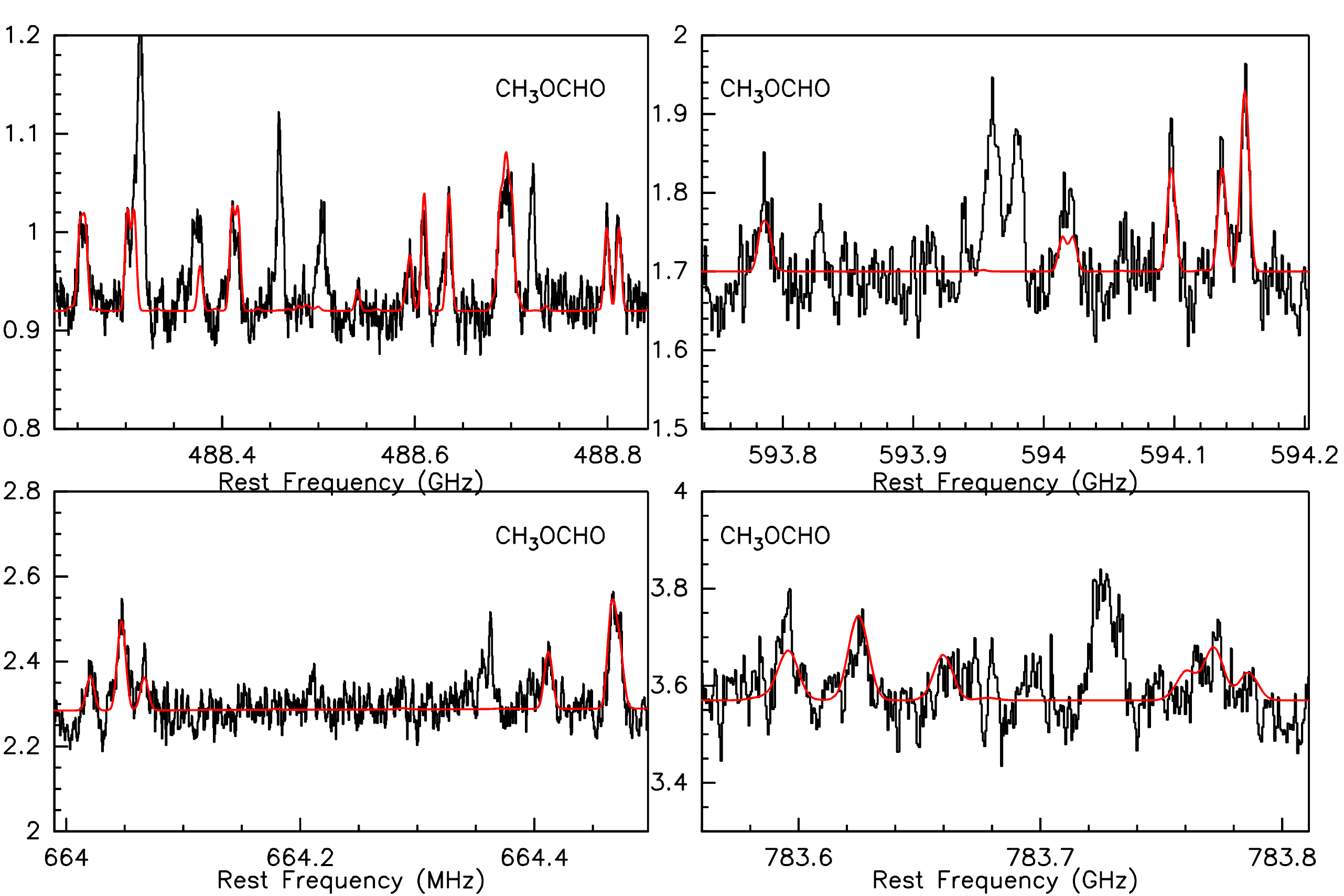}}\\
\subfloat[][\ce{C2H5OH}]{\label{fig:et}\includegraphics[angle=270,width=0.65\textwidth]{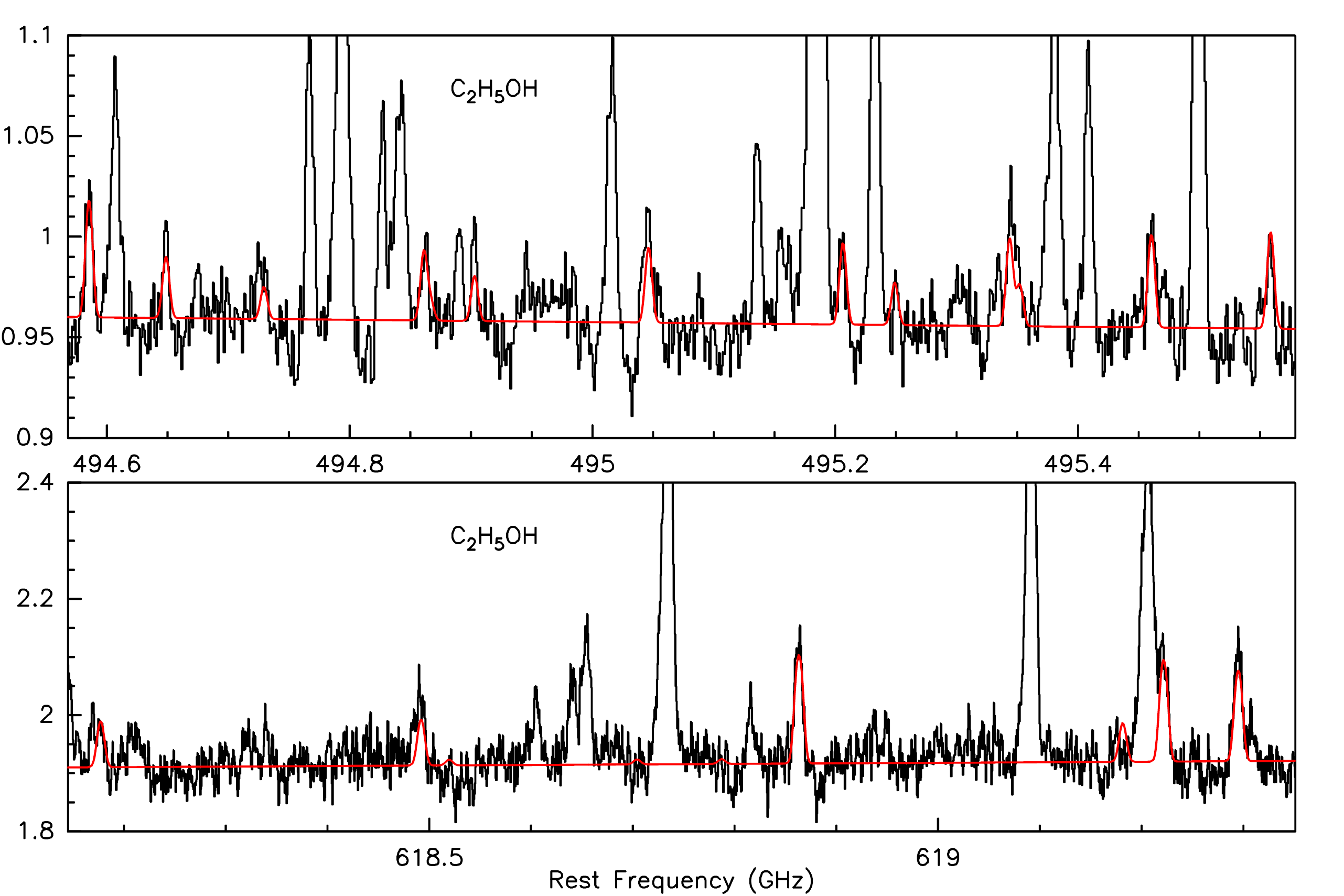}}
\subfloat[][\ce{H2CO}]{\label{fig:fa}\includegraphics[angle=270,width=0.65\textwidth]{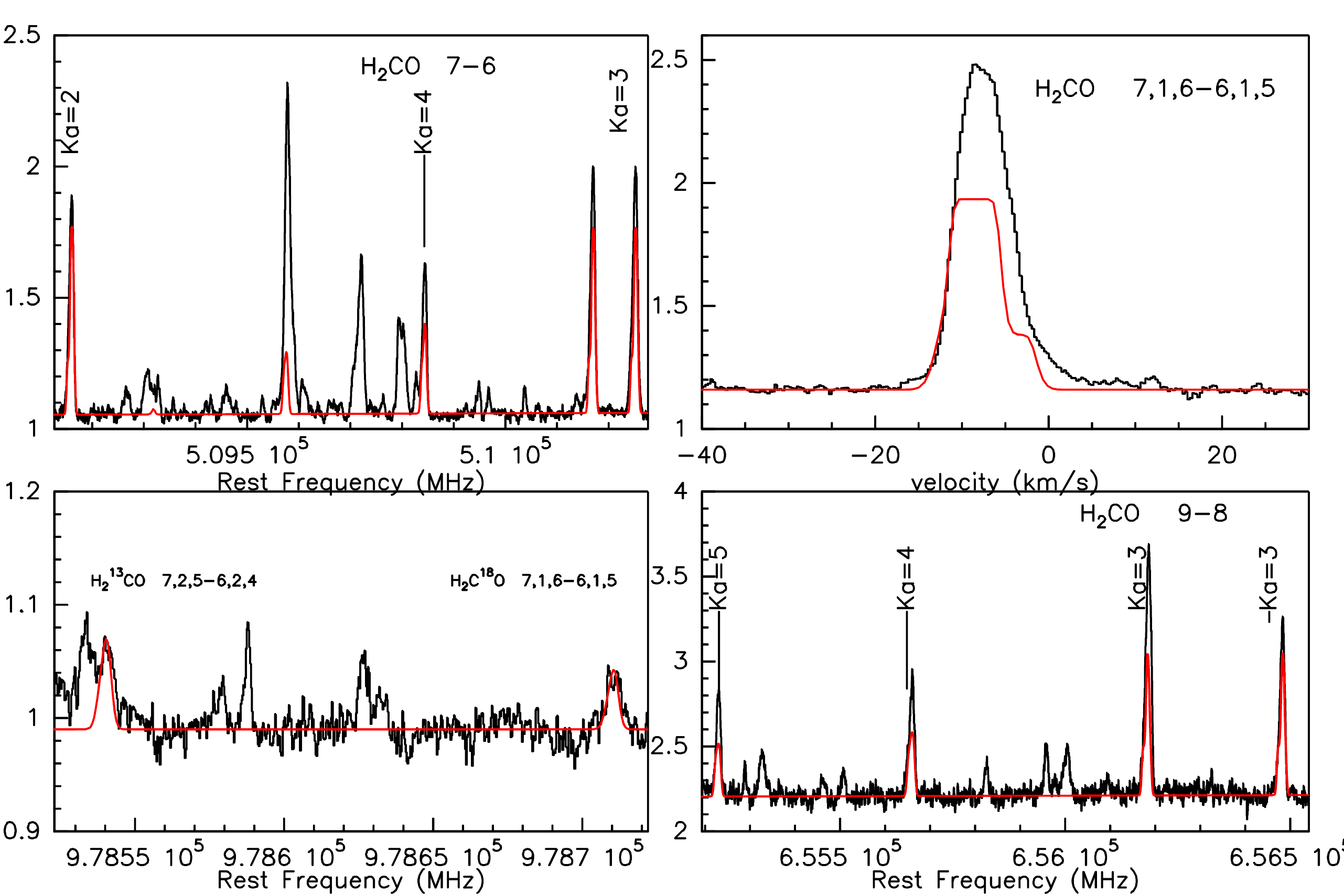}}
\caption{Continued}
\end{figure}
\end{landscape}

\begin{landscape}
\begin{figure}
 \ContinuedFloat
\centering
\subfloat[][\ce{CH3CN}]{\label{fig:mcn}\includegraphics[angle=270,width=0.65\textwidth]{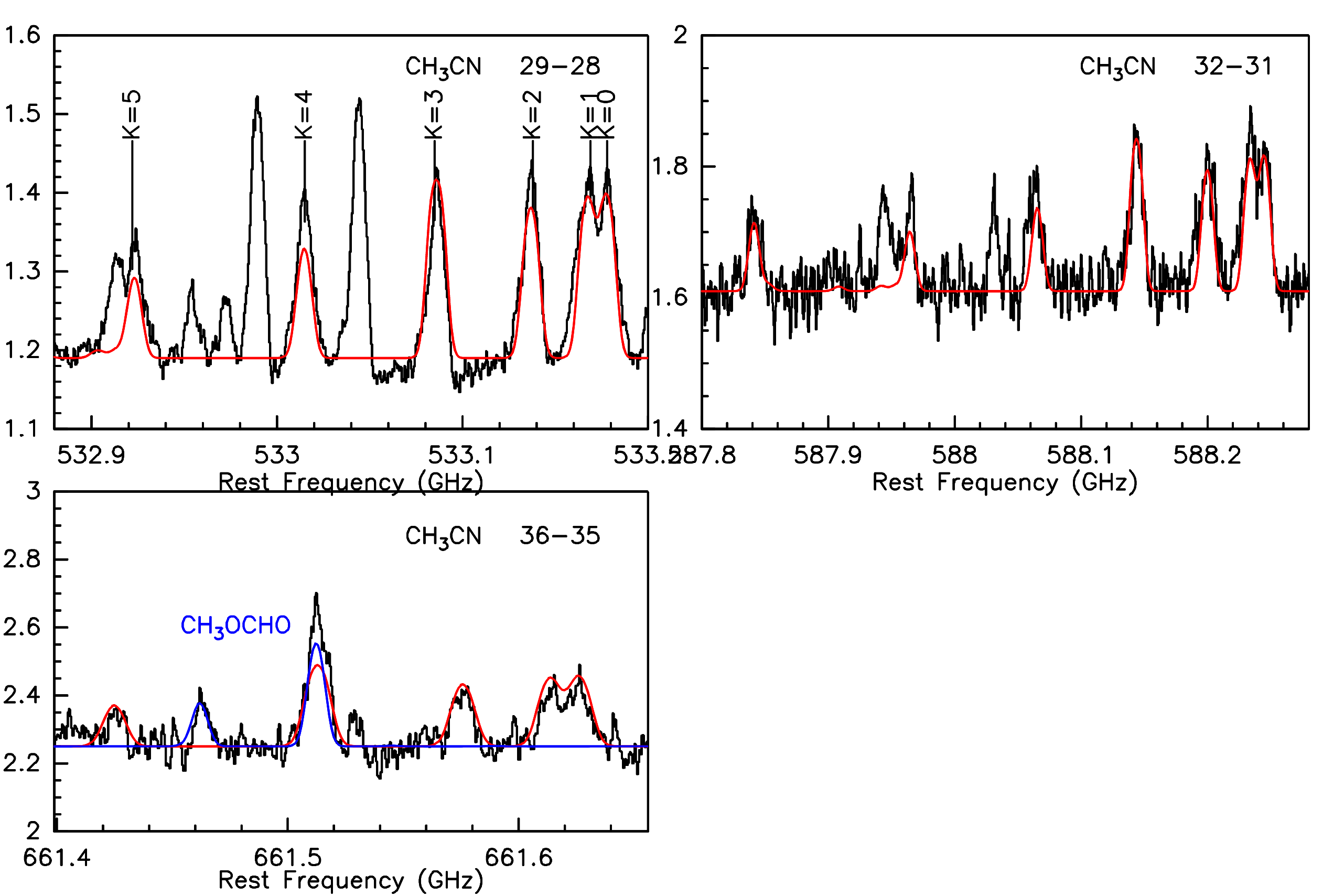}}
\subfloat[][\ce{C2H4O}]{\label{fig:eto}\includegraphics[angle=270,width=0.65\textwidth]{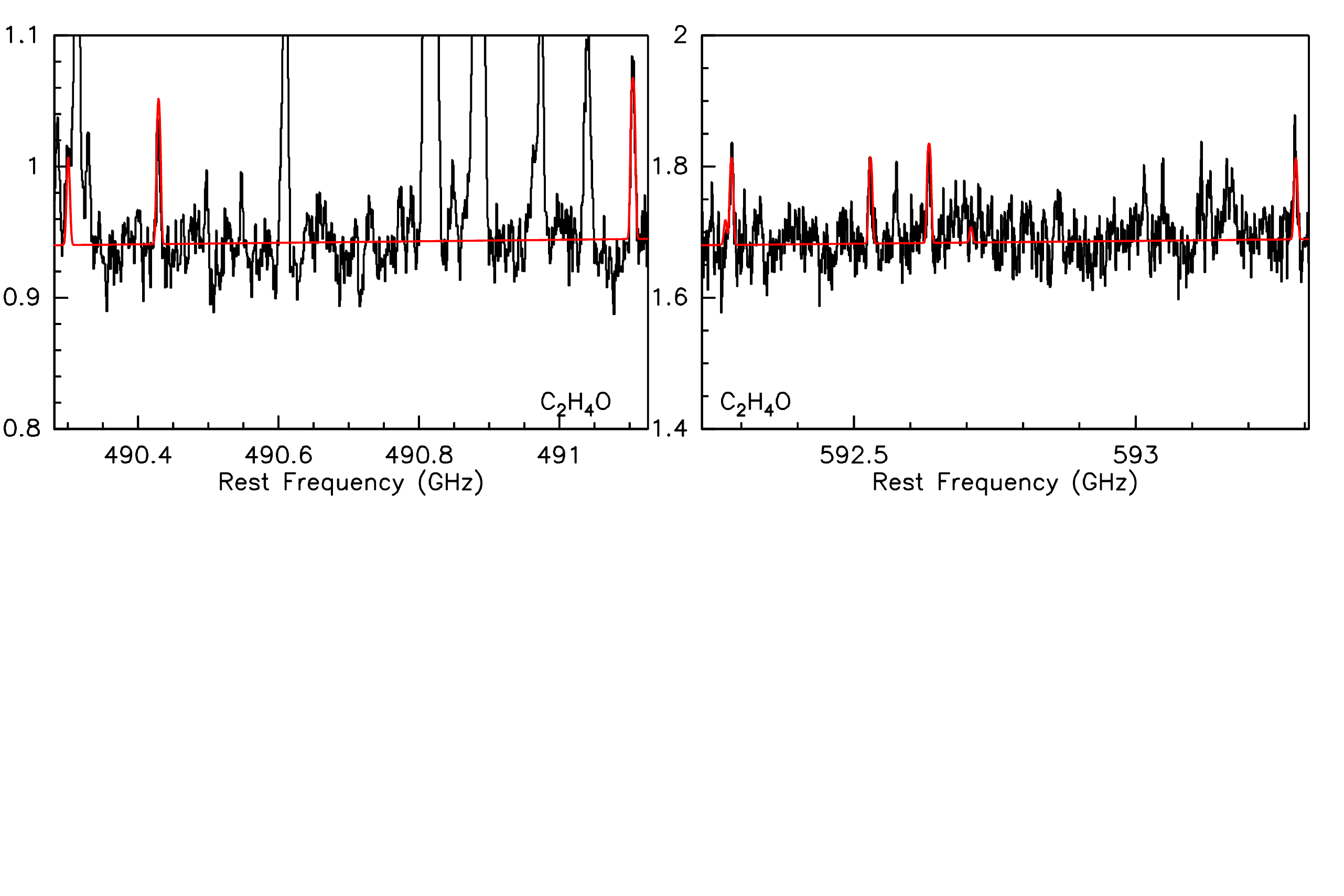}}\\
\subfloat[][\ce{H2CCO}]{\label{fig:h2cco}\includegraphics[angle=270,width=0.65\textwidth]{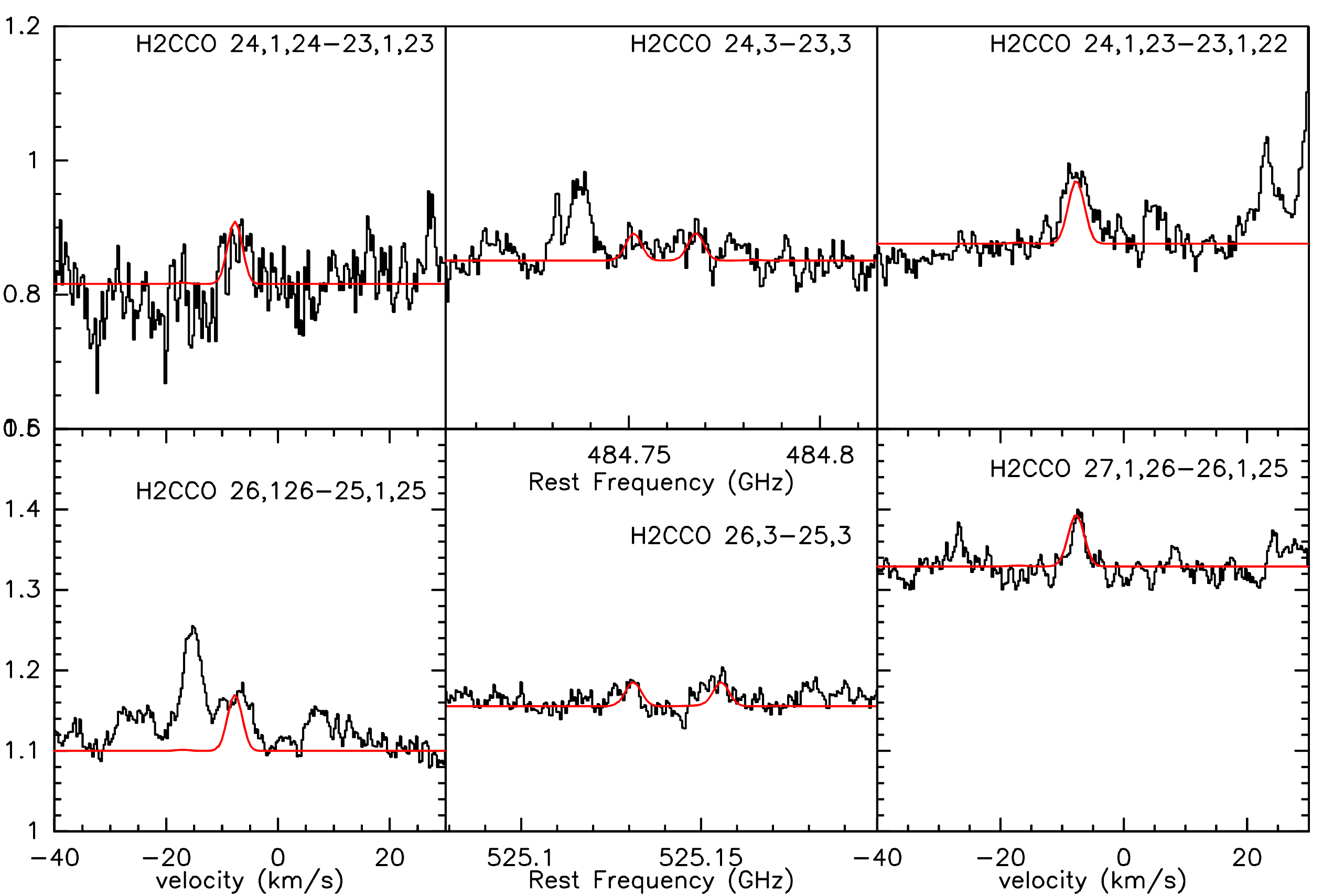}}
\subfloat[][\ce{HCO+}]{\label{fig:hco}\includegraphics[angle=270,width=0.65\textwidth]{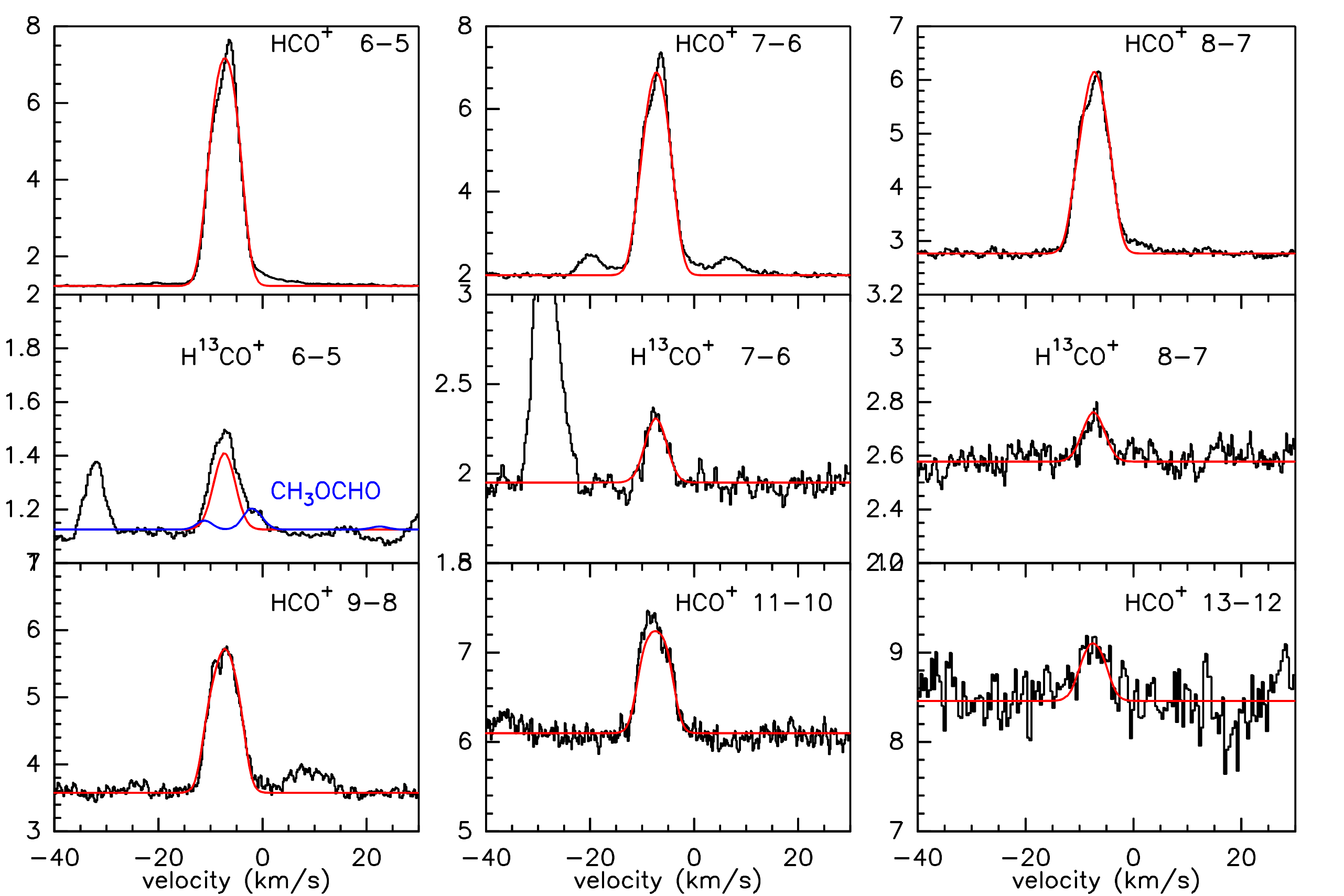}}
\caption{Continued}
\end{figure}
\end{landscape}

\begin{landscape}
\begin{figure}
 \ContinuedFloat
\centering
\subfloat[][\ce{CH2NH}]{\label{fig:ch2nh}\includegraphics[angle=270,width=0.65\textwidth]{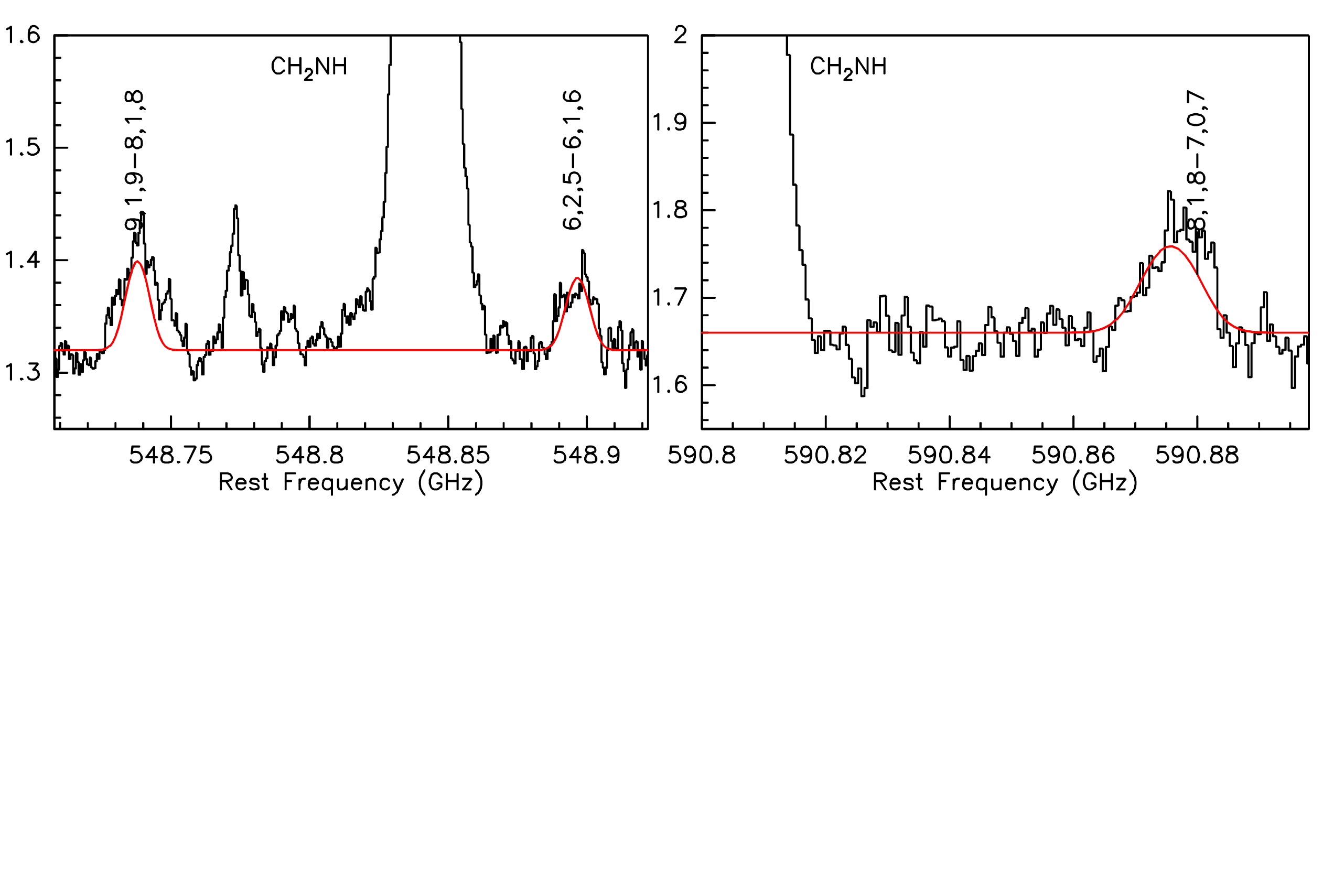}}
\subfloat[][HNCO]{\label{fig:hnco}\includegraphics[angle=270,width=0.65\textwidth]{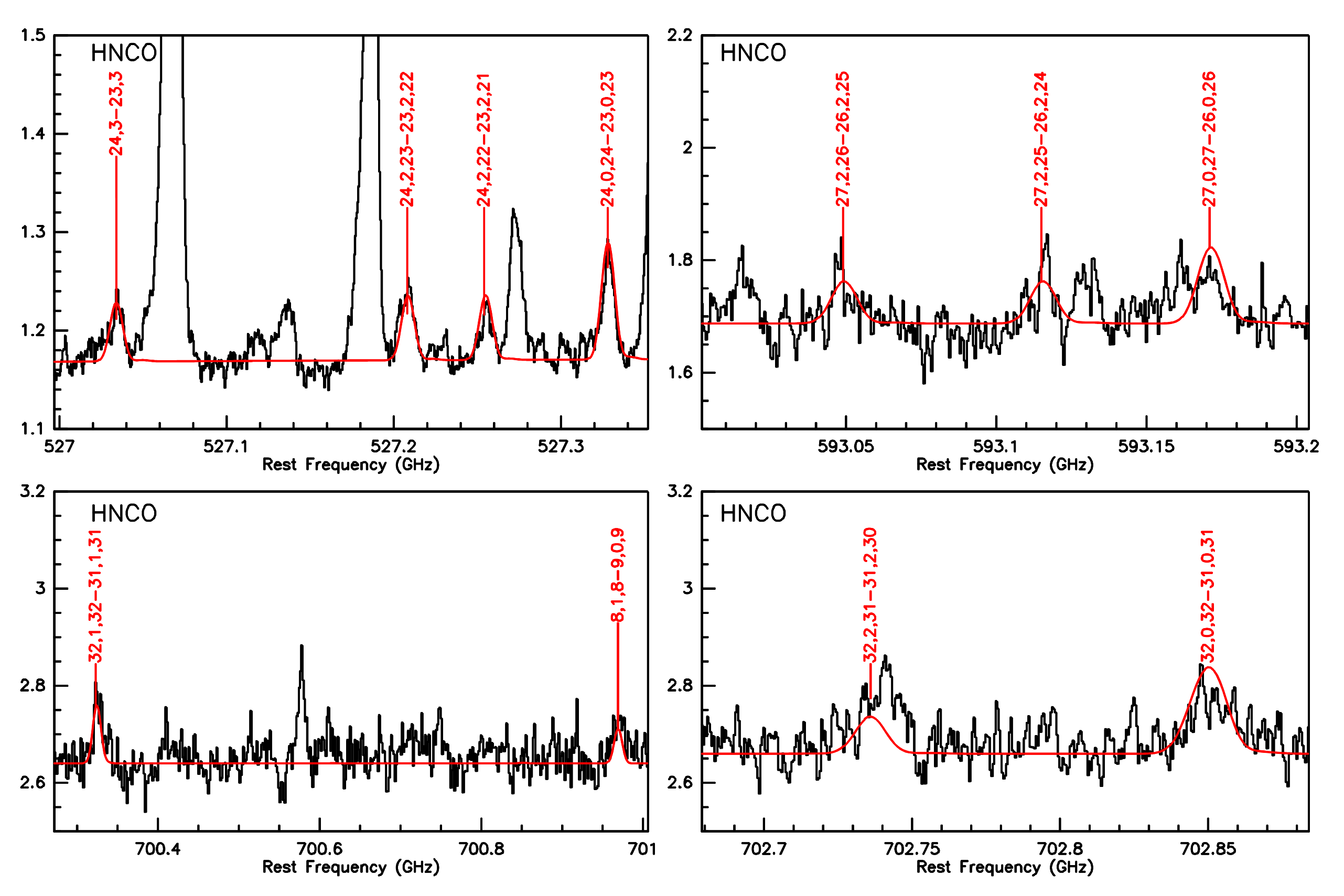}}\\
\subfloat[][\ce{NH2CHO}]{\label{fig:form}\includegraphics[angle=270,width=0.65\textwidth]{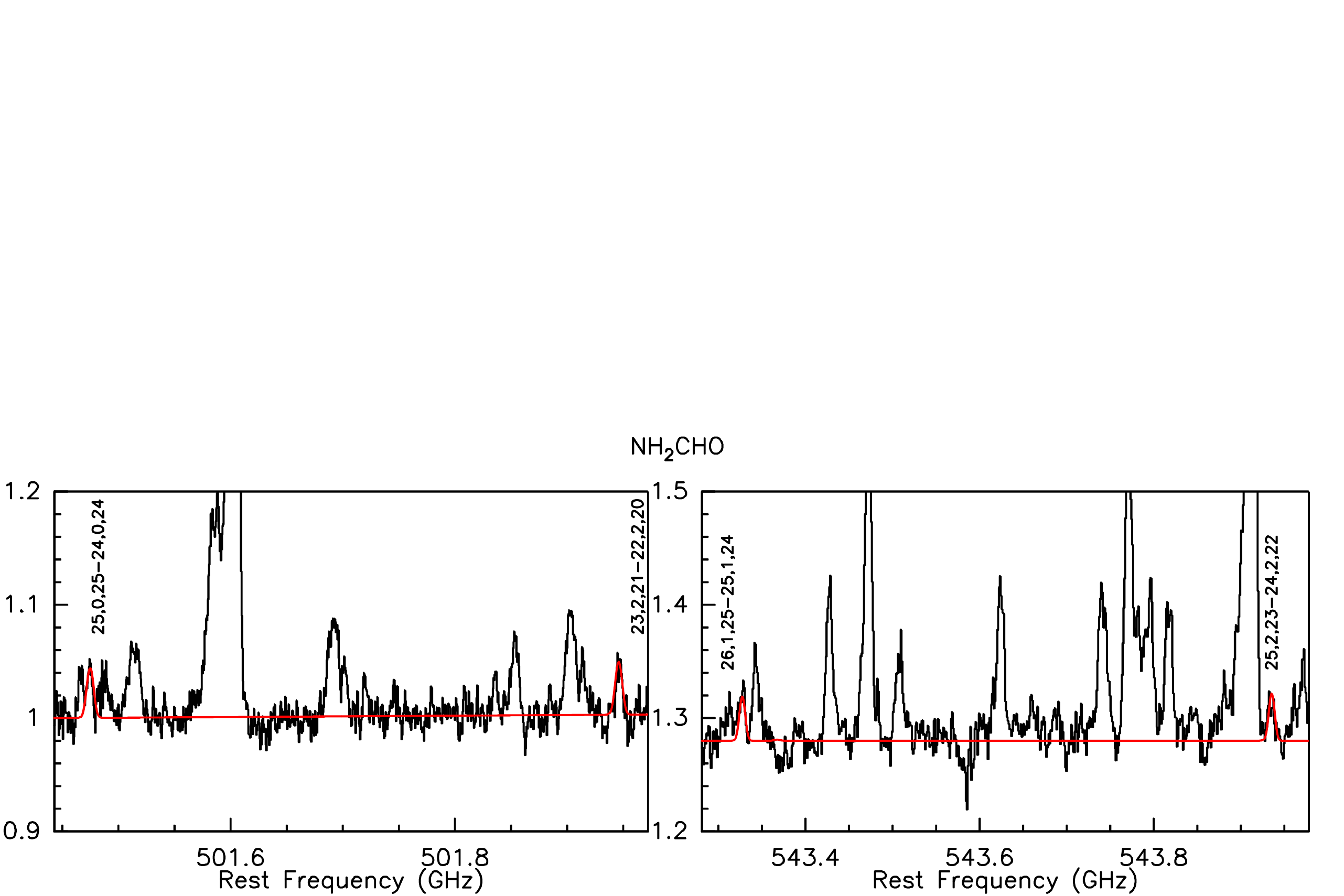}}
\subfloat[][\ce{HC3N}]{\label{fig:hc3n}\includegraphics[angle=270,width=0.65\textwidth]{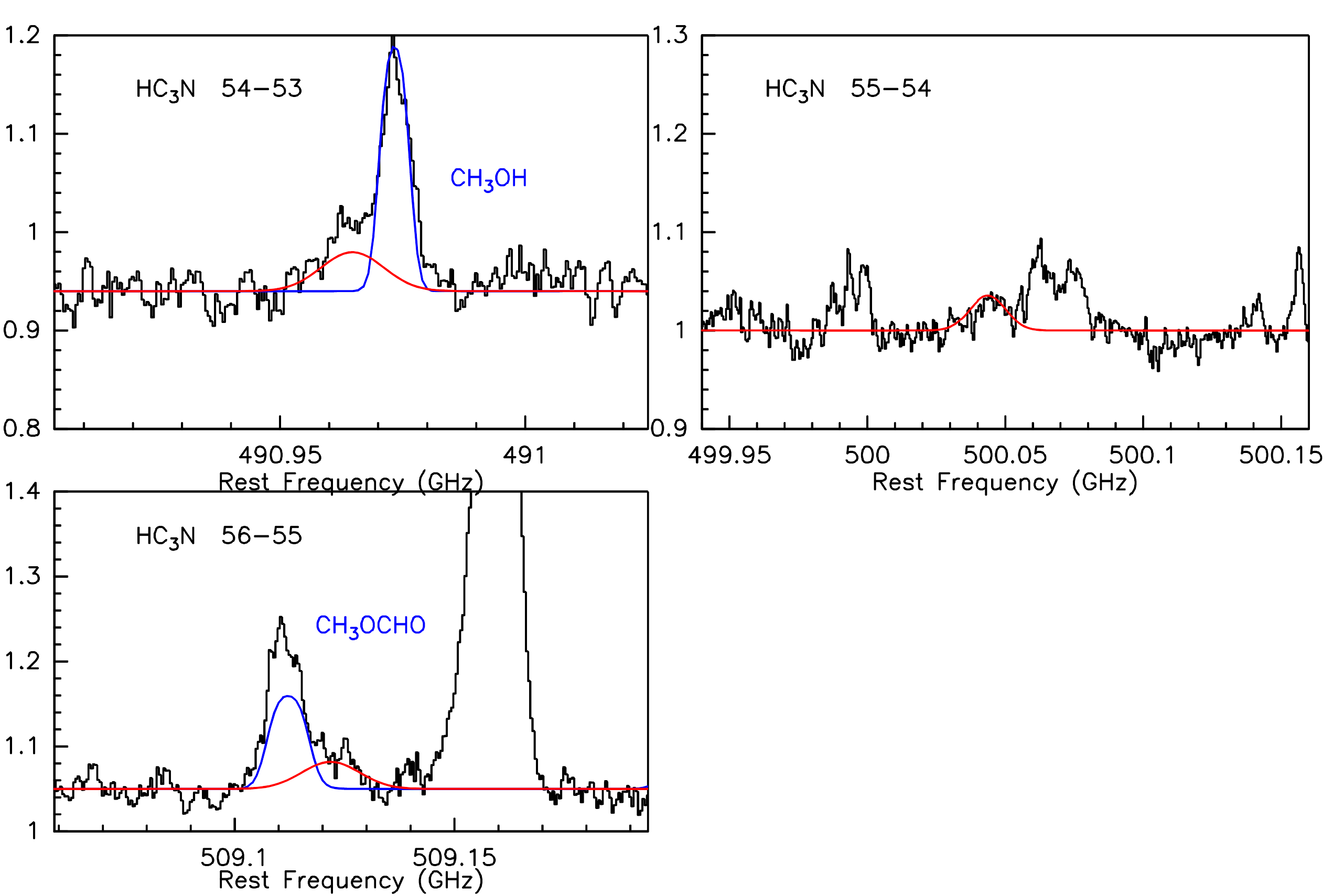}}
\caption{Continued}
\end{figure}
\end{landscape}

\end{appendix}
\end{document}